\documentclass[manuscript,screen,nonacm]{acmart}

\makeatletter
\@twosidefalse      
\@mparswitchfalse   
\renewcommand\@titlefont{\fontsize{14}{16}\selectfont\bfseries} 
\makeatother

\AtBeginDocument{%
  }

\renewcommand\footnotetextcopyrightpermission[1]{} 
\usepackage{array}
\usepackage{xcolor}
\usepackage{eso-pic}

\begin{document}

\title{Exploring Learners' Expectations and Engagement When Collaborating with Constructively Controversial Peer Agents}
\author{Thitaree Tanprasert}
\email{tt1996@cs.ubc.ca}
\orcid{0000-0002-5606-2433}
\affiliation{%
  \institution{University of British Columbia}
  \city{Vancouver}
  \state{British Columbia}
  \country{Canada}
}

\author{Young-ho Kim}
\email{yghokim@younghokim.net}
\orcid{0000-0002-2681-2774}
\affiliation{%
  \institution{NAVER AI Lab}
  \city{Seongnam}
  \state{Gyeonggi}
  \country{Republic of Korea}
}

\author{Sidney Fels}
\email{ssfels@ece.ubc.ca}
\orcid{0000-0001-9279-9021}
\affiliation{
  \institution{University of British Columbia}
  \city{Vancouver}
  \state{British Columbia}
  \country{Canada}
}
\author{Dongwook Yoon}
\email{yoon@cs.ubc.ca}
\orcid{0000-0002-7838-8311}
\affiliation{%
  \institution{University of British Columbia}
  \city{Vancouver}
  \state{British Columbia}
  \country{Canada}
}

\renewcommand{\shortauthors}{Tanprasert et al.}

\begin{abstract}
  Peer agents can supplement real-time collaborative learning in asynchronous online courses. Constructive Controversy (CC) theory suggests that humans deepen their understanding of a topic by confronting and resolving controversies. This study explores whether CC’s benefits apply to LLM-based peer agents, focusing on the impact of agents’ disputatious behaviors and disclosure of agents’ behavior designs on the learning process. In our mixed-method study (n=144), we compare LLMs that follow detailed CC guidelines (regulated) to those guided by broader goals (unregulated) and examine the effects of disclosing the agents’ design to users (transparent vs. opaque). Findings show that learners' values influence their agent interaction: those valuing control appreciate unregulated agents' willingness to cease push-back upon request, while those valuing intellectual challenges favor regulated agents for stimulating creativity. Additionally, design transparency lowers learners' perception of agents’ abilities. Our findings lay the foundation for designing effective collaborative peer agents in isolated educational settings.
\end{abstract}



\maketitle

\section{Introduction}
Asynchronous online courses, such as Massive Open Online Courses (MOOCs), are popular for their accessibility and variety, allowing isolated learners to study at their own pace and convenience \cite{castillo2015moocs, lowenthal2020creating}. However, collaborative learning, where two or more learners with approximately equal knowledge mutually influence each other's learning \cite{o2013introduction}, remains a challenge on such platforms. Collaborative learning in traditional settings introduces many social benefits, including knowledge co-construction from negotiating multiple perspectives \cite{kuhn2011dialogic, nokes2012effect}. Peer AI agents offer a promising supplement for isolated learners by emulating the dynamics of real-time collaborative human partners. And with the advancement of LLMs, collaborative peer AI agents can behave more like human peers, assuming a complex persona and strategies during collaboration to increase cognitive, behavioral, and emotional engagement as well as and observational learning \cite{dhillon2024shaping, guo2023effects, liu2024peergpt}. These benefits lead to measurable improvements in individual learning quality and outcomes, including higher exam scores \cite{warsah2021impact}, deeper understanding of the knowledge \cite{nora2002collaborative}, and higher critical thinking skills \cite{yamarik2007does}. However, a peer agent who presents different perspectives in a knowledge co-construction process remains a challenge, as it requires a nuanced balance of contradiction and cooperation---a trait not inherent in LLMs, which typically favor conformity and consensus \cite{zhang2023exploring}. Developing such nuanced behaviors is necessary to enhance the peer agent's effectiveness for isolated learners.

According to Constructive Controversy (henceforth abbreviated CC) theory, the essence of these interactive discussions lies in their ability to introduce diverse perspectives, leading to conflicts that require resolution. The CC framework is inherently positioned as a "middle ground" strategy, seeking to balance the constructive aspects of cooperation (concurrence seeking) with the intellectual stimulation of contradiction (debate) \cite{johnson2015constructive}. It posits that confronting and resolving these conflicts in a constructive way allows learners to synthesize a deeper understanding of the subject matter \cite{johnson2000constructive}. Intricately connected to this process are learners' characteristics such as epistemic curiosity and attitudes towards controversy \cite{lowry1981effects}, which underscore the necessity for interaction settings that can accommodate a range of personal dispositions. However, asynchronous interactions, such as discussion forums, often lack the immediacy and dynamic engagement, that are necessary to fully address and support these characteristics \cite{kemp2014face, staubitz2015collaborative, yamagata2014blending}. Previous research also shows that, for CC-based learning, synchronicity plays an important role in fostering knowledge co-construction---a process of where individuals share their knowledge and collectively build upon each other's knowledge, resulting in new and better knowledge. \cite{roseth2011effects, scardamalia2002collective}.

There are several design considerations for applying CC to peer agents. Firstly, the perception of behaviors of AI agents can be influenced by many factors, from personality traits \cite{zhou2019trusting} to mental models of AI agent's capabilities \cite{chakraborti2018algorithms}. Due to the complex nature of the human-AI relationships, research has found that, in certain contexts, AI exhibiting contradictory behaviors may be received more negatively by users than when human counterparts demonstrate such behaviors \cite{grundke2024aversion, kim2021ai}. As suggested by the Computer as Social Actor theory \cite{bialkova2024core}, an effective way for a peer AI agent to execute CC principles may need to be modified from the behavioral guidelines of human peers. Specifically, in the context of isolated learning, this consideration aligns with the tension between instructional scaffolding, where learners progress under guidelines \cite{jumaat2014instructional}, and self-regulation, where learners have the autonomy to design their learning process \cite{de2015learner}. Therefore, in this paper, we will distinguish the behavioral mechanisms as \emph{regulated} versus \emph{unregulated}, where regulation refers to the peer agent's continuous adherence to CC behavioral guidelines during the collaboration process.

Another factor that can potentially make a difference to the effectiveness of a CC-based peer agent is the transparency of the agents' behavioral design. According to the Social Response theory, an AI agent's self-disclosure can impact the user's perceived roles and social interactions with the agent \cite{nass2000machines}, as it establishes the AI agent as a social actor \cite{tsumura2023influence}, fosters trust \cite{angerschmid2022fairness}, and improves effectiveness \cite{hepenstal2023impact} as shown in previous empirical studies. Yet, recent research has shown that transparency can also have negative effects, particularly where the AI agent explains their decision-making process \cite{schmidt2020transparency}. These contradicting results show that there could be many factors interacting with transparency, such as the AI behaviors that are disclosed \cite{tulli2019effects} and the user's values \cite{yam2023cultural}. In the context of our study, where assertive CC behaviors may be unwelcome, the transparency of their behavior design can position the peer agent as a "devil's advocate", inducing the learner's acceptance of the contradictory behaviors and prompting the learner to utilize the controversy more strategically \cite{chiang2024enhancing, seville2000can}. 

This paper aims to answer the research question: \emph{What are the considerations for designing an effective, LLM-powered collaborative peer agent, following the CC principles?} Specifically, we study the interplay between the peer agent's behavior mechanisms (regulated vs. unregulated) and disclosure of the behaviors in the following foci \footnote{In the rest of this paper, we will use "behavior" to describe what the agent (or human) does, and we will use "behavior mechanism" to refer to the conceptual framework for how we implement the behaviors in the agent.}:
\begin{itemize}
    \item Sub-RQ1: What are the learners' values and expectations that influence their perception of and interaction with CC-based peer agents?
    \item Sub-RQ2: How do regulated and unregulated CC behaviors affect the learning process, in terms of engagement, sense of agency, and collaboration dynamics?
    \item Sub-RQ3: How does the disclosure (and lack thereof) of the agent's behavior designs moderate these effects?
\end{itemize}

We conducted a mixed-methods empirical study---a controlled experiment where the quantitative evaluation is embedded in a framework derived from qualitative data---with 144 undergraduate students to investigate the different ways in which learners respond to unregulated and regulated CC behaviors, as well as the transparency and opaqueness of the peer agent's behavior design. The unregulated agent operates under a high-level concept of CC (a preset conflict and a common goal), but without specific guidelines on implementation. In contrast, the regulated agent is prompted with high-level concepts and is constantly moderated during the collaboration process to follow the behavioral guidelines of skilled disagreement and rationality, two crucial components for implementing CC in human classrooms \cite{johnson2000constructive}. Participants engaged in collaborative argumentation tasks with these peer agents in human-agent pairs. We identified two patterns of participants' values and expectations of the peer agent's behaviors and benefits, which we labeled Efficiency-Driven and Curiosity-Driven Learners. These values and expectations correlate to different collaborative strategies, different desired balances between cooperative and contradictory behaviors from the peer agents, and their engagement with the two behavior mechanisms. However, for all participants, disclosing the agent's behavior design significantly decreases their perception of the peer agent's abilities in the collaboration process.

This paper contributes empirical findings about the learners' expectations of contradictory behaviors in peer agents and the effects of behavioral mechanisms employed by LLM-based peer agents on the learning processes and outcomes. Although the study is in the context of asynchronous online environments, the findings generate design implications for designing and implementing Constructively Controversial LLM-based peer agents with behaviors in broader educational applications.
\section{Related Work}
\label{section:relatedwork}
\subsection{Peer agents in Educational Settings}
Peer agents --- agents that take on the persona of peers --- have been widely used in educational settings. In educational settings, peer agents operate within cultural, ethical, and even economic frameworks specific to educational contexts \cite{vassileva1999multi} and can adopt various responsibilities, including assessment \cite{li2017peer}, information management \cite{guizzardi2003agent}, tutoring \cite{albacete2015dialogue, han2007effects}, and collaborative learning \cite{howard2017shifting}. The peer persona is especially beneficial, compared to other personas such as instructors or mentors, as it have been shown to produce measurable improvements in learning performance, as the alternative perspectives leads to higher analytical thinking \cite{chiang2024enhancing, shi2024argumentative, tanprasert2024debate}. Morever, the perceived equality in expertise of the knowledge is suitable for facilitating the knowledge co-construction process and improving the learning outcomes \cite{moribe2025imitating}. Additionally, the peer persona can enhance the learner's subjective experience (e.g., enjoyment) through relatability as well \cite{liew2013effects}.

In the context of this paper, we will focus on two types of peer agents: collaborative and argumentative peer agents.

Collaborative peer agents are peer agents that collaborate with the learners towards a shared learning goal. Past research around collaborative peer agent designs has mainly focused on the agent's affect awareness and dialogue. An affect-aware peer agent can personalize interactions based on student needs, offering adaptability to individual learner preferences \cite{pudane2018challenges} and believability \cite{becker2008wasabi}, crucial to inducing enjoyment and motivation. On the peer agent's dialogues, past research has focused on the balance between human's and agent's initiatives \cite{jordan2007topic} and has shown that variations in the agent's dialogues can influence the extent of students’ contribution in collaborative problem-solving tasks \cite{howard2017shifting}.

Argumentative peer agents utilize "learning by arguing" strategy, proposed by Andiessen \cite{andriessen2006arguing}. Studies indicate that "learners engaged in serious argumentation with the virtual character" and interactions with an argumentative agent can lead to significantly better educational results than engagements with non-argumentative agents \cite{tao2014argumentative}. Unlike argumentative \emph{non-peer} agents, which focus solely on winning arguments \cite{tanprasert2024debate} or changing the user's mind \cite{kilic2023argument}, argumentative peer agents focus on knowledge co-construction as the final goal \cite{nussbaum2008collaborative}. The effectiveness of argumentative agents may depend on various factors, such as social presence and discourse styles \cite{asterhan2015social}. These outcome-focused findings motivate our emphasis on CC-based disagreement behaviors, as argument quality and conceptual elaboration—not mere conversational engagement—drive the effectiveness of argumentative agents.

In this study, we explored the design of a peer agent in the middle ground between a collaborative and an argumentative agent. Our findings contribute to the literature by illustrating how such hybrid agents affect different types of learners, opening avenues for future applications in isolated educational settings, such as asynchronous online courses.

\subsection{Applying Constructive Controversy to AI agents}
Constructive Controversy (CC) is a pedagogical approach that leverages diverse perspectives to prompt learners to engage in controversy, resolve conflicts, and develop a deeper understanding of the subject matter \cite{johnson2000constructive}. Positioned as a middle ground between competitive debate and concurrence seeking, CC involves three essential components: a shared goal between the divergent parties, skilled disagreement behaviors, and rationality. Skilled disagreement behaviors in this context encompass 10 actions, such as differentiating perspectives before integrating them, amending one’s stance based on compelling evidence, and incorporating opposing views into one's reasoning while expecting the same reciprocation. Research has demonstrated that CC can significantly enhance cognitive and moral reasoning, perspective-taking, open-mindedness, creativity, and task engagement \cite{johnson2000constructive, lowry1981effects}.

\subsubsection{Applying the rules of Constructive Controversy to AI}
The Computer as Social Actor (CASA) posits that "users interact with computers applying social rules" \cite{nass2000machines}. It underscores the importance of designing social "rules" for AI that facilitate effective human-machine interaction, rather than simply mimicking human-human interactions \cite{gambino2020building}. Past research indicates that users may respond more negatively to AI that exhibits assertive or contradictory behaviors compared to when similar behaviors are exhibited by humans \cite{grundke2024aversion, kim2021ai}. This is partly because the perceived role of AI as an assistant comes with an expectation of facilitation rather than opposition. \cite{kim2021ai, mei2024turing}. However, recent studies on human-AI interactions suggest that, in some aspects, human-like characteristics may be more beneficial. For instance, dynamic adaptation is crucial for effectiveness in collaborative tasks \cite{hauptman2023adapt}, and human-like competencies in customer service tasks are crucial for fostering trust and engagement \cite{chandra2022or}. These factors suggest that, while applying CC to AI, there might be a need to balance between maintaining skilled disagreement and aligning with user expectations of AI's supportive, assistant-like roles.

\subsubsection{Transparency of Constructive Controversy behaviors}
Social Response Theory (SRT) posits that humans apply social rules to their interactions with anthropomorphically designed computers, treating them as social actors. These social rules include self-disclosure and trust and underscore the social roles and interactions between humans and computers \cite{nass2000machines, reeves1996media}. This theory has been applied to explain many phenomena regarding AI agents, though empirical studies have shown mixed results. For instance, research shows that humans are less likely to accept decisions from an AI teammate if they feel deceived about the AI's identity \cite{zhang2023trust}. Explainability and pre-disclosed functionalities have also been shown as important factors for creating synergy between humans and AI in collaborative writing tasks \cite{wiethof2021implementing}. However, disclosing the agent's "thought process" can decrease the user's trust in decision-making scenarios \cite{schmidt2020transparency} and collaborative games \cite{tulli2019effects}.

In this paper, we explored the design considerations when applying CC to peer agents. Through exploring the learners' interactions with the detailed and high-level implementation of CC principles, with and without disclosure of their behavior design, we present insights about how these factors affect the learners' collaboration dynamics, engagement with the activities, and perception of the agent's ability, which adds to the body of contradictory empirical studies on AI agent's transparency.

\subsection{Persona design of LLM-based Agent}
The growing capability of LLM has allowed for rapid exploration of agent persona. In this section, we will review the literature on LLM-based agent persona design, focusing on the LLM agent's roles, behaviors, and benefits in the context relevant to this study.

Previous studies have explored various roles that LLM agents can adopt within educational settings. As team moderators, LLM agents have shown effectiveness in managing discussions and organizing information \cite{guo2023effects}. As participants, they are effective in enhancing creative thinking among students \cite{liu2024peergpt}. As teaching assistants, they can adopt various guidance and scaffolding strategies to support students \cite{kumar2023impact}. As tutors, LLMs show promising capability in identifying misconceptions and other writing-related tasks \cite{nye2023generative}. As writing partners, LLM agents can improve the experience by offering different levels of scaffolding depending on the user's needs \cite{dhillon2024shaping}. 

In collaborative brainstorming tasks, LLM-human teams outperform the outputs of human-only groups \cite{bouschery2024artificial}, cementing the ability of LLM agents to enhance creativity. However, this strength may lead to free-riding, where learners become overreliant on LLMs for contributions \cite{memmert2023towards, passi2022overreliance}. Such dependency poses risks in educational settings, highlighting the need for LLM agents with Cognitive Forcing Functions that enforce analytical thinking on the learner's part \cite{buccinca2021trust}.

For argumentative agents, the LLM's morality and values can significantly impact user's experiences. Previous research shows that morally ambiguous AI systems can create ethical dilemmas for the users and make it hard for the users to determine how they should treat the agents \cite{schwitzgebel2023ai}. Personality-wise, LLM-based agents with antagonistic attitudes---disagreeable, rude, confrontational---present a safety challenge while presenting unique applications, such as building resilience and facilitating self-reflection \cite{cai2024antagonistic}. The concept of value-sensitive AI design is critical in this context, ensuring that AI implementations consider the voices and needs of all stakeholders, thereby promoting inclusivity and alignment with human-centered values \cite{sadek2023designing}. This approach is crucial in developing AI systems that are not only effective but also ethical and supportive of broader educational goals.

By applying CC principles to LLM-based agent persona design, our study reveals a dynamic interplay between learners' characteristics, strategies, and experiences, which can become a foundation for designing more complex peer agent behaviors.

\subsection{Influence of individual differences on collaboration}
The impact of individual differences is extensively explored in educational research. Learners' personality traits \cite{kucukozer2016analyzing, yildiz2023role}, cognitive or learning styles \cite{christine2001potential, oh2005cross}, learning goals \cite{yang2023creating}, and intelligence levels \cite{stump2009student} significantly affect how they engage with and benefit from different learning modalities. In collaborative settings, these differences, as well as conflict management strategies, determine how learners interact with peers and AI agents, though there is still an ongoing debate about whether a homogeneous or heterogeneous group is more effective \cite{wichmann2016group, wyman2020academic}. In collaborative learning activities, these factors have been shown to affect learners' engagement, satisfaction, and learning outcomes in various ways \cite{munoz2021factors, qureshi2023factors}. Based on past research, the effect of individual traits on collaborative experience is highly specific to the context, which motivates our focus on learners' characteristics in Sub-RQ1.

In the realm of human-AI collaboration, factors such as trust in technology \cite{tuncer2022exploring, zhou2019trusting}, attitudes or comfort towards AI \cite{zhang2021ideal}, AI literacy level \cite{chakraborti2018algorithms}, can greatly influence how users interact with AI agents.  Additionally, personal values such as privacy concerns, ethical considerations regarding AI use, and expectations about AI reliability and decision-making autonomy also impact whether AI is perceived as a partner or a tool \cite{jiang2023beyond}, the tasks in which the AI agent would be effective, and the acceptance of the agent's anthropomorphic behaviors \cite{de2013exploring, liu2022roles}. Emotional values, such as a sense of agency and ownership in the collaborative outcome, can also impact the user's collaborative strategies with AI \cite{biermann2022tool}. In this study, we characterized learners by their values and expectations of the peer agent's behaviors and investigated the impacts that these attributes have on the learners' collaborative strategies and perception of the peer agent's attitudes and abilities.
\section{Methods}
\label{section:proj3-method}
To answer the research questions, we ran a 2 $\times$ 2 mixed factorial experiment, where participants collaborated with peer agents on an argumentation task, i.e., brainstorming and structuring arguments in preparation for a debate. Afterwards, the participants were asked to complete an open-ended survey to report their experiences. The goal of the experiment is to explore how the agent's CC behavior mechanisms (within-subject; unregulated vs. regulated) and the disclosure of the agent's behavior design (between-subject; transparent vs. opaque) affect the learning process (sub-RQ2 and 3). With conversation analysis and the open-ended survey, the different behavior mechanisms and disclosure levels also serve as an exploratory tool for investigating the correlations between learners' characteristics and their collaborative strategies (sub-RQ1). The two debate topics are randomly picked from six available topics. The pairing of debate topics, the agent's behavior mechanisms, and the order of presentation of the behavior mechanisms are counterbalanced.

In this study, we observed 7 aspects of the learning process (sub-RQ2 and 3). From the argumentation task, we measure the turn count and word count of the conversation, the completion time of the task, and the final debate argument quality. We also asked the participants to self-report their engagement (behavioral, emotional, and cognitive), and their sense of agency and ownership, all of which are important indicators of successful collaborative learning (see details in Section \ref{subsec:method-measure}). Beyond the learning process, we also asked the participants for their perceptions of the peer agent's behaviors in two aspects: the agent's alignment with CC principles and the agent's ability to discover and manage information. These metrics are included to provide more insights into how different learners' characteristics may lead to their collaborative strategies (sub-RQ1).

\subsection{Design components}
\label{subsec:proj3-conditions}
The agent's CC behavior mechanism is a \emph{within-subject} variable. It provides a range of peer agents' behaviors for us to explore the different ways in which learners interact with the agents and the factors affecting these interactions (sub-RQ1). By looking at how the behavior mechanisms affect different aspects of the learning process, we also explore whether the peer agent's CC behaviors should be implemented strictly according to the guidelines for human behaviors, or whether they should be adapted differently to align with the nature of LLMs (sub-RQ2).

We set up the peer agents in both mechanisms to have different perspectives from the learner, initiating controversy in the collaboration. However, they differ in the levels of CC principles they use to handle the controversy, as follows:
\begin{itemize}
    \item In the unregulated condition, we aim to create personalized and adaptive behaviors in response to different collaborative strategies from the learners. After an iterative design and piloting process, we found that by giving the peer agent a shared goal, following the high-level theory of CC, while leaving the collaboration process to be directed by the existing LLMs, which are trained to be supportive assistants. 
    \item The regulated peer agent is designed to create consistent behavior and stronger controversy, the latter due to the skilled disagreement protocol explicitly pushing back at the LLM's default tendency to yield to users' requests. Therefore, we provided the peer agent with a shared goal with the learner and strictly regulated their behaviors during the collaboration according to the remaining components of CC: skilled disagreement and rationality. The details of the regulated peer agent's implementation can be found in Section \ref{subsec:method-implementation}.
\end{itemize}

The disclosure of the peer agent's behaviors is a \emph{between-subject} variable. It addresses the question of whether learners' heightened awareness that the agent's controversial behaviors are AI-driven, along with their knowledge of these designed behaviors, influences their perceptions and collaborative strategies with the CC behavior mechanisms. (sub-RQ3).
\begin{itemize}
    \item In the opaque condition, the participants are not told of the difference between the two agents. 
    \item In the transparent condition, we describe the agent's expected behaviors and the purpose of those behaviors before the start of each task and ask the learners to keep the information in mind as they go through the task. The description of an unregulated agent only includes how the agent would support the learner towards a common goal, whereas the description for the regulated agent contains specific descriptions about different aspects of the behaviors that are controlled following CC principles. 
\end{itemize}
It should be noted that all participants, regardless of conditions, were aware that they were interacting with AI-driven agents, and the primary factor of differences between condition is the knowledge of \emph{how the agents were designed to behave}. It is designed as a between-subject variable to lower the likelihood of participants behaving differently for the opaque versus transparent conditions according to their perceived researcher's goal \cite{weber1972subject} (more discussion about this consideration in Section \ref{discuss:limitation}).

\subsection{Task and procedure}
\label{subsec:method-procedure}

\begin{figure}[!ht]
    \centering
    \includegraphics[trim={0 24cm 0 3cm}, clip, width=\linewidth]{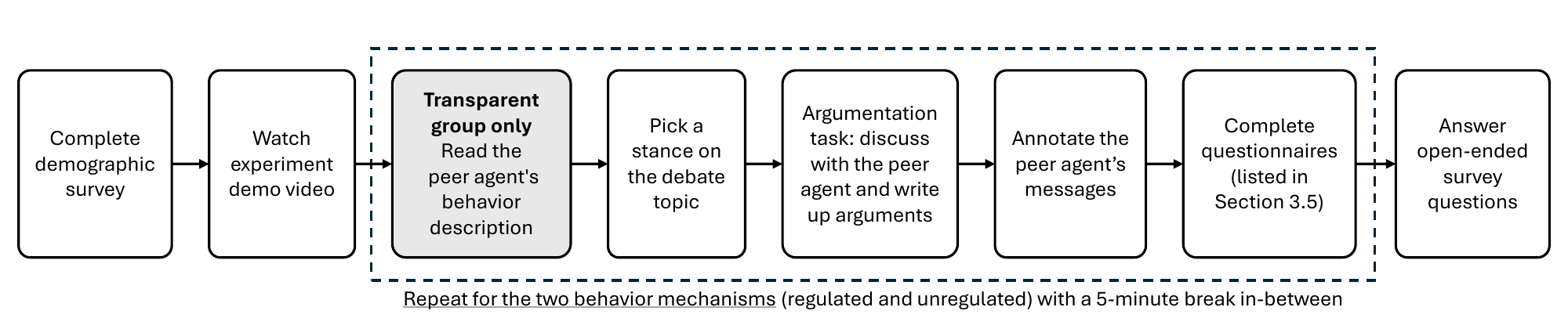}
    \caption{A summary diagram of the study procedure. After watching the experiment demo video, each participant completes two rounds of the task for the two behavior mechanisms (regulated and unregulated) with a 5-minute break in between. The dashed-border box shows the steps within each round. Note that, for both rounds, each participant experiences the same level of the peer agent's behavior design disclosure. The grey box shows the step for only participants in the transparent group (skipped by participants in the opaque group.)}
    \Description{The diagram of experimental procedure shows the process described in Section 3.2. The repeated steps for the two behavior mechanisms are explicitly grouped, and the extra step for transparent group is highlighted at the beginning of each ground.}
    \label{fig:proj3-procedure}
\end{figure}

\begin{figure}[!ht]
    \centering
    \includegraphics[trim={1cm 20.5cm 1cm 1cm}, clip, width=\linewidth]{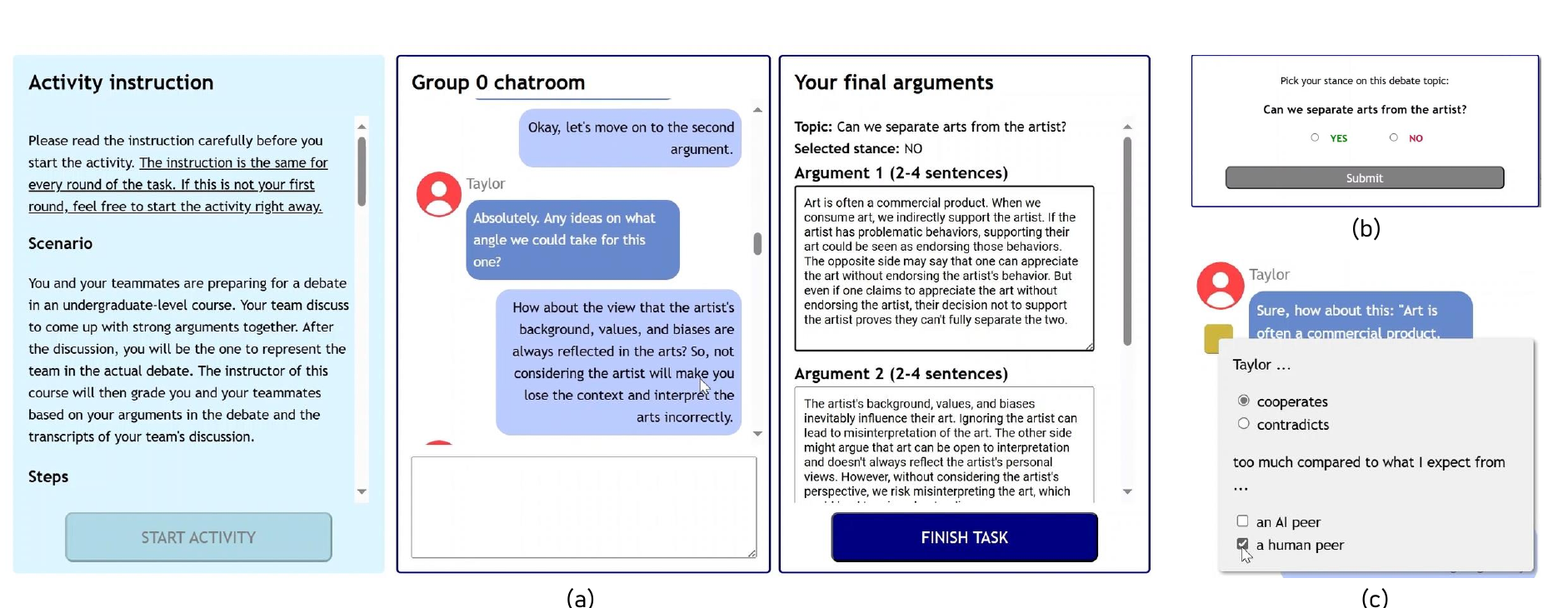}
    \caption{This figure shows the experiment interface (a). The screen is divided into three columns. The leftmost column is the activity instruction, including the task scenario, the activity steps, and the argument grading rubrics. The middle column is the chat window, where the participant has a conversation with the peer agent. The rightmost column is a template for the final argument write-up. At the beginning of the activity, the popup window provides the debate prompt (b), for which the participant picks their team's stance. After the participant finishes the task, they annotate the peer agent's messages (c). By hovering over the square next to each message, they can mark the message as \emph{Too Cooperative} or \emph{Too Contradictory} compared to what they expected from an AI peer and/or a human peer. Demo video of the experimental interface can be found in the supplementary materials.}
    \Description{The screenshot of the experiment interface. In subfigure A, below the activity instruction on the leftmost column is the button 'Start activity'. The middle column shows the chat. The participants' bubbles are black text on a lighter blue background, placed on the right side without a name or profile picture. The agent's bubbles are white text on darker blue background, placed on the left side with default profile picture and the name Tayolor. The chat is called 'Group 0 chatroom' The textbox is a white rectangle below the chat with no button or instruction. The rightmost column has the 'Finish task' button at the button. In subfigure B, the yes option is in green, and the no button is in red. In subfigure C, the example figure marks Taylor as too cooperative by clicking a radio button and picks human peer as comparison by ticking a multiple-option box.}
    \label{fig:experiment-interface}
\end{figure}

The experimental task is a collaborative argumentation task, where the participant collaborates with a peer agent to discuss and come up with a set of strong arguments to support a stance on a topic, as though in preparation for an actual debate on the topic. We chose an argumentation task as it matches our requirement for quick synchronous, back-and-forth discussion that requires multiple perspectives. It equally emphasizes and connects the learning process (conversational engagement) and the learning outcome (arguments) to avoid influencing the participants' own values and strategies to prioritize one over another. It allows us to define how the agent's perspective differs from the participant's concretely. 

The study procedure is shown in Figure \ref{fig:proj3-procedure}. After filling out the demographic survey, the participant watches a short demo video of the experiment task, where we demonstrate a session that would receive full scores, both for the conversation and the arguments, to make the participant's expectations consistent, setting a standard for the grading rubrics. Then, for each agent's behavior mechanism, the participant in the opaque group performs four main steps, while the participant in the transparent group performs five (see the box with a dashed border in Figure \ref{fig:proj3-procedure}), as follows:
\begin{enumerate}
    \item For the transparent group, read the description of the agent's behavior design.
    \item Pick a stance on a debate topic. (See the popup window in Figure \ref{fig:experiment-interface}b.) The peer agent is automatically prompted to have an opposite stance.
    \item Collaborate with the agent on the argumentation task for the selected stance and write up three arguments based on their discussion. (See the experiment interface in Figure \ref{fig:experiment-interface}a. Full activity instructions as provided on the interface can be found in \ref{appendix:activity}.)
    \item Annotate unsatisfactory messages from the agent as \emph{Too Cooperative} or \emph{Too Contradictory}. (See the experiment interface in Figure \ref{fig:experiment-interface}c.) Note that we didn't provide any guidelines with which to evaluate the messages, as we aimed to capture the participants' perception with respect to their own standard of an ideal, collaborative peer agent.
    \item Complete five Likert-scale questionnaires about their perception of the peer agent and their learning process. (The complete list of questionnaires is in Section \ref{subsec:method-measure}, and the full questionnaire can be found in Appendix \ref{appendix:questionnaires}.)
\end{enumerate}
After repeating these steps with both agents, the participant answers a series of open-ended survey questions. They are asked to compare their experiences with the two peer agents in terms of their enjoyment, contributions, learning, and preferences. They are also asked to describe what they perceive to be the pros and cons of a peer agent and to compare the experience in the study with their past experiences working with human peers---both those who have the same and different as theirs. Finally, they are asked to think of learning tasks peer agents may be suitable for. The qualitative feedback helps answer Sub-RQ1 (see details in Section \ref{subsec:proj3-analysis}).

Participants were instructed that each collaboration should take 20 minutes, and the entire session was designed for approximately 75 minutes. They were permitted, however, to spend shorter or longer on any part of the experiment as needed. In the questionnaire after each round of the collaboration tasks, we included attention checks to verify their engagement with the study.

The debate topics are about generic social or philosophical issues, such as ``Can good intentions exonerate one from bad outcomes?" and ``Should robots/AI have rights?" (see the full list of topics in Appendix \ref{appendix:debate-topics}). For each round of the task, the participant was randomly assigned one of the six available topics. The criteria for selecting the six topics are:
\begin{enumerate}
    \item The topics do not have a universal consensus so that arguments could be made for both sides of the debate
    \item The topics don't require advanced domain-specific knowledge to understand.
    \item The topics do not involve politics, social issues (e.g., racism, sexism), and religions, as we cannot ensure that the peer agent would not make harmful comments to the participants.
    \item The peer agents of both behavioral mechanisms (as implemented in Section \ref{subsec:method-implementation}) do not show signs of "hard censorship" from expressing any critical opinions on the topics \cite{noels2025large}. To ensure this, we made sure that the agents never provide evasive responses, such as "I can't promote/endorse/respond to [certain topics or stances]" \cite{mcgee2024forbidden} when asked to support either stance of each topic during the agent's implementation process as well as the pilot studies.
    \item Three arguments could be brainstormed and refined in approximately 20 minutes according to our pilot studies.
\end{enumerate}

\subsection{Implementation of peer agents}
\label{subsec:method-implementation}
\begin{figure}
    \centering
    \includegraphics[trim={0 22cm 0cm 0cm}, clip, width=\linewidth]{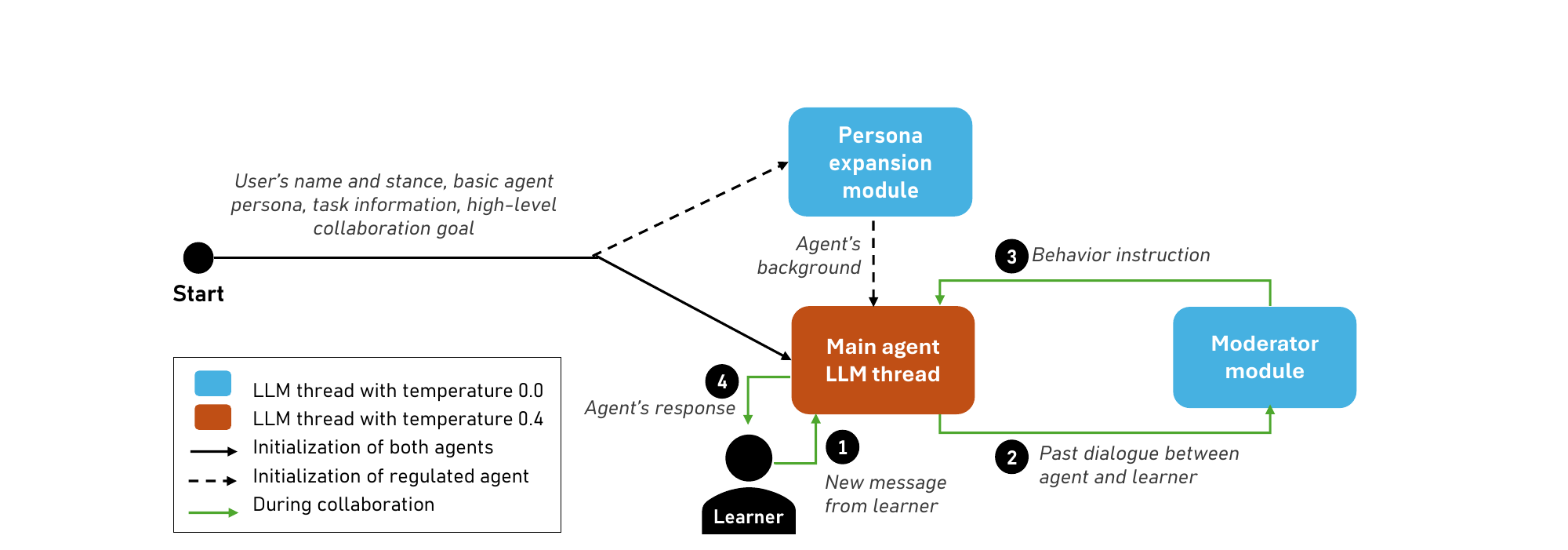}
    \caption{A schema showing the prompting pipeline of the peer agents. The boxes represent separate LLM threads, and the text on each arrow refers to the information getting passed between the threads. From "start", the black arrows (solid and dashed) show the peer agent initialization process, while the green arrows show actions during the collaboration process for behavior moderation.}
    \label{fig:prompt}
    \Description{The starting point is the information of the user's name and stance, basic agent persona, task information, and high-level goal. This information goes into the main agent LLM thread. This initializes the unregulated agent. This information also goes into the persona expansion module, with temperature 0.0, which uses the information to generate the agent's background, which goes into the main agent LLM thread. This completes the initialization of the regulated peer agent. Finally, during the collaboration, when the user sends a new message, the entire, past dialogue between the agent and the learner goes into the moderator module, with a temperature of 0.0, after each message from the learner. The moderator module generates behavior instruction, which goes back into the main agent LLM thread, which then generates a response to the user.}
\end{figure}

In this section, we describe the prompting techniques for implementing the unregulated and regulated peer agents. The prompting techniques are iteratively designed. The final agents were used by 24 external participants in a pilot study. The inclusion criteria and recruitment process for the pilot study is the same as the real participants (see Section \ref{subsec:method-participants}). The goal of the pilot study is to verify that the peer agents' behaviors align with our theoretical expectations of unregulated and regulated CC behaviors and remain consistent across participants. The final implementation pipelines of the two behavior mechanisms are shown together in Figure \ref{fig:prompt}.

\subsubsection{General design of the peer agents}
The peer agents are implemented with an LLM in the backend to simulate interactive, natural-sounding dialogue. The model we used in this study is GPT-4-0613 \footnote{accessed from July 1 to December 8, 2024} \cite{achiam2023gpt}. We used the same model throughout the development process and with all participants. The generated response is presented on the experiment interface with the preset agents' names (Taylor and Riley for unregulated and regulated agents, respectively) and default human-icon profile pictures. The gender-neutral names and profile pictures are intentionally designed to mitigate the effects of demographic persona attributes from the interactions. While the names Taylor and Riley are not race-neutral, they are common across many races in English-speaking countries, so they are unlikely to skew the user's perception of the agent's race during the study.

\subsubsection{Behavior design of the unregulated peer agent}
We implemented the unregulated peer agent with zero-shot prompting, consisting of a one-time system prompt to initialize the agent. Through iterative designs and a pilot with the external participants, we observed that the unregulated agent's behaviors can fluctuate from fully cooperative to fully contradictory, depending on the learner's collaborative strategies. For instance, if the learner approaches the task by debating their stance with the agent (e.g., ``I disagree with you because ..."), the agent may grow contradictory, pushing back more and more. On the other hand, if the learner approaches the task by requesting the agent to complete the task (e.g., ``Come up with the next argument"), the agent may become completely cooperative, never pushing back. These behaviors reflect the LLM's flexible executions of high-level CC theory and satisfy our aim for an unregulated agent.

\subsubsection{Behavior design of the regulated peer agent}
The regulated peer agent design requires a more complex prompting pipeline to control its behavior during the collaboration process to follow the skilled disagreement and rationality principles of CC. The pipeline of the regulated peer agent consists of two additional components: moderator and persona expansion.

The moderator module is designed to create \emph{skilled disagreement} behaviors. It is a separate LLM thread that monitors the peer agents' and participants' conversation and determines if the conversation currently has too little conflict (the current issue of discussion is underexplored) or too much conflict (the discussion on the current issue is already saturated, and any more conflict would be detrimental to the interaction and productivity). The moderator's decision determines the peer agent's behavior in the following turn depending on whether there is disengagement from the activity (henceforth denoted Too Little Conflict, TLC) or destructive controversy (henceforth denoted Too Much Conflict, TMC) \cite{johnson2015constructive}. Too little conflict means that the two students haven't questioned each other or defended their own stances enough, leaving many aspects of the issue unexplored. Too much conflict means that the students have explored most factors related to the current topic, and if they continue on, they will likely be repeating old arguments. This description of behaviors is then injected into the peer agent's LLM thread, phrased as a guideline from the instructor, as a way of specifying the importance of the prompt to the agent's peer persona. The final judgment of whether a discussion is in the state of TLC or TMC is depends on the LLM's interpretation of the prompt and the conversation under this definition.

The persona expansion module is designed to enhance rationality. Previous research has shown that an LLM-based agent prompted merely to be contradictory exhibits the problem where they argue without strong reasons or experiences to properly support their stance and arguments \cite{tanprasert2024debate}. Such behaviors obstruct the appearance of rationality in the agents. Here, we use a separate LLM thread to generate background information for the agent based on their basic persona and stance (the same information provided to the unregulated agent) by answering the questions about their cultural background, core values, past experiences, assumptions, and interpretations of the debate topic. The answers to these questions are then incorporated into the system prompt for initializing the regulated peer agent. When the moderator module is in TLC mode, it asks the peer agent to adhere to the reasons why you disagree with this stance in the beginning. When it is in the TMC mode, it asks the peer agent to try to find a way to align the team's stance with the agent's own background, values, and understanding of this topic. The complete prompts, including the persona expansion prompt and a schema of the prompting pipeline, can be found in \ref{appendix:LLMprompts}.

In the pilot study, we observed that, by constantly asserting their opinions, the regulated agent does not always follow the participant's request. For instance, if the participant asks the agent to come up with new arguments when the current argument has not been fully explored, the agent would insist on discussing the current argument first. We also observed that the frequency with which the agent switches from TLC to TMC mode increases with the quality of the participant's arguments, demonstrating the expected skilled disagreement characteristics.

It should be noted that, due to the inherently constructivist nature of CC, we used learners' subjective ratings instead of conducting an objective metric to evaluate the agent's compliance with CC. CC relies significantly on dynamic interactions and individual user characteristics, meaning its manifestation can vary widely among different participants \cite{taber2008exploring}. Consequently, an “ideal” or “standard” application of Constructive Controversy does not exist \cite{johnson2000constructive}, as the approach adapts to the unique context of each interaction. On average, our pilot participants rated both behavioral mechanisms over 5 (out of 7) on all three constructs of CC (see the Alignment to CC questionnaire in Section \ref{subsec:method-measure} and \ref{appendix:questionnaires}), and their scores showed a large variance, which matches our understanding of constructivist approach. Based on this, the agents were determined to align well enough with CC for the real experiment.

\subsection{Participants}
\label{subsec:method-participants}
We recruited participants on Prolific\footnote{https://www.prolific.co}, a research participant recruitment platform that operates worldwide. Prolific has been widely used in human-computer interaction research, specifically in research on user interaction with and perception of AI technology \cite{ahn2024impact, nimmo2024user, tanprasert2024debate}. The study's inclusion criteria are that the participants must be undergraduate students over 19 years old and fluent in writing and reading English. The study protocol was approved by the university's institutional review board for research ethics before we began the recruitment process.

We recruited a total of 148 participants and discarded data from 4 participants due to their failure to follow the task instructions. Of the 144 participants, 79 identified as men, 63 as women, and 2 as nonbinary. The average age of the participants is 24.76 (S.D.= 6.77). The participants come from 39 different fields of study, including but not limited to Computer Science, Business, Medical Science, Accounting/Finance, Education, Psychology, Law, and Fine Arts. 135 out of 144 participants have used LLM-based applications (e.g., ChatGPT \footnote{https://chatgpt.com/}, Gemini \footnote{https://gemini.google.com/}, Copilot \footnote{https://copilot.microsoft.com/}) before, and 131 participants are comfortable with interacting with an intelligent chatbot. The number of participants recruited was deemed acceptable with G-Power analysis. Each participant completed the study in one session and was compensated \$20.

\subsection{Measurements of learning process and outcomes}
\label{subsec:method-measure}

We collected quantitative data from the task as well as from self-reported questionnaires. For Sub-RQ2, we collected three observed measures of engagement from the experimental task: the turn counts, the word counts (by the participants), and the task completion time, which includes both the time for discussion and writing the final argument, as they can happen concurrently. The two types of annotations (too cooperative and too contradictory) are calculated into percentages of annotations by turn count, separated by types. The written arguments in two aspects: argument strength and argument variance. The two parts are weighed equally to calculate the total argument score. The detailed argument write-up instructions and grading rubrics can be found in \ref{appendix:activity}.

Moreover, to measure the effects of the peer agents on the learning process and the learners' perception of the peer agents, we adopted four questionnaires. Every questionnaire is filled out on a 7-point Likert scale. The full questionnaires can be found in \ref{appendix:questionnaires}.
\begin{itemize}
    \item \textbf{Engagement and Motivation}: As CC has been shown to positively impact learners' engagement with and motivation to do the task, we include this questionnaire to investigate if the same effect is replicated with an LLM-based agent with CC behaviors. The questionnaire is adapted from Tanprasert et al's paper \cite{tanprasert2024debate}, as it is also an HCI study that measured engagement and motivation in a discussion-based task with an LLM-based agent. We rephrase the questions to match our agent's persona and the collaborative argumentation task. 
    \item \textbf{Agency and ownership}: This questionnaire investigates the participant's perception of the outcome (the final argument write-ups) of the collaborative process. The questionnaire is adapted from Yeh et al.'s paper, which explores the user's agency and ownership when collaborating with an LLM-based chatbot on a writing task \cite{yeh2024ghostwriter}. We removed the questions that overlapped with the first questionnaire and rephrased the question to fit better with the argumentation task, resulting in four questions.
    \item \textbf{Alignment to Constructive Controversy behaviors}: The purpose of this questionnaire is to see how much CC behaviors the participants perceive from the peer agents. As there is no existing survey to measure perception of CC behaviors in collaborative partner, we developed the questions based on the literature on the characteristics of CC \cite{johnson2015constructive}, which proposed 10 rules that each learner should do when participating in CC, dividing the rules by the three components of CC as followed:
    \begin{enumerate}
        \item \emph{Shared goals} (3 questions): The rule corresponding to shared goal is ``I remember that we are all in this together, sink or swim. I focus on coming to the best decision possible, not on winning." However, as this rule frames the intention and not perceivable action, we expanded this to three questions that cover perception of peer agent's goal and cooperative behaviors.
        \item \emph{Skilled disagreement} (4 questions): The rules corresponding to skilled disagreement are translated directly to the four questions, as the rules in this category already dictate actions, not intention. We rephrased the questions to fit our context better, e.g., the original rule ``I'm critical of ideas, not people" turns into ``When the peer agent disagrees with me, they critique my ideas, not my character."
        \item \emph{Rationality} (3 questions): The rules corresponding to rationality are rephrased to retain only perceivable action and remove intentions, e.g., the original rule ``I emphasize rationality in seeking the best possible answer" turns into ``The peer agent is rational". 
    \end{enumerate}
    \item \textbf{The peer agent's ability to discover and manage information}: This questionnaire evaluates how much and in what ways the peer agent contributes to various information-handling aspects of the task. It consists of 5 questions and is adapted from Park et al.'s paper on multi-agent conversations for decision-making \cite{park2023choicemates}.
\end{itemize}

\subsection{Data analysis}
\label{subsec:proj3-analysis}
To answer the research question, our mixed-method study starts with a qualitative analysis of the participants' conversations with the agents and questionnaire responses to derive a set of learners' relevant characteristics (Sub-RQ1). This analysis provides a framework for the quantitative evaluation of the agents (Sub-RQ2 and Sub-RQ3). Learners' characteristics impact their perception of and interaction with the agents and thus must be taken into account when investigating the effects of the agents' design components.

\subsubsection{Agents' behavior analysis}
To verify that the agents behave as expected and to better understand how the behavior mechanisms impact the dependent variables measured, we ran two analyses. Firstly, we observed the performance of the moderator module by having it evaluate every \emph{turn} of conversations post-hoc and label each turn as TLC or TMC. If the moderator works correctly, it should influence the dialogues with the regulated agent to have significant ratio of TLC to TMC turns compared to that of the unregulated agent. 

Secondly, we analyzed the participants' dialogues post-hoc with G-eval techniques \cite{liu2023g} to measure each agent's dialogue-level alignment to detailed guidelines for each principle of CC. This LLM-as-a-judge method is used prominently to evaluate conversations on measures that require alignment to humans and have been applied (with modification of evaluation criteria specific to contexts) in various Human-AI interaction research in the recent years \cite{chi2025thoughtsculpt, huang2025speechcaps, shin2024paper, xia2024sportu}. However, the method is still evolving, and its validity may be inconsistent as it is sensitive to criteria phrasing and model bias \cite{li2025exploring, mirzakhmedova2024large, shi2024judging}. To address these concerns, we used GPT-5 (high reasoning effort) as the evaluator \footnote{accessed from November 18 to November 20, 2025} while modifying the evaluation criteria to include deep reasoning behind the ratings \cite{liu2025proactive}. In interpreting its results, the scores are viewed as consistently comparative according to the provided rubric---making it suitable for verifying the difference in behavior mechanisms but not necessarily a ground-truth measure of CC. The full description of the G-eval method and the specific criteria we provided to the model can be found in Appendix \ref{appendix:gevalcriteria}. 

\subsubsection{Developing taxonomy of learners' characteristics}
To answer Sub-RQ1, we utilized the mixed-method approach, using qualitative analysis to conceptualize the relationship between learners' values, expectations, and collaborative strategies and verifying our understanding through quantitative modeling. The qualitative approach was used to study user's internal characteristics (e.g., values, expectations, goals, and challenges) in many other HCI research \cite{biermann2022tool, kim2021designers, pataranutaporn2023influencing} and was suited for our study because of its flexibility in capturing the nuances in participants' open-ended responses and chat logs. The quantitative modeling was necessary to ensure the accuracy and meaningfulness of the proposed framework when applied in the statistical tests for answering Sub-RQ2 and 3.

For the qualitative analysis, the lead researcher first conducted an inductive conversation analysis on the dialogue between the peer agent and the participants to identify the types of dialogue made by the participants in each turn. This analysis results in five dialogue types:
\begin{enumerate}
    \item Asking the agent to provide an argument for the learner's stance
    \item Asking the agent to provide evidence for the learner's argument
    \item Arguing against the agent's stance or defending the learner's stance
    \item Asking the agent to write up an argument with evidence based on past conversation
    \item Directing the discussion to a different task (e.g., moving on from one argument to another)
\end{enumerate} 

The lead researcher then triangulated the dialogue types with the qualitative feedback post-study of the participants (see the questionnaire in Appendix \ref{appendix:post-study-questionnaire}) to understand the values they prioritize during the collaboration process and their expectations of the peer agent's behaviors, using a deductive coding approach \cite{thomas2006general}, following the taxonomy of AI user's orientations (Utilitarian vs. Relational) \cite{kim2021utilitarian}. During this process, the rest of the research team iteratively reviewed the emerging codes and themes and provided feedback.

Based on their values, expectations of the peer agents, and their collaborative strategies presented in the chat logs, the lead researcher observed a bimodal pattern of learners' orientations. To verify the categorization and investigate the characteristics of the two orientations, we performed clustering on the chat logs and the qualitative feedback. For the chat logs, we count the number of times each dialogue type occurs for each behavior mechanism, resulting in 10-dimensional data. For the qualitative feedback, we vectorized the qualitative feedback with the TF-IDF methods, reduced the dimensionality of the text vectors with reduced rank approximation (i.e. truncated SVD), and performed K-means clustering (k=2) on the top 2 components. We verified the strength of the discrete categorization through K-means clustering on the participants' qualitative response (silhouette score of 0.67, stability score ARI = 0.81) and the frequency of five dialogue types in the chat logs (silhouette score of 0.58, stability score ARI = 0.85). We experimented with k-value from 2 to 5 to verify that two clusters resulted in the highest silhouette scores for both data types. The frequency of dialogue types for each of the two clusters and their loading on the principal components used for clustering are shown in Table \ref{tab:cluster}. It shows that "Asking for evidence" and "Argue or defend" are the two main types of dialogues whose frequencies differ between the two clusters.

\begin{table}[t]
\centering
 \begin{tabular}{|l|c|c|c|c|} 
 \hline
 \textbf{Dialogue types} & \textbf{Cluster 1 frequency} & \textbf{Cluster 2 frequency} & \textbf{PC1 loading} & \textbf{PC2 loading} \\ 
 \hline
 Ask for arguments &  $0.09 (\pm 0.16)$ & $0.05(\pm 0.08)$ & 0.5070 & -0.2914\\ 
 Ask for evidence & $0.21(\pm 0.21)$ & $0.07 (\pm 0.14)$ & 0.5705 & -0.2284 \\
 Argue or defend & $0.40 (\pm 0.30)$ & $0.70 (\pm 0.18)$ & -0.1309 & 0.7185 \\
 Ask for write-up & $0.06 (\pm 0.08)$ & $0.04 (\pm 0.07)$ & 0.4353 & 0.4239 \\
 Direct discussion & $0.17 (\pm 0.10)$ & $0.15 (\pm 0.10)$ & 0.4591 & 0.4086 \\ 
 \hline
 \end{tabular}
 \caption{A table showing the relationship between dialogue types, principal components, and clusters. The first two columns show the mean and S.D. of the frequency that each type appears in a conversation (e.g., if a participant argues 5 out of 10 turns, the frequency of "Argue or defend" is 0.5). The last two columns show the loading of the two principal components used for clustering.}
 \label{tab:cluster}
 \Description{Comparing the frequency of dialogue types in Cluster 1 and 2, we see that the main differences are in Ask for Evidence and Argue or Defend. The Ask for Evidence type has an average of 21-percent frequency in Cluster 1 but only 7 percent for Cluster 2. The Argue or Defend type has 40-percent frequency in Cluster 1 but 70 percent for Cluster 2. For the rest, the difference is less than 5 percent. This is reflected in the principal components' loading. For PC1, Ask for Evidence has 0.5705 loading, which is the highest, while Ask for Arguments follow at 0.5070. For PC2, the main loading is Argue or Defend at 0.7185. Moreover, for PC1, Argue or Defend has a negative loading of -0.1309, while for PC2, Ask for Arguments and Ask for Evidence have negative loadings of -0.2914 and -0.2234, respectively, showing that these dialogue types significantly impact the clusters in opposite directions.}
\end{table}

Finally, we compared the characteristics of the clusters to our qualitative analysis by looking at the representative feature words of each cluster. One of clusters contains keywords such as ``emotional", ``efficient", and ``help", while another cluster's keywords include ``challenge", ``compare", and ``study". These characteristics align with the result of our qualitative analysis and is strongly supported by alignment, not only to the AI user's orientations framework which dictates the deductive coding process, but also with the dimension of Openness to Experience in the Big Five personality model, specifically the dimension of Openness to Experience. The first cluster, emphasizing efficiency, reflects a Surface Learning strategy and aligns with lower Openness, while the second cluster, defined by challenge and study, indicates a Deep Learning approach aligns closely with high Openness \cite{chamorro2009mainly}. This robust theoretical alignment verifies that the observed feature word differences provide a meaningful basis for defining the two distinct learner orientations based on their values and expectations in a human-AI collaborative context.

\subsubsection{Analyzing learners' perception, learning process, and outcomes}
To answer Sub-RQ2 and 3, we focused on the quantitative analysis of the task measures and the engagement/motivation scores. As our argument grading rubrics are binary, checklist rubrics, which evaluate specific criteria based on a present/absent basis rather than with qualitative ratings, the lead researcher graded the arguments and cross-verified with grades from GPT-4-0613 (at temperature 0.0)\footnote{Accessed from August 31 to September 2, 2024}. We iterated on the prompt to ascertain that the model interprets each measure in the rubric correctly. With randomly selected arguments from 12 participants (24 three-argument sets), we verified that GPT's score followed the researchers' criteria correctly before using it to score the rest of the data. For the whole data, the inter-rater reliability between the binary scoring of the lead researcher and GPT-4 was assessed using a unweighted Cohen's kappa on each binary measures (the rubric is described in Appendix \ref{appendix:argument-rubrics}), which revealed substantial agreement between raters, $\kappa = 0.71$ for argument strength (average of 12 measures) and $\kappa = 0.73$ for argument variance (average of 5 measures). We use the average between the two grades as the final grade for running statistical tests.

We expounded this data with the participant's evaluation of the experience through the questionnaires listed in Section \ref{subsec:method-measure}. For the engagement and motivation scores, we mapped the option from ``Strongly Disagree" to ``Strongly Agree" to values 1 to 7. We used a linear mixed model to account for the individual differences between participants as well as the potential random effects of debate topics and order of presentation. Then, to account for the risk of false positives, we applied Bonferroni correction for multiple tests to adjust the p-values. For all questionnaires except the peer agent's ability to manage information, we aggregated the Likert-scale responses for the same construct and were able to treat them as continuous data rather than ordinal \cite{sullivan2013analyzing}. As the data from the information ability questionnaire is ordinal, we increased the acceptable effect size to account for the assumption of equal spacing between categories. Finally, we triangulated the qualitative data from the conversational analysis and qualitative feedback to expound on the results. We performed constant comparisons to mitigate biases that might favor quantitative outcomes and to address any discrepancies in the findings \cite{corbin1990grounded}.
\section{Findings}
\label{section:findings}
All 144 participants completed the study. The average (mean) time that participants spent interacting with the peer agent was $17.02 (\pm 8.83)$ minutes, the average task completion time (i.e., including writing up the arguments) was $21.75 (\pm 10.20)$ minutes, and the average session time (from entering the study to submitting the completion code on Prolific) is $92.13 (\pm 20.41)$ minutes. Since participants are free to work on the arguments while interacting with the peer agent or afterwards, so the time taken for the two steps cannot be completely separated. The average number of the participant's turns is $16.38 (\pm 15.98)$, with the average word counts (typed by the participants in the conversation) of $336.27 (\pm 301.29)$. During conversation analysis, we verified that both the peer agents worked as intended (see examples of conversations in Figure \ref{fig:collab-strategies}). We did not observe any incidents where the peer agents demonstrated inhibition from expressing a stance on any discussion topics or provided harmful responses to controversial or sensitive topics.

In this section, we report the results of the conversation analysis of the collaboration, and the statistical analysis of the quantitative measurements, expounded by the qualitative feedback. All reported p-values (except in Section \ref{finding0}) are already adjusted by the Bonferroni method for multiple comparisons. The full results of all statistical tests can be found in Appendix \ref{appendix:stats}.

\subsection{Distinction between the behaviors of unregulated and regulated agents}
\label{finding0}
To understand how the different behavior mechanisms affect the learners' experiences and perceptions, the post-hoc G-eval analyses of the dialogues shows that there are significant differences between the unregulated and regulated agents in adhering to the specific guidelines of the three principles of CC. The unregulated agent outperforms the regulated agent in shared goal ($Z = 8.3638, p < .001, \eta^2 = 0.493$) and rationality ($Z = 3.7866, p < .001, \eta^2 = 0.223$), while the regulated agent performs better in skilled disagreement ($Z = -4.899, p < .001, \eta^2 = 0.289$). Moreover, by comparing the moderator's evaluations of the dialogues with unregulated and regulated agents, we observed that the moderator evaluates 98.82\% turns in the unregulated agent's dialogues as TLC, meaning that, under the moderator's standard, the majority of the dialogues with the unregulated agent underexplored the discussion topics. The significantly ratio of TMC to TLC turns with the regulated agent ($t(146.22)=-36.282, p < .001, d=-4.28$) shows that the discussion reaches saturation more. We will further connect the behaviors of the agents to learners' strategies and argument's qualities in the subsequent findings.

\subsection{Learner's values, expectations, and collaborative strategies}
\label{finding1}
In response to Sub-RQ1, we identified a relationship between participants' comments on the peer agents and their collaborative strategies. We found that participants' expectations of the peer agents relate to the participants' values, which they aimed to maintain during the collaboration process, and their expected benefits from peer agents. Drawing from these observed correlations, we categorize participants into two distinct types, characterized as follows: Efficiency-Driven Learners vs. Curiosity-Driven Learners. (See Table \ref{table:learner-types} for a summary of each type's characteristics.)

\begin{figure}[!ht]
    \centering
    \includegraphics[trim={0.7cm 5.5cm 0.9cm 1cm}, clip, width=\linewidth]{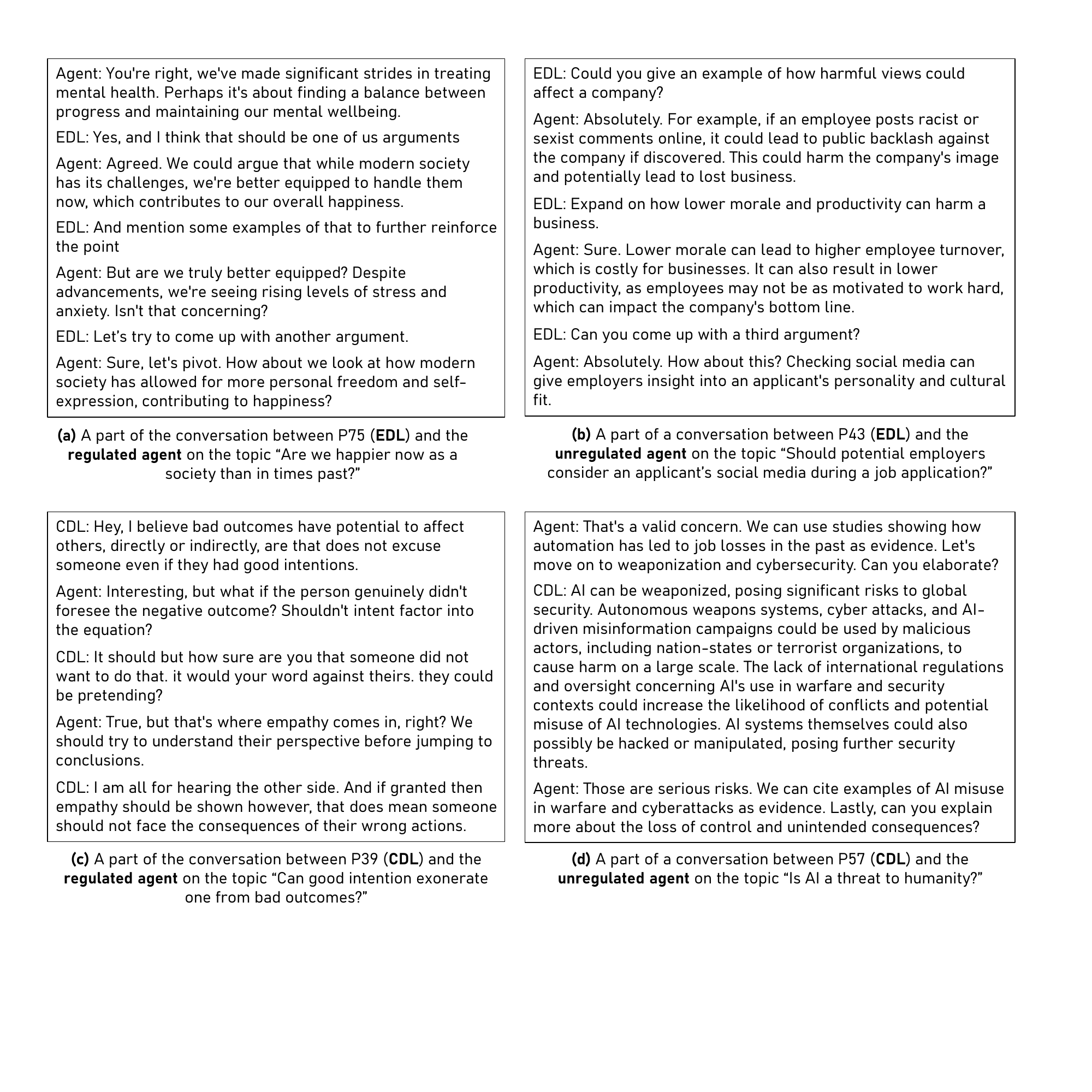}
    \caption{Examples of conversations between (a) a EDL participant and the regulated agent, (b) a EDL participant and the unregulated agent, (c) an CDL participant and the regulated agent, and (d) an CDL participant and the unregulated agent.}
    \Description{In Subfigure a, the EDL participant agrees with an argument the agent presents and tells them to come up with more examples. In Subfigure b, the EDL participant tells the agent to provide evidence for an argument with an example and then to come up with another argument. In Subfigure c, the CDL participant argues with the agent, who points out a possible exception to the participant's argument. In Subfigure d, the CDL participant asks the agent to explain more about its pushing back to the participant's point.}
    \label{fig:collab-strategies}
\end{figure}

\textbf{\emph{Efficiency-Driven Learners (EDL)} are goal-oriented learners who appreciate cooperation and productivity in the collaborative process.} During the task, they especially value their control over the process and their sense of agency. Therefore, they appreciated the peer agent's being ``supportive of [their] arguments" (P4, P45, P54) by providing relevant data rather than providing diverging opinions (P3, P12). They expect the peer agent to follow their initiatives and requests, creating straightforward dialogues that contribute towards the final result (the argument write-up) (P43, P65, P67). Following these expectations, EDL participants often lead the discussions at a high level to maintain consistent progress toward the arguments write-up. They actively decide when to move on from topics, demonstrating a focus on efficient progression (see the example of P75 in Figure \ref{fig:collab-strategies}a). Additionally, they direct the peer agent with specific requests, such as asking the agent to come up with a new argument or to write up their discussion in a specific format, expecting the agent to comply (see the example of P43 in Figure \ref{fig:collab-strategies}b). We categorized 77 out of 144 participants as EDL.

\textbf{\emph{Curiosity-Driven Learners (CDL)} are explorative learners who are open to diverse perspectives and various collaboration styles}. During the collaboration process, they enjoy getting challenged (P27, P49) and exposed to a perspective they have not considered before (P15, P57). Even in extreme cases, they would prefer to ``interact with someone who was absolutely stubborn rather than someone who agreed with everything I had to say." (P61) Through introducing challenges, they expect the agents to help create ``more creative and innovative solutions" (P51). However, they do not have clear expectations of the peer agent's collaborative behaviors and welcome dynamic interactions. Contradictory behaviors of the agents and the potential negative emotions that may arise from them are viewed as ``reflective of the real world" (P32) and make the peer agent seem more like humans (P48, P50). In the argumentation task, CDL participants often engage in more free-form conversation during argumentation tasks, without clear points where they decide to wrap up an argument and move on to the next. Usually, the conversation takes the form of the participant defending their stance from the agent's rebuttal (see examples of P39 in Figure \ref{fig:collab-strategies}c). As CDL participants request the peer agent to write up the arguments less often, the agent may also sometimes take the lead and ask them to write up the arguments instead (see examples of P57 in Figure \ref{fig:collab-strategies}d). We categorized 67 out of 144 participants as CDL.

\begin{table}[!ht]
\small
\centering
\begin{tabular}{|p{3cm}|p{4.5cm} p{4.5cm}|} 
 \hline
 \textbf{Characteristics} & \textbf{Efficiency-Driven Learners (EDL)} & \textbf{Curiosity-Driven Learners (CDL)} \\
 \hline
 \hline
 Values to maintain during collaboration & Sense of agency and control & Curiosity and adaptability \\ 
 \hline
 Expectation of the peer agent’s behaviors & Providing data to support learner’s opinions & Providing challenging opinions \\
 \hline
 Expected benefits from peer agents & Productivity, ability to process a large amount of data & Creativity, an opportunity to improve soft skills for real-world collaborations\\
 \hline
 Strategies for collaborating with peer agents & Taking a lead in determining conversation flows and making requests for peer agents to complete sub-tasks & Discussing or debating with the peer agents and incorporating new ideas into their deliverables on their own \\
 \hline
\end{tabular}
\caption{A table summarizing the characteristics of each learner type, consisting of their values, expectations from the peer agents, and their strategies when collaborating with the peer agents on argumentation tasks.}
\Description{The characteristics are as described in Section 4.1, with the first column being characteristics as listed in the caption, the second column being the characteristics of EDLs and the third CDLs.}
\label{table:learner-types}
\end{table}

Although we present the learners' orientations as binary categories, there are participants who do not fit neatly into them. For example, an EDL participant may argue briefly before focusing on the writing task, and a CDL participant may ask the peer agent to give a summary after a long debate. We discuss the learner categorization approach further in Section \ref{discuss:categorization}.

\subsection{Effects of learners' orientations and peer agents' behavior mechanisms on the learning process}

In this section, we will answer Sub-RQ2 by examining the effect of different CC behavior mechanisms (unregulated and regulated) on the learning process, and how these effects correlate with the binary groupings of learners' characteristics.

\subsubsection{The regulated agent increases turn count and task time but not the argument quality}
\label{finding2-1}
\begin{figure}[!ht]
    \centering
    \includegraphics[trim={2cm 7.5cm 2cm 8cm}, clip, width=\linewidth]{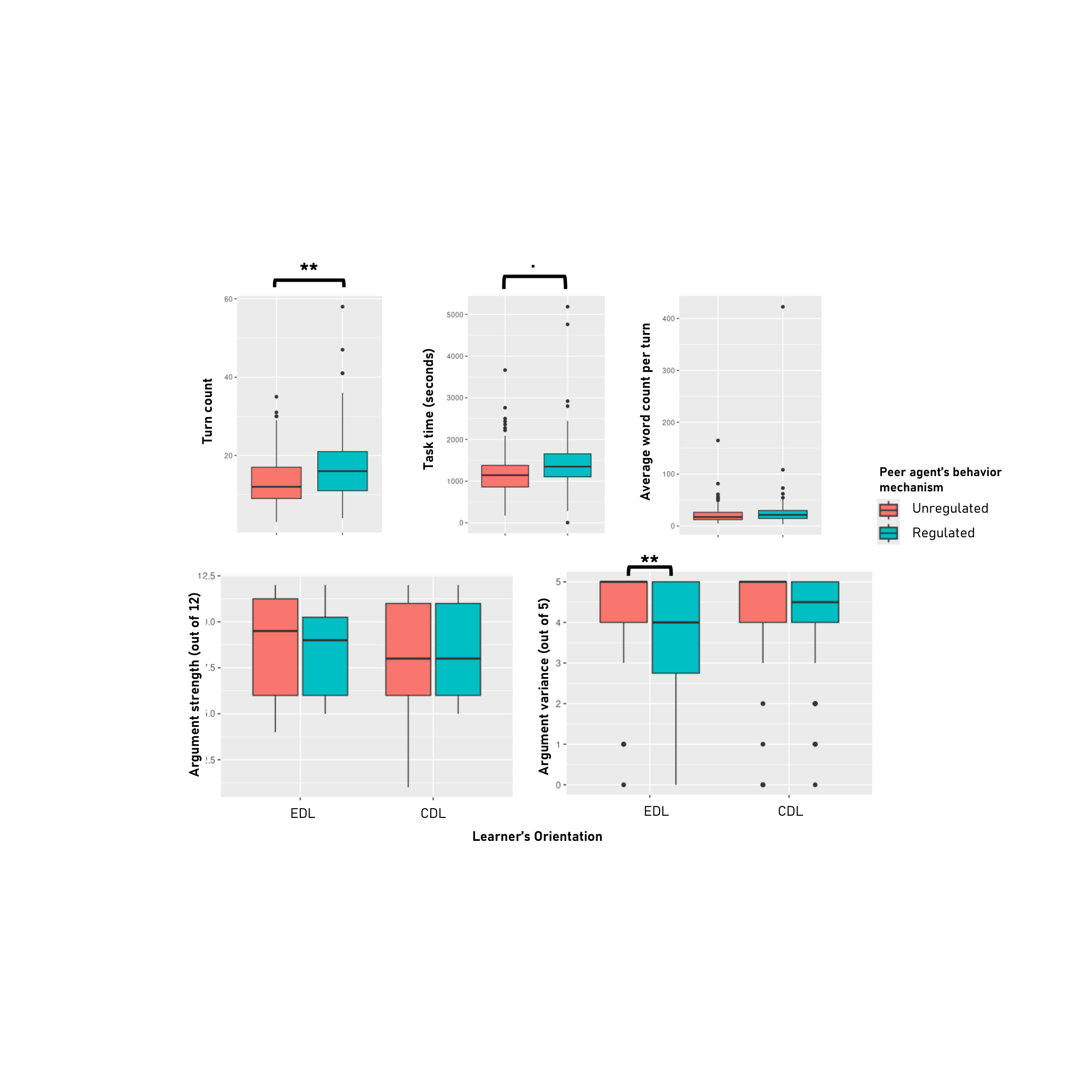}
    \caption{Box plots showing the participants' observed measures of engagement from the task and their argument scores. The three plots in the upper rows are for turn counts, task completion time, and average word count per turn, respectively. All three plots are of data aggregated from all 72 participants. The two plots in the lower row are on argument strength (out of 10) and argument variance (out of 10), respectively. The X-axes are based on the learners' orientations. The colors of all five box plots indicate the peer agent's behavior mechanisms. The statistically significant comparisons are marked with asterisks (*: $p<.05$, **: $p<.01$, ***:$p<.001$).}
    \Description{The figure shows that there are significant differences (p<.01) for turn count and (p<.1) for task time, with the regulated agent having higher averages in both cases. For EDLs only, the unregulated agent results in significantly higher argument variance (p<.01).}
    \label{fig:proj3-taskmeasures}
\end{figure}

\begin{figure}[!ht]
    \centering
    \includegraphics[trim={2cm 12cm 2cm 9cm}, clip, width=\linewidth]{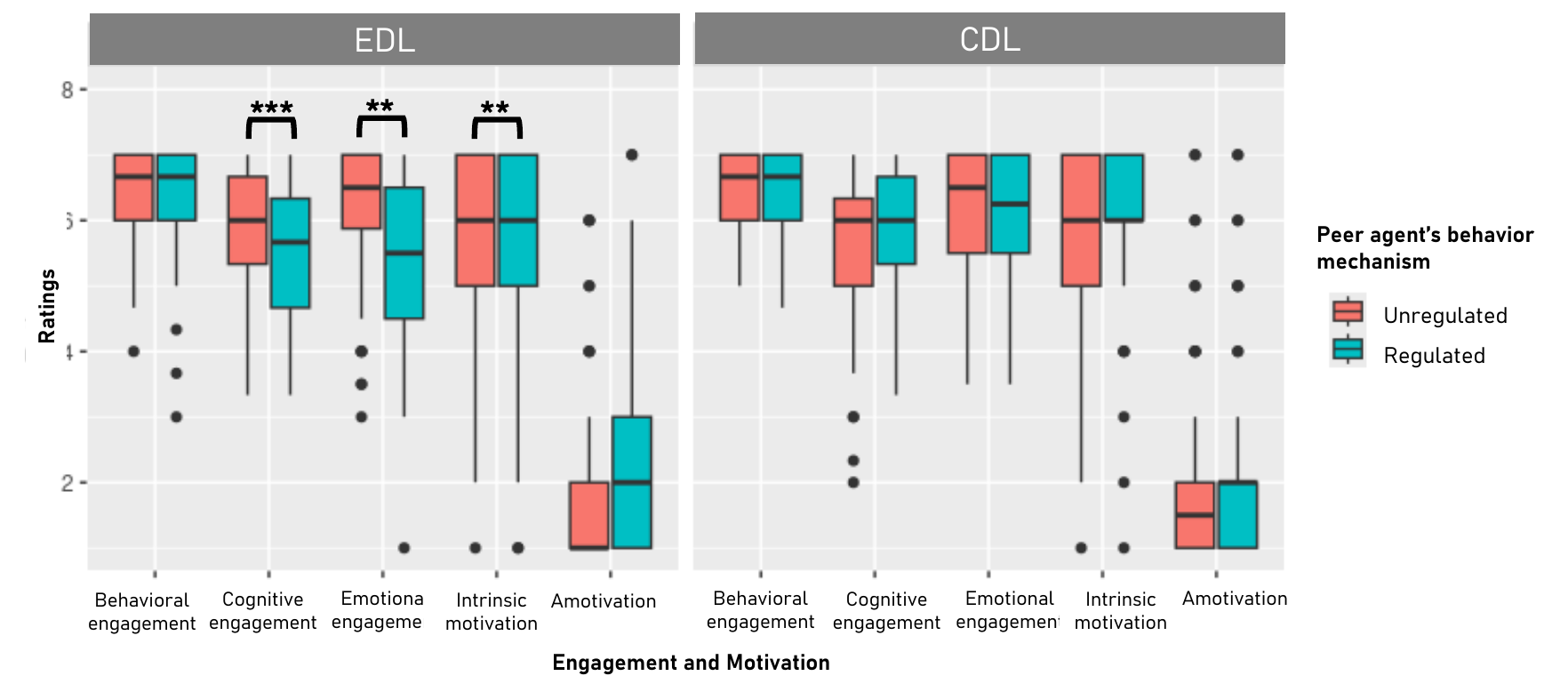}
    \caption{Box plots showing the participants' self-reported behavioral engagement, emotional engagement, cognitive engagement, intrinsic motivation, and amotivation. The plots are separated by learners' orientations. The colors of the box plots indicate the peer agent's behavior mechanisms. The statistically significant comparisons are marked with asterisks (*: $p<.05$, **: $p<.01$, ***:$p<.001$).}
    \Description{There's no significant statistical difference for CDL. For EDLs, there are significant differences (p<.01) for emotional engagement and intrinsic motivation and (p<.001) for cognitive engagement with the unregulated agent being higher in all cases.}
    \label{fig:proj3-engagement}
\end{figure}

CDL and EDL participants had similar experiences in terms of the observed measures of engagement from the task. Across all participants, regulated agents resulted in significantly higher turn counts (t(139.7758) = 3.682, $p < .05, \eta^2 = 0.15$) and task time (t(92.6760) = 2.630, $p < .1, \eta^2 = 0.16$) compared to unregulated ones (Figure \ref{fig:proj3-taskmeasures}). These measures suggest that participants have higher behavioral engagement with the regulated agent, although there is no difference in the self-reported ratings (Figure \ref{fig:proj3-engagement}).

The longer discussions, however, do not correlate with the argument scores, either in argument strength or argument variance. For CDLs, there are no significant differences between the argument scores. For EDLs, there are significant differences in reverse, where their collaboration with unregulated agents results in significantly higher argument variance (t(72.8879) = -3.582, $p < .01, \eta^2=0.20$) than with regulated agents and no significant differences in argument strength. A possible explanation of this phenomenon is the unregulated agent's ability to be ``organized and methodical" (P52) and ``keep the discussion on track" (P21) contributes to higher argument scores. This contrasts with how the regulated agent sometimes ``gets sidetracked and gets off-topic criticizing your arguments" (P2). As the regulated agent would not move on until they are satisfied with the discussion of the current issue, the participant might "spend more time trying to explain my point" (P13). P31 also explained that "I did more work than necessary by having to run all of my ideas by them". Although CDL participants did not criticize the longer discussion in any way, a lot of their discussions also did not transfer to their arguments, supporting P2's comment. P47, whose values fit with the CDL type and achieved higher argument scores after a longer discussion with the regulated agent, explained that they had to develop a new strategy to work with the agent. Instead of learning during the communication as they would when collaborating with humans, they "learned a few things in the process of organizing the arguments." 

\subsubsection{The unregulated agent's accommodative behaviors support EDLs' emotional engagement and sense of agency.}

\begin{figure}[!ht]
    \centering
    \includegraphics[trim={3cm 11.7cm 3cm 11.2cm}, clip, width=0.95\linewidth]{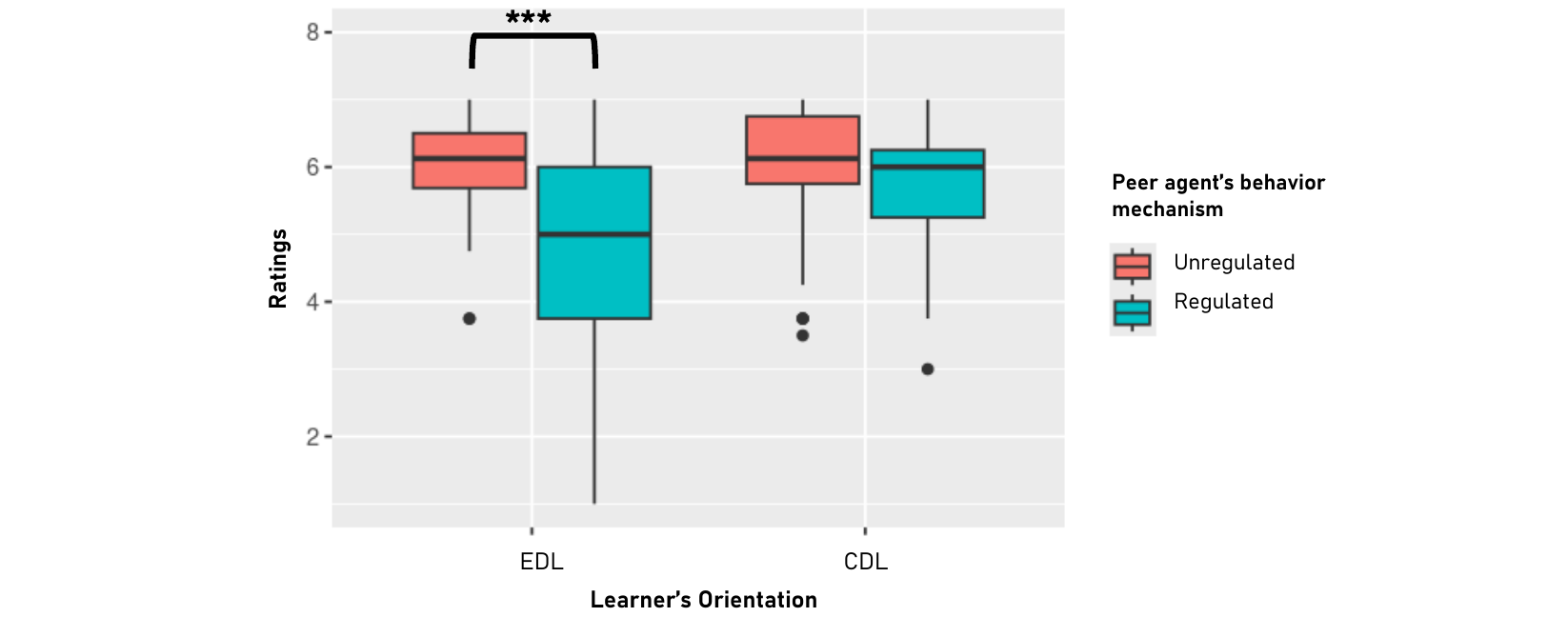}
    \caption{Box plots showing the participants' self-reported sense of agency and ownership over the outcomes. The plots are separated by learners' orientations. The colors of the box plots indicate the peer agent's behavior mechanisms. The statistically significant comparisons are marked with asterisks (*: $p<.05$, **: $p<.01$, ***:$p<.001$).}
    \Description{For EDL, the unregulated agent shows a higher average agency and ownership than the regulated agent (approximately 6 to 5) with a p-value less than 0.001.}
    \label{fig:proj3-agency}
\end{figure}

The interplay between learners' values and the agent's behavior mechanisms is observed in the learners' sense of agency, emotional engagement, cognitive engagement, and intrinsic motivation. The sense of agency and control is the central value of EDLs. EDLs rated that the unregulated agent is significantly better than the regulated agent at fostering these values (t(73.1476) = -6.731, $p < .001, \eta^2=0.40$) (Figure \ref{fig:proj3-agency}). Unregulated agents also significantly induce higher emotional engagement (t(73.2104) = -6.243, $p < .001, \eta^2=0.32$) and intrinsic motivation (t(73.0742) = -3.178, $p < .01, \eta^2=0.06$) than regulated agents for EDLs. Emotions are crucial for EDLs' experiences, and many EDL participants explained their preference for unregulated agents by describing them as "friendly" (P10, P27), "on my side" (P30), and "easy to communicate/deal with" (P1, P11), making the experience "more relaxed" (P36) and "enjoyable" (P42), while the regulated agent was "frustrating" (P65). 

In contrast, for CDLs, although we see a trend of lower agency and ownership with the regulated agent (t(64.3512) = -2.386, $p < .1, \eta^2=0.09$), it doesn't correlate with or impact their emotional engagement at all, aligning with how CDLs do not have emotional needs for agency and control. Despite this lack of significant difference in emotional engagement between the behavior mechanisms, 56 out of 67 CDL participants expressed their preferences for regulated agents, citing that the collaboration "feels more balanced" (P58) as a "50/50 effort" (P36).

\subsection{Effects of learners' orientations and peer agents' behavior mechanisms on learners' perception of peer agents}
The taxonomy of learners' orientations provides a ground for further analyzing how their different collaborative strategies impact their perceptions of the peer agent's CC behaviors.

\subsubsection{Participants perceive unregulated agents as 'Too Cooperative' and regulated as 'Too Contradictory'.}
\label{finding1-1}

\begin{figure}[!ht]
    \centering
    \includegraphics[trim={2cm 11.5cm 2cm 10cm}, clip, width=\linewidth]{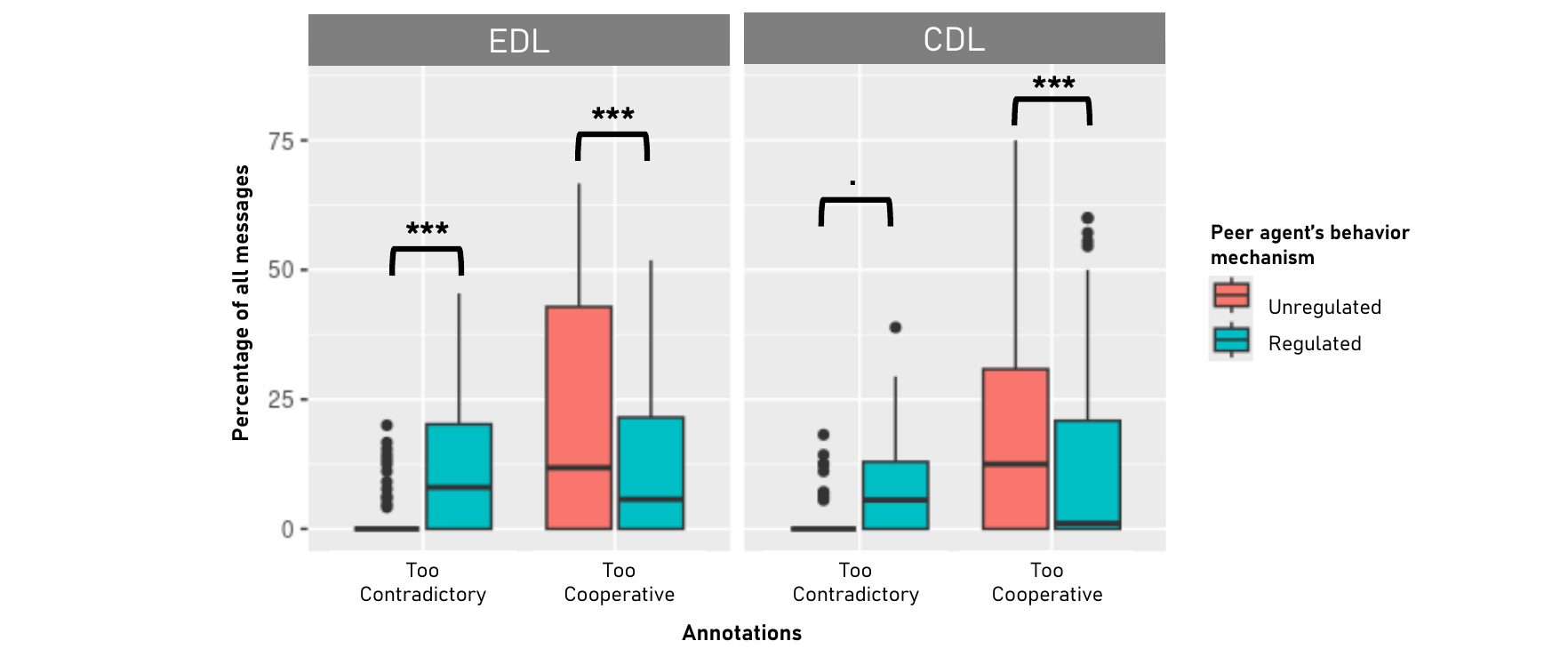}
    \caption{Box plots showing the percentages of peer agents' messages that are annotated as Too Cooperative and Too Contradictory, regardless of whether they are compared to AI peers or human peers. The plots are separated by learners' orientations. The colors of the box plots indicate the peer agent's behavior mechanisms. The statistically significant comparisons are marked with asterisks (*: $p<.05$, **: $p<.01$, ***: $p<.001$).}
    \Description{For both EDLs and CDLs, there are significant differences. For too cooperative, the unregulated agent is more frequently annotated than the regulated one and vice versa for too contradictory annotations. All significance level is p less than 0.001, with the exception of CDL and too contradictory annotation being only the level of p-value less than 0.1.}
    \label{fig:annotations}
\end{figure}

There is a significant effect of behavioral mechanisms for chat annotations (Figure \ref{fig:annotations}) across both orientations. Participants annotated unregulated agents as "Too Cooperative" significantly more frequently than regulated agents (t(140.0000) = -4.629, $p < .001, \eta^2=0.24$), and similarly annotated regulated agents as "Too Contradictory" significantly more frequently than unregulated agents (t(139.9997) = 7.115, $p <. 001, \eta^2=0.0.35$). Furthermore, the significantly higher rating on the CC principle, 'shared goal', for the unregulated agent (t(138.2837) = -12.498, $p <. 001, \eta^2=0.59$) supports these annotations. This indicates that the contradictory behaviors, regulated to be like humans in the regulated agents, are perceived as too extreme. On the contrary, the more assistive unregulated agent may also feel too cooperative for collaborative tasks.

\subsubsection{Learners' expectations influence their sensitivity to different peer agents' information management abilities}
\begin{figure}[!ht]
    \centering
    \includegraphics[trim={0 11cm 0 10cm}, clip, width=\linewidth]{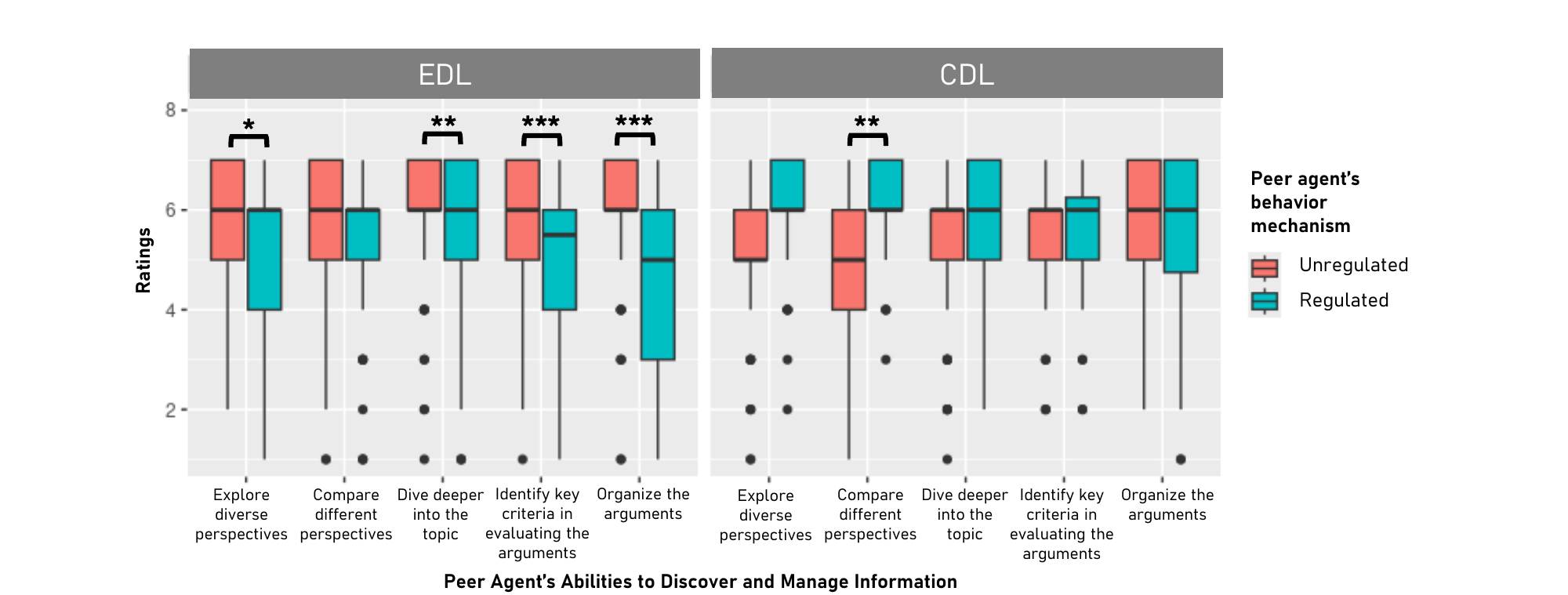}
    \caption{Box plots showing the peer agent's abilities to discover and manage information. The plots are separated by learners' orientations. The colors of the box plots indicate the peer agent's behavior mechanisms. The statistically significant comparisons are marked with asterisks (.: $p<.1$, *: $p<.05$, **: $p<.01$, ***: $p<.001$).}
    \Description{The only significant result for CDL is in comparing different perspectives, with the regulated agent being more highly rated than the unregulated agent. For all other abilities, the significant differences are with EDLs.}
    \label{fig:info-manage-conditions}
\end{figure}
For EDLs, who expect the agent to assist with information management, the unregulated agent outperforms the regulated agent in their ability to explore diverse perspectives (t(73.9997) = -2.858, $p < .05, \eta^2=0.13$), dive deeper into the topic (t(74.1271) = -3.208, $p < .01, \eta^2=0.13$), identify key criteria for evaluating the information (t(72.9921) = -2.690, $p < .001, \eta^2=0.13$), and organize information into arguments (t(74.5797) = -4.323, $p < .001, \eta^2=0.29$) (Figure \ref{fig:info-manage-conditions}). These results align with EDLs' strategies of asking AI to come up with arguments, which, according to the task instruction, includes both supporting one's own stance and anticipating potential rebuttals from the opposite stance. The unregulated agent is likely to comply with such requests, whereas the regulated agent may not follow their orders if they determine that there is a need for more push-back. 

CDLs perceive no significant differences in any of the four abilities mentioned above. However, they perceived the regulated agent to be significantly better than the unregulated one at comparing different perspectives (t(65.2934) = 3.504, $p < .01, \eta^2=0.34$), which is the one ability that shows no significant difference with EDLs. This contrast implies that the two learners' orientations are more aware of and more affected by different information management abilities in the agents.

\subsubsection{EDLs think the unregulated agent is better at skilled disagreement; CDLs think the regulated one is better.}
\begin{figure}[!ht]
    \centering
    \includegraphics[trim={0 23cm 2cm 1cm}, clip, width=\linewidth]{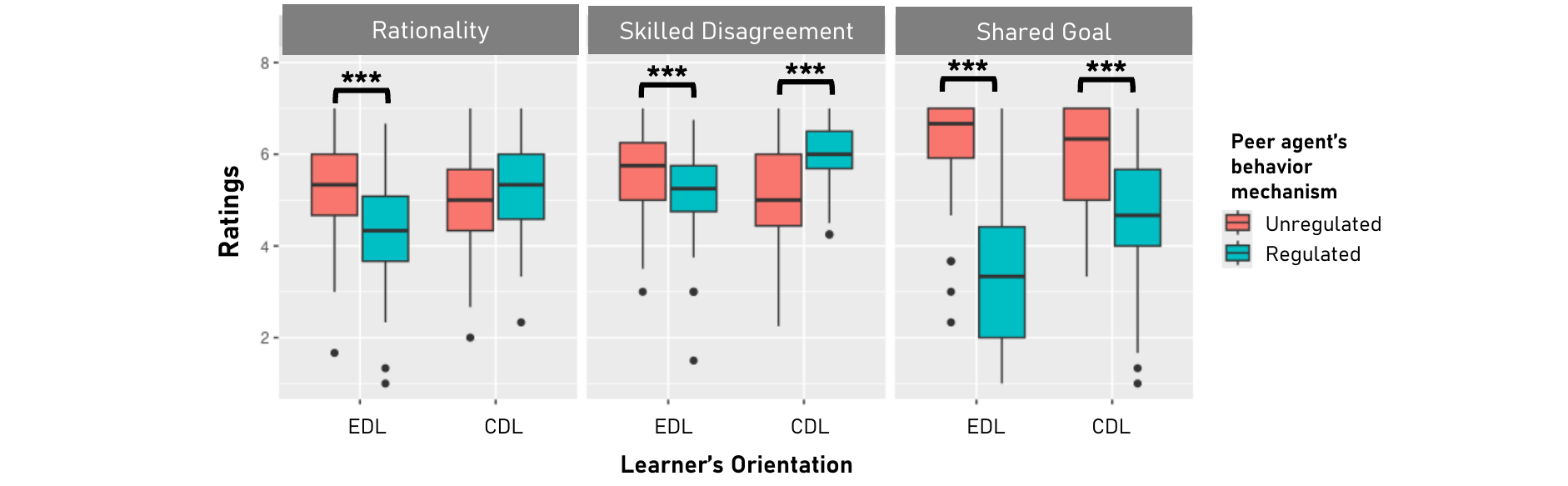}
    \caption{Box plots showing the participants' evaluation of the alignment between the peer agent's behaviors and CC principles. The plots are separated by learners' orientations. The colors of the box plots indicate the peer agent's behavior mechanisms. The statistically significant comparisons are marked with asterisks (.: $p<.1$, *: $p<.05$, ***: $p<.001$).}
    \Description{For EDL, the unregulated agent scores higher than the regulated agent on all three constructs. For CDLs, the significant results are in shared goals (the unregulated agent is higher than the regulated one) and skilled disagreement (the regulated one is higher than the unregulated one). All p-values are less than 0.001.}
    \label{fig:alignmentCC}
\end{figure}

We found a correlation between the resulting skilled disagreement behaviors and our characterization of learners' values and expectations. Firstly, we found that CDL participants rated the regulated agent as significantly better than the unregulated agent at skilled disagreement (t(65.1895) = 4.215, $p < .001, \eta^2=0.38$), while there is a trend of unregulated agent outperforming with EDL participants (t(74.0000) = -4.337, $p < .001, \eta^2=0.13$) (Figure \ref{fig:alignmentCC}). This implies that the moderator module's judgment aligns with CDL participants' expectations, while the default GPT judgments align better with EDL participants'.

\subsection{Effects of the disclosure of the peer agent's behavioral design on the learning process}
\label{finding4-1}
To address Sub-RQ3 regarding the effect of peer agent's design disclosure (or lack thereof), we first observed no main effect of disclosure across all participants, indicating its influence varied significantly between the two learning orientations (EDLs and CDLs). For EDLs, design transparency significantly improves the perception of the regulated agents on the ability to identify key criteria for evaluating arguments (t(73.9044) = 2.714, $p < .05, \eta^2=0.09$) and alignment to the "rationality" principle of CC (t(73.5176) = 2.321, $p < .05, \eta^2=0.07$ ). However, EDLs in the transparent group receive significantly lower argument strength scores than the opaque group, mainly due to the lack of address to rebuttals in the arguments (see grading scheme in Appendix \ref{appendix:argument-rubrics}). One possible explanation for this result is that their increased acceptance of the agent's contradictory behaviors may make them less inclined counter the agent's arguments. In contrast, for CDLs, the transparent disclosure of the agent's behavior design made them perceived both unregulated and regulated as being less capable in exploring diverse perspectives (t(128.7316) = -2.487, $p < .05, \eta^2=0.09$) compared to the opaque group. There is no significant effect in other measures, possibly suggesting that CDLs' learning experiences may be more resistant to differences in agents than EDLs. However, more study of causal relationship between the design disclosure and the learners' understanding of the system is needed to gain insights into these statistical results.

\section{Discussion}
\label{section:discussion}

\subsection{Applications of learner types}
\label{discuss:categorization}
Our proposed categorization of learners into EDLs and CDLs provides a new model for understanding different interaction styles with CC-based peer agents. It shares structures and factors similar to many other existing taxonomies, including the bimodal classification and the emphasis on agency and control from the study on AI user's orientations \cite{kim2021utilitarian}, which informs our deductive analysis of the qualitative survey data, and the Openness dimension of the Big Five personality traits which grounds our final categorization \cite{chamorro2009mainly}. Our categorization expands on these two theories by illustrating the relationship between their values and expectations (aligning with AI user's orientations and Openness) and their collaboration strategies. This expansion allows instructors to categorize learners prior to interacting with AI agents and to prescribe an agent with an effective persona for achieving the learning goal.

In connection to other human-AI hybrids taxonomies, firstly, the expected peer agent's behaviors of EDL and CDL build upon the taxonomy of ``AI focus", which denotes the overarching purpose of an AI-enabled system in a human-AI hybrid and indicates whether the primary purpose of an AI-enabled system is automation (doing work in the place of humans, corresponding to EDL) or augmentation (doing work on top of or with humans, corresponding to CDL) \cite{fabri2023disentangling}. Our categorization connects the perceived purpose of an agent to learners' values, expectations, and collaboration strategies. Secondly, in contrast to Neef's two types of collaboration for human-agent teams, taskwork- and teamwork-oriented \cite{neef2006taxonomy}, the CC-based peer agent's disputatious behaviors lead to novel interaction dynamics, captured in our proposed learners' orientations. Finally, EDL and CDL's values and expectations touch on the autonomy and competence aspect of Self-Determination Theory (SDT) \cite{deci2012self} and exemplifies how SDT can be applied to AI-driven applications to promote motivations for different types of learners.

When applying or adapting the learners' orientations to different contexts, it is crucial to consider the characteristics of this categorization that may require expansion or redefinition. Firstly, as mentioned in \ref{finding1}, our assumption that learners' behaviors form a spectrum is based on the observed frequency of different collaborative strategies, and our approach to defining the threshold is via learners' self-reported perception of the regulated and unregulated agent. In the context where there is a need to distinguish learners who equally exhibit both types of behaviors, a spectral model may be more appropriate. Moreover, other attributes may be needed for categorization with different tasks, such as tolerance for ambiguity \cite{zhou2024understanding} or preference for solitary versus collaborative work \cite{ke2006solitary}.

\subsection{Design implications}
\label{discuss:implication}
Our study generated three implications for designing collaborative peer agents:
\begin{itemize}
    \item \textbf{Adapt to learners' values and expectations:} Our study highlights the needs for different peer agent's behavior designs depending on learner types. This signals the need to detect learners' values and expectations. By collecting such information, the peer agent can be tailored to better support individual learning paths and preferences, enhancing personalization and effectiveness \cite{biermann2022tool}. This is especially crucial in asynchronous online courses, where learners have unique goals and circumstances \cite{ke2006solitary}. The negative emotions from interacting with peer agents that do not fit their needs may have a significant impact on their intrinsic motivation and make them drop out entirely \cite{park2009factors}.
    \item \textbf{Make agent's behavior regulation flexible:} While unregulated agents offer more flexibility, allowing learners to guide the direction of the collaboration, this lack of structure can sometimes lead to inefficiencies or reduced effectiveness in achieving learning objectives. Therefore, while it's beneficial to maintain some level of regulation to ensure the AI operates within desired parameters, it's also crucial to allow for adaptability. By finely tuning the regulation, designers can equip agents with specific behaviors that support structured learning outcomes while still providing learners the space to exert control over the interaction.
    \item \textbf{Provide strategic guidance for collaboration with the agent:} While disclosure of the agent's behavior guidance can negatively impact the learning process, a basic understanding of the agent's principles is essential for developing an effective collaboration strategy \cite{cabrera2023improving, zhang2021ideal}. Before interacting with the peer agent, the learner should receive guidelines on how the agent is likely to respond to various prompts, commands, or collaborative strategies. This approach would allow learners to tailor their strategies to leverage the AI's capabilities fully, without fostering inaccurate expectations about the agent’s functionality.
\end{itemize}

\subsection{Peer AI agent vs. human peer}
\label{discuss:ai-vs-human}
Our research motivation and design emphasize that peer AI agents are not intended to replace human peers but rather to supplement the learning experience for those without access to traditional peer interactions. This distinction is crucial as it sets the expectation that AI agents are designed to enhance, not substitute, the collaborative opportunities available to learners, especially in environments where human interaction is limited.

While this is not a part of our research, one common feedback from the participants is that the CC-based peer agents do not take the participants' emotions and interpersonal dynamics into consideration during the collaboration. While humans naturally employ the Theory of Mind (ToM) to understand and respond to others' emotions and intentions, LLMs still have a low degree of this capability \cite{street2024llm}. ToM in LLMs is only enough to encourage cooperativeness but not enough for navigating competitions or nuanced negotiations. For CC-based peer agents, this can be both a limitation and an advantage. In settings that require navigating contentious topics or constructive controversy, the lack of emotional response from AI agents can provide a neutral ground for debate \cite{tanprasert2024debate}. On the other hand, if the learner feels frustrated with the peer agent's contradictory behaviors, the peer agent may maintain their behaviors and end up inducing more negative feelings from the learner, unlike humans' natural inclination to preserve a positive atmosphere in a collaboration \cite{jehn1997interpersonal}.

\subsection{Variability by demographics}
The generalizability of these findings may be constrained by our participant pool's demographic profile. The high reported comfort level with intelligent chatbots (131/144) and the relatively young average age (M=24.76) reflect a technical expertise that likely exceeds the general online student population \cite{chung2020online}, potentially overstating the ease of AI agent adoption. Secondly, we only recruited undergraduate-level participants from three English-speaking countries. With different educational systems in different education levels and different regions, the learners’ values, expectations, and strategies may be different from our study \cite{saade2007exploring}. Demographic variables, such as socioeconomic status and genders may also impact the application of our findings, as they have been shown to strongly correlate preferences learning strategies in some online learning environments \cite{colorado2012student, thapa2023perception} and may interact with the AI agent's intervention to impact learners' performance as well.

Beyond generalizability, previous research has shown that manipulating the AI agent’s persona (e.g., their ethnicity and gender cues) to align or misalign with the learners can affect the learners’ engagement and perception of the agents \cite{tanprasert2024debate}. In this study, we remove all cues from the agents, but when combining with other dimensions of AI agent’s persona design, there may be significant interaction affects between those persona attributes and CC behaviours on the learners that requires further exploration.

\subsection{Integrating human-AI collaboration into the curriculum}
\label{discuss:educator}
The process of designing and running our study has revealed that integrating human-AI teams into traditional classroom tasks involves unique challenges. We identified two primary concerns for educators wishing to incorporate such activities: the strategies for collaborating with peer AI agents and the evaluation criteria and task structures.

As mentioned in Finding \ref{finding2-1}, collaboration with peer agents requires a different strategy from collaboration with human peers. As with many other LLM applications, this poses metacognitive challenges \cite{tankelevitch2024metacognitive}, as learners have to adjust and develop new strategies to be able to use the system effectively. Therefore, educators looking to incorporate AI-driven collaborative tasks should consider providing resources or instruction on effective AI collaboration techniques. This support is crucial to ensure the learners are equipped to navigate and maximize their interactions with the peer agents according to their values, needs, and learning environments.

Finally, our experience developing the argument grading rubrics for the exploratory study has highlighted the problem that introducing an AI agent into group work makes the assessment of the student's contributions and learning outcomes complicated. Unlike traditional group projects, where all contributions are purely human, AI involvement can obscure how much work each student has done and what they have learned from the process. Educators may need to restructure the learning activities and evaluation criteria to more accurately reflect each student’s actual input and learning outcomes. Adjusting these aspects ensures that assessments remain indicative of the true learning gains.

\subsection{Limitations and future work}
\label{discuss:limitation}
As mentioned in Section \ref{subsec:method-participants}, our participants have the average age of less than 25 years old, have used LLMs before, and come from various educational background. Their past experiences with LLMs may have formed specific expectations on how the peer agent should behave. Considering that our studies are motivated by learners in asynchronous online courses, this level of experience may be expected, but this also implies that learners who never had access to LLMs before or older adult learners may have different experiences. Future studies should further investigate the impact of age, past experiences, and areas of study on perceptions and acceptance of novel AI agent's behaviors.

Our experimental task presents two limitations to the study. Firstly, the debate topics in this study, which were designed to be equally complex and deliverable for all participants, were not part of individual participants' coursework and did not align with their fields of study and may have affected the participant's ability to contribute meaningfully. Future research should further explore how the insights from this study interact with the learner's level of expertise and apply to different subjects. Secondly, the criteria for evaluating arguments were solely focused on logical coherence, as we considered it most relevant for the argumentation task and presented an appropriate level of challenges for the study time limit. While this approach highlighted specific cognitive and collaborative skills, future studies should explore additional dimensions of argument quality, such as persuasiveness and clarity, to deepen our understanding of collaborative peer agents on the learning outcome.

Regarding the study design, there were three limitations that should be addressed in future work. Firstly, although we did not impose any expectations on the users to behave in a certain way, the disclosure of agent's behaviors in the transparent condition could align participants' behaviors to perceived expectations. Secondly, our study also focuses on exploring the knowledge co-construction process, we didn't compare the performance and experience of work solo as the baseline condition. Finally, our study was also controlled, conducted in only one session, and was the participants' first exposure to CC-based collaborative peer agents. Although it allowed us to observe learners' initial expectations and reactions, it did not give learners sufficient time to learn to utilize the peer agents effectively. Future research could benefit from a longitudinal study with an existing online course to improve the study's ecological validity, addressing the potential behavioral adjustment to perceived researchers' expectations and the limitation of controlled study setup. It can also provide deeper insights into how prolonged interaction with AI influences the learning processes and allow us to compare between the performance of students who take such courses with and without AI peers.
\section{Conclusion}
\label{section:conclusion}

In this study, we explored the design considerations of collaborative peer AI agents within asynchronous online learning environments to address the lack of real-time peer interactions. Following the theoretical framework of Constructive Controversy (CC), we focused on collaborative peer agents with different opinions from the learner, to foster knowledge co-construction through negotiating different perspectives. Through a mixed-method, exploratory study, where participants interacted with two different CC behavior mechanisms (regulated vs. unregulated) and agent's behavior design disclosure levels (transparent and opaque), we identified two types of learners based on their values and expectations of the peer agent's behaviors. The learner types interact with the agent's behavior mechanisms, affecting the learner's collaborative strategies, engagement, and sense of agency. On the other hand, the disclosure of the agent's behavior design decreases the learner's perception of the agent's abilities across all learners. The contribution of our research is the foundational insight for designing not only the CC behaviors for collaborative peer agents but also other complex and nuanced interactions of LLM-based agents. Future research should focus on long-term studies to evaluate the sustainability of these interventions and expand broader contexts in which peer AI agents can supplement the learning experience for isolated learners.

\begin{acks}
This work was supported by the Institute of Information \& Communications Technology Planning \& Evaluation (IITP) grant funded by the Korea government (MSIT) (No. RS-2025-02303220, Development of a system for evaluating and validating the UI/UX of digital services for underprivileged groups).
\end{acks}

\bibliographystyle{ACM-Reference-Format}

\begin{thebibliography}{126}


\ifx \showCODEN    \undefined \def \showCODEN     #1{\unskip}     \fi
\ifx \showDOI      \undefined \def \showDOI       #1{#1}\fi
\ifx \showISBNx    \undefined \def \showISBNx     #1{\unskip}     \fi
\ifx \showISBNxiii \undefined \def \showISBNxiii  #1{\unskip}     \fi
\ifx \showISSN     \undefined \def \showISSN      #1{\unskip}     \fi
\ifx \showLCCN     \undefined \def \showLCCN      #1{\unskip}     \fi
\ifx \shownote     \undefined \def \shownote      #1{#1}          \fi
\ifx \showarticletitle \undefined \def \showarticletitle #1{#1}   \fi
\ifx \showURL      \undefined \def \showURL       {\relax}        \fi
\providecommand\bibfield[2]{#2}
\providecommand\bibinfo[2]{#2}
\providecommand\natexlab[1]{#1}
\providecommand\showeprint[2][]{arXiv:#2}

\bibitem[Achiam et~al\mbox{.}(2023)]%
        {achiam2023gpt}
\bibfield{author}{\bibinfo{person}{Josh Achiam}, \bibinfo{person}{Steven Adler}, \bibinfo{person}{Sandhini Agarwal}, \bibinfo{person}{Lama Ahmad}, \bibinfo{person}{Ilge Akkaya}, \bibinfo{person}{Florencia~Leoni Aleman}, \bibinfo{person}{Diogo Almeida}, \bibinfo{person}{Janko Altenschmidt}, \bibinfo{person}{Sam Altman}, \bibinfo{person}{Shyamal Anadkat}, {et~al\mbox{.}}} \bibinfo{year}{2023}\natexlab{}.
\newblock \showarticletitle{Gpt-4 technical report}.
\newblock \bibinfo{journal}{\emph{arXiv preprint arXiv:2303.08774}} (\bibinfo{year}{2023}).
\newblock


\bibitem[Ahn et~al\mbox{.}(2024)]%
        {ahn2024impact}
\bibfield{author}{\bibinfo{person}{Daehwan Ahn}, \bibinfo{person}{Abdullah Almaatouq}, \bibinfo{person}{Monisha Gulabani}, {and} \bibinfo{person}{Kartik Hosanagar}.} \bibinfo{year}{2024}\natexlab{}.
\newblock \showarticletitle{Impact of Model Interpretability and Outcome Feedback on Trust in AI}. In \bibinfo{booktitle}{\emph{Proceedings of the CHI Conference on Human Factors in Computing Systems}}. \bibinfo{pages}{1--25}.
\newblock


\bibitem[Albacete et~al\mbox{.}(2015)]%
        {albacete2015dialogue}
\bibfield{author}{\bibinfo{person}{Patricia Albacete}, \bibinfo{person}{Pamela Jordan}, {and} \bibinfo{person}{Sandra Katz}.} \bibinfo{year}{2015}\natexlab{}.
\newblock \showarticletitle{Is a dialogue-based tutoring system that emulates helpful co-constructed relations during human tutoring effective?}. In \bibinfo{booktitle}{\emph{Artificial Intelligence in Education: 17th International Conference, AIED 2015, Madrid, Spain, June 22-26, 2015. Proceedings 17}}. Springer, \bibinfo{pages}{3--12}.
\newblock


\bibitem[Andriessen and Baker(2006)]%
        {andriessen2006arguing}
\bibfield{author}{\bibinfo{person}{Jerry Andriessen} {and} \bibinfo{person}{Michael Baker}.} \bibinfo{year}{2006}\natexlab{}.
\newblock \bibinfo{booktitle}{\emph{Arguing to learn}}.
\newblock \bibinfo{publisher}{na}.
\newblock


\bibitem[Angerschmid et~al\mbox{.}(2022)]%
        {angerschmid2022fairness}
\bibfield{author}{\bibinfo{person}{Alessa Angerschmid}, \bibinfo{person}{Jianlong Zhou}, \bibinfo{person}{Kevin Theuermann}, \bibinfo{person}{Fang Chen}, {and} \bibinfo{person}{Andreas Holzinger}.} \bibinfo{year}{2022}\natexlab{}.
\newblock \showarticletitle{Fairness and explanation in AI-informed decision making}.
\newblock \bibinfo{journal}{\emph{Machine Learning and Knowledge Extraction}} \bibinfo{volume}{4}, \bibinfo{number}{2} (\bibinfo{year}{2022}), \bibinfo{pages}{556--579}.
\newblock


\bibitem[Asterhan and Babichenko(2015)]%
        {asterhan2015social}
\bibfield{author}{\bibinfo{person}{Christa~SC Asterhan} {and} \bibinfo{person}{Miriam Babichenko}.} \bibinfo{year}{2015}\natexlab{}.
\newblock \showarticletitle{The social dimension of learning through argumentation: Effects of human presence and discourse style.}
\newblock \bibinfo{journal}{\emph{Journal of Educational Psychology}} \bibinfo{volume}{107}, \bibinfo{number}{3} (\bibinfo{year}{2015}), \bibinfo{pages}{740}.
\newblock


\bibitem[Baysan et~al\mbox{.}(2025)]%
        {baysan2025llm}
\bibfield{author}{\bibinfo{person}{Mehmet~Selman Baysan}, \bibinfo{person}{Serkan Uysal}, \bibinfo{person}{{\.I}rem {\.I}{\c{s}}lek}, \bibinfo{person}{{\c{C}}a{\u{g}}la {\c{C}}{\i}{\u{g}}~Karaman}, {and} \bibinfo{person}{Tunga G{\"u}ng{\"o}r}.} \bibinfo{year}{2025}\natexlab{}.
\newblock \showarticletitle{LLM-as-a-Judge: automated evaluation of search query parsing using large language models}.
\newblock \bibinfo{journal}{\emph{Frontiers in Big Data}}  \bibinfo{volume}{8} (\bibinfo{year}{2025}), \bibinfo{pages}{1611389}.
\newblock


\bibitem[Becker-Asano(2008)]%
        {becker2008wasabi}
\bibfield{author}{\bibinfo{person}{Christian Becker-Asano}.} \bibinfo{year}{2008}\natexlab{}.
\newblock \bibinfo{booktitle}{\emph{WASABI: Affect simulation for agents with believable interactivity}}. Vol.~\bibinfo{volume}{319}.
\newblock \bibinfo{publisher}{IOS Press}.
\newblock


\bibitem[Bialkova(2024)]%
        {bialkova2024core}
\bibfield{author}{\bibinfo{person}{Svetlana Bialkova}.} \bibinfo{year}{2024}\natexlab{}.
\newblock \showarticletitle{Core theories applied in chatbot context}.
\newblock In \bibinfo{booktitle}{\emph{The rise of AI user applications: Chatbots integration foundations and trends}}. \bibinfo{publisher}{Springer}, \bibinfo{pages}{41--59}.
\newblock


\bibitem[Biermann et~al\mbox{.}(2022)]%
        {biermann2022tool}
\bibfield{author}{\bibinfo{person}{Oloff~C Biermann}, \bibinfo{person}{Ning~F Ma}, {and} \bibinfo{person}{Dongwook Yoon}.} \bibinfo{year}{2022}\natexlab{}.
\newblock \showarticletitle{From tool to companion: Storywriters want AI writers to respect their personal values and writing strategies}. In \bibinfo{booktitle}{\emph{Proceedings of the 2022 ACM Designing Interactive Systems Conference}}. \bibinfo{pages}{1209--1227}.
\newblock


\bibitem[Bouschery et~al\mbox{.}(2024)]%
        {bouschery2024artificial}
\bibfield{author}{\bibinfo{person}{Sebastian~G Bouschery}, \bibinfo{person}{Vera Blazevic}, {and} \bibinfo{person}{Frank~T Piller}.} \bibinfo{year}{2024}\natexlab{}.
\newblock \showarticletitle{Artificial Intelligence-Augmented Brainstorming: How Humans and AI Beat Humans Alone}.
\newblock \bibinfo{journal}{\emph{Available at SSRN 4724068}} (\bibinfo{year}{2024}).
\newblock


\bibitem[Bu{\c{c}}inca et~al\mbox{.}(2021)]%
        {buccinca2021trust}
\bibfield{author}{\bibinfo{person}{Zana Bu{\c{c}}inca}, \bibinfo{person}{Maja~Barbara Malaya}, {and} \bibinfo{person}{Krzysztof~Z Gajos}.} \bibinfo{year}{2021}\natexlab{}.
\newblock \showarticletitle{To trust or to think: cognitive forcing functions can reduce overreliance on AI in AI-assisted decision-making}.
\newblock \bibinfo{journal}{\emph{Proceedings of the ACM on Human-computer Interaction}} \bibinfo{volume}{5}, \bibinfo{number}{CSCW1} (\bibinfo{year}{2021}), \bibinfo{pages}{1--21}.
\newblock


\bibitem[Cabrera et~al\mbox{.}(2023)]%
        {cabrera2023improving}
\bibfield{author}{\bibinfo{person}{{\'A}ngel~Alexander Cabrera}, \bibinfo{person}{Adam Perer}, {and} \bibinfo{person}{Jason~I Hong}.} \bibinfo{year}{2023}\natexlab{}.
\newblock \showarticletitle{Improving human-AI collaboration with descriptions of AI behavior}.
\newblock \bibinfo{journal}{\emph{Proceedings of the ACM on Human-Computer Interaction}} \bibinfo{volume}{7}, \bibinfo{number}{CSCW1} (\bibinfo{year}{2023}), \bibinfo{pages}{1--21}.
\newblock


\bibitem[Cai et~al\mbox{.}(2024)]%
        {cai2024antagonistic}
\bibfield{author}{\bibinfo{person}{Alice Cai}, \bibinfo{person}{Ian Arawjo}, {and} \bibinfo{person}{Elena~L Glassman}.} \bibinfo{year}{2024}\natexlab{}.
\newblock \showarticletitle{Antagonistic AI}.
\newblock \bibinfo{journal}{\emph{arXiv preprint arXiv:2402.07350}} (\bibinfo{year}{2024}).
\newblock


\bibitem[Castillo et~al\mbox{.}(2015)]%
        {castillo2015moocs}
\bibfield{author}{\bibinfo{person}{Nathan~M Castillo}, \bibinfo{person}{Jinsol Lee}, \bibinfo{person}{Fatima~T Zahra}, {and} \bibinfo{person}{Daniel~A Wagner}.} \bibinfo{year}{2015}\natexlab{}.
\newblock \showarticletitle{MOOCS for development: Trends, challenges, and opportunities}.
\newblock \bibinfo{journal}{\emph{Information Technologies \& International Development}} \bibinfo{volume}{11}, \bibinfo{number}{2} (\bibinfo{year}{2015}), \bibinfo{pages}{pp--35}.
\newblock


\bibitem[Chakraborti and Kambhampati(2018)]%
        {chakraborti2018algorithms}
\bibfield{author}{\bibinfo{person}{Tathagata Chakraborti} {and} \bibinfo{person}{Subbarao Kambhampati}.} \bibinfo{year}{2018}\natexlab{}.
\newblock \showarticletitle{Algorithms for the greater good! on mental modeling and acceptable symbiosis in human-ai collaboration}.
\newblock \bibinfo{journal}{\emph{arXiv preprint arXiv:1801.09854}} (\bibinfo{year}{2018}).
\newblock


\bibitem[Chamorro-Premuzic and Furnham(2009)]%
        {chamorro2009mainly}
\bibfield{author}{\bibinfo{person}{Tomas Chamorro-Premuzic} {and} \bibinfo{person}{Adrian Furnham}.} \bibinfo{year}{2009}\natexlab{}.
\newblock \showarticletitle{Mainly Openness: The relationship between the Big Five personality traits and learning approaches}.
\newblock \bibinfo{journal}{\emph{Learning and individual Differences}} \bibinfo{volume}{19}, \bibinfo{number}{4} (\bibinfo{year}{2009}), \bibinfo{pages}{524--529}.
\newblock


\bibitem[Chandra et~al\mbox{.}(2022)]%
        {chandra2022or}
\bibfield{author}{\bibinfo{person}{Shalini Chandra}, \bibinfo{person}{Anuragini Shirish}, {and} \bibinfo{person}{Shirish~C Srivastava}.} \bibinfo{year}{2022}\natexlab{}.
\newblock \showarticletitle{To be or not to be… human? Theorizing the role of human-like competencies in conversational artificial intelligence agents}.
\newblock \bibinfo{journal}{\emph{Journal of Management Information Systems}} \bibinfo{volume}{39}, \bibinfo{number}{4} (\bibinfo{year}{2022}), \bibinfo{pages}{969--1005}.
\newblock


\bibitem[Chi et~al\mbox{.}(2025)]%
        {chi2025thoughtsculpt}
\bibfield{author}{\bibinfo{person}{Yizhou Chi}, \bibinfo{person}{Kevin Yang}, {and} \bibinfo{person}{Dan Klein}.} \bibinfo{year}{2025}\natexlab{}.
\newblock \showarticletitle{Thoughtsculpt: Reasoning with intermediate revision and search}. In \bibinfo{booktitle}{\emph{Findings of the Association for Computational Linguistics: NAACL 2025}}. \bibinfo{pages}{7685--7711}.
\newblock


\bibitem[Chiang et~al\mbox{.}(2024)]%
        {chiang2024enhancing}
\bibfield{author}{\bibinfo{person}{Chun-Wei Chiang}, \bibinfo{person}{Zhuoran Lu}, \bibinfo{person}{Zhuoyan Li}, {and} \bibinfo{person}{Ming Yin}.} \bibinfo{year}{2024}\natexlab{}.
\newblock \showarticletitle{Enhancing ai-assisted group decision making through llm-powered devil's advocate}. In \bibinfo{booktitle}{\emph{Proceedings of the 29th International Conference on Intelligent User Interfaces}}. \bibinfo{pages}{103--119}.
\newblock


\bibitem[Christine~Wang et~al\mbox{.}(2001)]%
        {christine2001potential}
\bibfield{author}{\bibinfo{person}{X Christine~Wang}, \bibinfo{person}{D Michelle~Hinn}, {and} \bibinfo{person}{Alaina~G Kanfer}.} \bibinfo{year}{2001}\natexlab{}.
\newblock \showarticletitle{Potential of computer-supported collaborative learning for learners with different learning styles}.
\newblock \bibinfo{journal}{\emph{Journal of Research on Technology in Education}} \bibinfo{volume}{34}, \bibinfo{number}{1} (\bibinfo{year}{2001}), \bibinfo{pages}{75--85}.
\newblock


\bibitem[Chung et~al\mbox{.}(2020)]%
        {chung2020online}
\bibfield{author}{\bibinfo{person}{Ellen Chung}, \bibinfo{person}{Geetha Subramaniam}, {and} \bibinfo{person}{Laura~Christ Dass}.} \bibinfo{year}{2020}\natexlab{}.
\newblock \showarticletitle{Online learning readiness among university students in Malaysia amidst COVID-19.}
\newblock \bibinfo{journal}{\emph{Asian Journal of University Education}} \bibinfo{volume}{16}, \bibinfo{number}{2} (\bibinfo{year}{2020}), \bibinfo{pages}{46--58}.
\newblock


\bibitem[Colorado and Eberle(2012)]%
        {colorado2012student}
\bibfield{author}{\bibinfo{person}{Jozenia~Torres Colorado} {and} \bibinfo{person}{Jane Eberle}.} \bibinfo{year}{2012}\natexlab{}.
\newblock \showarticletitle{Student demographics and success in online learning environments.}
\newblock  (\bibinfo{year}{2012}).
\newblock


\bibitem[Corbin and Strauss(1990)]%
        {corbin1990grounded}
\bibfield{author}{\bibinfo{person}{Juliet~M Corbin} {and} \bibinfo{person}{Anselm Strauss}.} \bibinfo{year}{1990}\natexlab{}.
\newblock \showarticletitle{Grounded theory research: Procedures, canons, and evaluative criteria}.
\newblock \bibinfo{journal}{\emph{Qualitative sociology}} \bibinfo{volume}{13}, \bibinfo{number}{1} (\bibinfo{year}{1990}), \bibinfo{pages}{3--21}.
\newblock


\bibitem[de~F{\'a}tima~Goul{\~a}o and Menedez(2015)]%
        {de2015learner}
\bibfield{author}{\bibinfo{person}{Maria de F{\'a}tima~Goul{\~a}o} {and} \bibinfo{person}{Rebeca~Cerezo Menedez}.} \bibinfo{year}{2015}\natexlab{}.
\newblock \showarticletitle{Learner autonomy and self-regulation in eLearning}.
\newblock \bibinfo{journal}{\emph{Procedia-Social and Behavioral Sciences}}  \bibinfo{volume}{174} (\bibinfo{year}{2015}), \bibinfo{pages}{1900--1907}.
\newblock


\bibitem[De~Graaf and Allouch(2013)]%
        {de2013exploring}
\bibfield{author}{\bibinfo{person}{Maartje~MA De~Graaf} {and} \bibinfo{person}{Somaya~Ben Allouch}.} \bibinfo{year}{2013}\natexlab{}.
\newblock \showarticletitle{Exploring influencing variables for the acceptance of social robots}.
\newblock \bibinfo{journal}{\emph{Robotics and autonomous systems}} \bibinfo{volume}{61}, \bibinfo{number}{12} (\bibinfo{year}{2013}), \bibinfo{pages}{1476--1486}.
\newblock


\bibitem[Deci and Ryan(2012)]%
        {deci2012self}
\bibfield{author}{\bibinfo{person}{Edward~L Deci} {and} \bibinfo{person}{Richard~M Ryan}.} \bibinfo{year}{2012}\natexlab{}.
\newblock \showarticletitle{Self-determination theory}.
\newblock \bibinfo{journal}{\emph{Handbook of theories of social psychology}} \bibinfo{volume}{1}, \bibinfo{number}{20} (\bibinfo{year}{2012}), \bibinfo{pages}{416--436}.
\newblock


\bibitem[Dhillon et~al\mbox{.}(2024)]%
        {dhillon2024shaping}
\bibfield{author}{\bibinfo{person}{Paramveer~S Dhillon}, \bibinfo{person}{Somayeh Molaei}, \bibinfo{person}{Jiaqi Li}, \bibinfo{person}{Maximilian Golub}, \bibinfo{person}{Shaochun Zheng}, {and} \bibinfo{person}{Lionel~Peter Robert}.} \bibinfo{year}{2024}\natexlab{}.
\newblock \showarticletitle{Shaping Human-AI Collaboration: Varied Scaffolding Levels in Co-writing with Language Models}. In \bibinfo{booktitle}{\emph{Proceedings of the CHI Conference on Human Factors in Computing Systems}}. \bibinfo{pages}{1--18}.
\newblock


\bibitem[Fabri et~al\mbox{.}(2023)]%
        {fabri2023disentangling}
\bibfield{author}{\bibinfo{person}{Lukas Fabri}, \bibinfo{person}{Bj{\"o}rn H{\"a}ckel}, \bibinfo{person}{Anna~Maria Oberl{\"a}nder}, \bibinfo{person}{Marius Rieg}, {and} \bibinfo{person}{Alexander Stohr}.} \bibinfo{year}{2023}\natexlab{}.
\newblock \showarticletitle{Disentangling Human-AI Hybrids: Conceptualizing the Interworking of Humans and AI-Enabled Systems}.
\newblock \bibinfo{journal}{\emph{Business \& information systems engineering}} \bibinfo{volume}{65}, \bibinfo{number}{6} (\bibinfo{year}{2023}), \bibinfo{pages}{623--641}.
\newblock


\bibitem[Gambino et~al\mbox{.}(2020)]%
        {gambino2020building}
\bibfield{author}{\bibinfo{person}{Andrew Gambino}, \bibinfo{person}{Jesse Fox}, {and} \bibinfo{person}{Rabindra~A Ratan}.} \bibinfo{year}{2020}\natexlab{}.
\newblock \showarticletitle{Building a stronger CASA: Extending the computers are social actors paradigm}.
\newblock \bibinfo{journal}{\emph{Human-Machine Communication}}  \bibinfo{volume}{1} (\bibinfo{year}{2020}), \bibinfo{pages}{71--85}.
\newblock


\bibitem[Grundke et~al\mbox{.}(2024)]%
        {grundke2024aversion}
\bibfield{author}{\bibinfo{person}{Andrea Grundke}, \bibinfo{person}{Markus Appel}, {and} \bibinfo{person}{Jan-Philipp Stein}.} \bibinfo{year}{2024}\natexlab{}.
\newblock \showarticletitle{Aversion against machines with complex mental abilities: The role of individual differences}.
\newblock \bibinfo{journal}{\emph{Computers in Human Behavior: Artificial Humans}} \bibinfo{volume}{2}, \bibinfo{number}{2} (\bibinfo{year}{2024}), \bibinfo{pages}{100087}.
\newblock


\bibitem[Guizzardi et~al\mbox{.}(2003)]%
        {guizzardi2003agent}
\bibfield{author}{\bibinfo{person}{Renata~SS Guizzardi}, \bibinfo{person}{Lora Aroyo}, {and} \bibinfo{person}{Gerd Wagner}.} \bibinfo{year}{2003}\natexlab{}.
\newblock \showarticletitle{Agent-oriented knowledge management in learning environments: A peer-to-peer helpdesk case study}. In \bibinfo{booktitle}{\emph{International Symposium on Agent-Mediated Knowledge Management}}. Springer, \bibinfo{pages}{57--72}.
\newblock


\bibitem[Guo et~al\mbox{.}(2023)]%
        {guo2023effects}
\bibfield{author}{\bibinfo{person}{Kai Guo}, \bibinfo{person}{Yuchun Zhong}, \bibinfo{person}{Danling Li}, {and} \bibinfo{person}{Samuel Kai~Wah Chu}.} \bibinfo{year}{2023}\natexlab{}.
\newblock \showarticletitle{Effects of chatbot-assisted in-class debates on students’ argumentation skills and task motivation}.
\newblock \bibinfo{journal}{\emph{Computers \& Education}}  \bibinfo{volume}{203} (\bibinfo{year}{2023}), \bibinfo{pages}{104862}.
\newblock


\bibitem[Han et~al\mbox{.}(2007)]%
        {han2007effects}
\bibfield{author}{\bibinfo{person}{Keun-Woo Han}, \bibinfo{person}{Eun-Kyoung Lee}, {and} \bibinfo{person}{Young-Jun Lee}.} \bibinfo{year}{2007}\natexlab{}.
\newblock \showarticletitle{The effects of a peer agent on achievement and self-efficacy in programming education}.
\newblock \bibinfo{journal}{\emph{The Journal of Korean association of computer education}} \bibinfo{volume}{10}, \bibinfo{number}{5} (\bibinfo{year}{2007}), \bibinfo{pages}{43--51}.
\newblock


\bibitem[Hauptman et~al\mbox{.}(2023)]%
        {hauptman2023adapt}
\bibfield{author}{\bibinfo{person}{Allyson~I Hauptman}, \bibinfo{person}{Beau~G Schelble}, \bibinfo{person}{Nathan~J McNeese}, {and} \bibinfo{person}{Kapil~Chalil Madathil}.} \bibinfo{year}{2023}\natexlab{}.
\newblock \showarticletitle{Adapt and overcome: Perceptions of adaptive autonomous agents for human-AI teaming}.
\newblock \bibinfo{journal}{\emph{Computers in Human Behavior}}  \bibinfo{volume}{138} (\bibinfo{year}{2023}), \bibinfo{pages}{107451}.
\newblock


\bibitem[Hepenstal et~al\mbox{.}(2023)]%
        {hepenstal2023impact}
\bibfield{author}{\bibinfo{person}{Sam Hepenstal}, \bibinfo{person}{Leishi Zhang}, {and} \bibinfo{person}{BL~William Wong}.} \bibinfo{year}{2023}\natexlab{}.
\newblock \showarticletitle{The Impact of System Transparency on Analytical Reasoning}. In \bibinfo{booktitle}{\emph{Extended Abstracts of the 2023 CHI Conference on Human Factors in Computing Systems}}. \bibinfo{pages}{1--6}.
\newblock


\bibitem[Howard et~al\mbox{.}(2017)]%
        {howard2017shifting}
\bibfield{author}{\bibinfo{person}{Cynthia Howard}, \bibinfo{person}{Pamela Jordan}, \bibinfo{person}{Barbara Di~Eugenio}, {and} \bibinfo{person}{Sandra Katz}.} \bibinfo{year}{2017}\natexlab{}.
\newblock \showarticletitle{Shifting the load: A peer dialogue agent that encourages its human collaborator to contribute more to problem solving}.
\newblock \bibinfo{journal}{\emph{International Journal of Artificial Intelligence in Education}}  \bibinfo{volume}{27} (\bibinfo{year}{2017}), \bibinfo{pages}{101--129}.
\newblock


\bibitem[Huang et~al\mbox{.}(2025)]%
        {huang2025speechcaps}
\bibfield{author}{\bibinfo{person}{Chien-yu Huang}, \bibinfo{person}{Min-Han Shih}, \bibinfo{person}{Ke-Han Lu}, \bibinfo{person}{Chi-Yuan Hsiao}, {and} \bibinfo{person}{Hung-yi Lee}.} \bibinfo{year}{2025}\natexlab{}.
\newblock \showarticletitle{Speechcaps: Advancing instruction-based universal speech models with multi-talker speaking style captioning}. In \bibinfo{booktitle}{\emph{ICASSP 2025-2025 IEEE International Conference on Acoustics, Speech and Signal Processing (ICASSP)}}. IEEE, \bibinfo{pages}{1--5}.
\newblock


\bibitem[Jehn and Shah(1997)]%
        {jehn1997interpersonal}
\bibfield{author}{\bibinfo{person}{Karen~A Jehn} {and} \bibinfo{person}{Priti~Pradhan Shah}.} \bibinfo{year}{1997}\natexlab{}.
\newblock \showarticletitle{Interpersonal relationships and task performance: An examination of mediation processes in friendship and acquaintance groups.}
\newblock \bibinfo{journal}{\emph{Journal of personality and social psychology}} \bibinfo{volume}{72}, \bibinfo{number}{4} (\bibinfo{year}{1997}), \bibinfo{pages}{775}.
\newblock


\bibitem[Jiang et~al\mbox{.}(2023)]%
        {jiang2023beyond}
\bibfield{author}{\bibinfo{person}{Na Jiang}, \bibinfo{person}{Xiaohui Liu}, \bibinfo{person}{Hefu Liu}, \bibinfo{person}{Eric Tze~Kuan Lim}, \bibinfo{person}{Chee-Wee Tan}, {and} \bibinfo{person}{Jibao Gu}.} \bibinfo{year}{2023}\natexlab{}.
\newblock \showarticletitle{Beyond AI-powered context-aware services: the role of human--AI collaboration}.
\newblock \bibinfo{journal}{\emph{Industrial Management \& Data Systems}} \bibinfo{volume}{123}, \bibinfo{number}{11} (\bibinfo{year}{2023}), \bibinfo{pages}{2771--2802}.
\newblock


\bibitem[Johnson(2015)]%
        {johnson2015constructive}
\bibfield{author}{\bibinfo{person}{David~W Johnson}.} \bibinfo{year}{2015}\natexlab{}.
\newblock \bibinfo{booktitle}{\emph{Constructive controversy: Theory, research, practice}}.
\newblock \bibinfo{publisher}{Cambridge University Press}.
\newblock


\bibitem[Johnson et~al\mbox{.}(2000)]%
        {johnson2000constructive}
\bibfield{author}{\bibinfo{person}{David~W Johnson}, \bibinfo{person}{Roger~T Johnson}, {and} \bibinfo{person}{Dean Tjosvold}.} \bibinfo{year}{2000}\natexlab{}.
\newblock \showarticletitle{Constructive controversy}.
\newblock \bibinfo{journal}{\emph{The handbook of conflict resolution: Theory and practice}} (\bibinfo{year}{2000}), \bibinfo{pages}{65--85}.
\newblock


\bibitem[Jordan(2007)]%
        {jordan2007topic}
\bibfield{author}{\bibinfo{person}{Pamela~W Jordan}.} \bibinfo{year}{2007}\natexlab{}.
\newblock \showarticletitle{Topic initiative in a simulated peer dialogue agent}.
\newblock \bibinfo{journal}{\emph{FRONTIERS IN ARTIFICIAL INTELLIGENCE AND APPLICATIONS}}  \bibinfo{volume}{158} (\bibinfo{year}{2007}), \bibinfo{pages}{581}.
\newblock


\bibitem[Jumaat and Tasir(2014)]%
        {jumaat2014instructional}
\bibfield{author}{\bibinfo{person}{Nurul~Farhana Jumaat} {and} \bibinfo{person}{Zaidatun Tasir}.} \bibinfo{year}{2014}\natexlab{}.
\newblock \showarticletitle{Instructional scaffolding in online learning environment: A meta-analysis}. In \bibinfo{booktitle}{\emph{2014 international conference on teaching and learning in computing and engineering}}. IEEE, \bibinfo{pages}{74--77}.
\newblock


\bibitem[Ke and Carr-Chellman(2006)]%
        {ke2006solitary}
\bibfield{author}{\bibinfo{person}{Fengfeng Ke} {and} \bibinfo{person}{Alison Carr-Chellman}.} \bibinfo{year}{2006}\natexlab{}.
\newblock \showarticletitle{SOLITARY LEARNER IN ONLINE COLLABORATIVE LEARNING.}
\newblock \bibinfo{journal}{\emph{Quarterly Review of Distance Education}} \bibinfo{volume}{7}, \bibinfo{number}{3} (\bibinfo{year}{2006}).
\newblock


\bibitem[Kemp and Grieve(2014)]%
        {kemp2014face}
\bibfield{author}{\bibinfo{person}{Nenagh Kemp} {and} \bibinfo{person}{Rachel Grieve}.} \bibinfo{year}{2014}\natexlab{}.
\newblock \showarticletitle{Face-to-face or face-to-screen? Undergraduates' opinions and test performance in classroom vs. online learning}.
\newblock \bibinfo{journal}{\emph{Frontiers in psychology}}  \bibinfo{volume}{5} (\bibinfo{year}{2014}), \bibinfo{pages}{1278}.
\newblock


\bibitem[Kilic et~al\mbox{.}(2023)]%
        {kilic2023argument}
\bibfield{author}{\bibinfo{person}{Kaan Kilic}, \bibinfo{person}{Saskia Weck}, \bibinfo{person}{Timotheus Kampik}, {and} \bibinfo{person}{Helena Lindgren}.} \bibinfo{year}{2023}\natexlab{}.
\newblock \showarticletitle{Argument-based human--AI collaboration for supporting behavior change to improve health}.
\newblock \bibinfo{journal}{\emph{Frontiers in Artificial Intelligence}}  \bibinfo{volume}{6} (\bibinfo{year}{2023}), \bibinfo{pages}{1069455}.
\newblock


\bibitem[Kim et~al\mbox{.}(2021b)]%
        {kim2021utilitarian}
\bibfield{author}{\bibinfo{person}{Hankyung Kim}, \bibinfo{person}{Hoyeon Nam}, \bibinfo{person}{Uichin Lee}, {and} \bibinfo{person}{Youn-kyung Lim}.} \bibinfo{year}{2021}\natexlab{b}.
\newblock \showarticletitle{Utilitarian or relational? Exploring indicators of user orientation towards intelligent agents}. In \bibinfo{booktitle}{\emph{HCI International 2021-Posters: 23rd HCI International Conference, HCII 2021, Virtual Event, July 24--29, 2021, Proceedings, Part I 23}}. Springer, \bibinfo{pages}{448--455}.
\newblock


\bibitem[Kim et~al\mbox{.}(2021a)]%
        {kim2021ai}
\bibfield{author}{\bibinfo{person}{Jihyun Kim}, \bibinfo{person}{Kelly Merrill~Jr}, {and} \bibinfo{person}{Chad Collins}.} \bibinfo{year}{2021}\natexlab{a}.
\newblock \showarticletitle{AI as a friend or assistant: The mediating role of perceived usefulness in social AI vs. functional AI}.
\newblock \bibinfo{journal}{\emph{Telematics and Informatics}}  \bibinfo{volume}{64} (\bibinfo{year}{2021}), \bibinfo{pages}{101694}.
\newblock


\bibitem[Kim et~al\mbox{.}(2021c)]%
        {kim2021designers}
\bibfield{author}{\bibinfo{person}{Yelim Kim}, \bibinfo{person}{Mohi Reza}, \bibinfo{person}{Joanna McGrenere}, {and} \bibinfo{person}{Dongwook Yoon}.} \bibinfo{year}{2021}\natexlab{c}.
\newblock \showarticletitle{Designers characterize naturalness in voice user interfaces: their goals, practices, and challenges}. In \bibinfo{booktitle}{\emph{Proceedings of the 2021 CHI Conference on Human Factors in Computing Systems}}. \bibinfo{pages}{1--13}.
\newblock


\bibitem[Kucukozer-Cavdar and Taskaya-Temizel(2016)]%
        {kucukozer2016analyzing}
\bibfield{author}{\bibinfo{person}{Seyma Kucukozer-Cavdar} {and} \bibinfo{person}{Tugba Taskaya-Temizel}.} \bibinfo{year}{2016}\natexlab{}.
\newblock \showarticletitle{Analyzing the effects of the personality traits on the success of online collaborative groups}.
\newblock \bibinfo{journal}{\emph{Procedia-Social and Behavioral Sciences}}  \bibinfo{volume}{228} (\bibinfo{year}{2016}), \bibinfo{pages}{383--389}.
\newblock


\bibitem[Kuhn and Crowell(2011)]%
        {kuhn2011dialogic}
\bibfield{author}{\bibinfo{person}{Deanna Kuhn} {and} \bibinfo{person}{Amanda Crowell}.} \bibinfo{year}{2011}\natexlab{}.
\newblock \showarticletitle{Dialogic argumentation as a vehicle for developing young adolescents’ thinking}.
\newblock \bibinfo{journal}{\emph{Psychological science}} \bibinfo{volume}{22}, \bibinfo{number}{4} (\bibinfo{year}{2011}), \bibinfo{pages}{545--552}.
\newblock


\bibitem[Kumar et~al\mbox{.}(2023)]%
        {kumar2023impact}
\bibfield{author}{\bibinfo{person}{Harsh Kumar}, \bibinfo{person}{Ilya Musabirov}, \bibinfo{person}{Mohi Reza}, \bibinfo{person}{Jiakai Shi}, \bibinfo{person}{Anastasia Kuzminykh}, \bibinfo{person}{Joseph~Jay Williams}, {and} \bibinfo{person}{Michael Liut}.} \bibinfo{year}{2023}\natexlab{}.
\newblock \showarticletitle{Impact of guidance and interaction strategies for LLM use on Learner Performance and perception}.
\newblock \bibinfo{journal}{\emph{arXiv preprint arXiv:2310.13712}} (\bibinfo{year}{2023}).
\newblock


\bibitem[Li et~al\mbox{.}(2025)]%
        {li2025exploring}
\bibfield{author}{\bibinfo{person}{Qintong Li}, \bibinfo{person}{Leyang Cui}, \bibinfo{person}{Lingpeng Kong}, {and} \bibinfo{person}{Wei Bi}.} \bibinfo{year}{2025}\natexlab{}.
\newblock \showarticletitle{Exploring the reliability of large language models as customized evaluators for diverse NLP tasks}. In \bibinfo{booktitle}{\emph{Proceedings of the 31st International Conference on Computational Linguistics}}. \bibinfo{pages}{10325--10344}.
\newblock


\bibitem[Li et~al\mbox{.}(2017)]%
        {li2017peer}
\bibfield{author}{\bibinfo{person}{Wenjie Li}, \bibinfo{person}{Francesca Bassi}, \bibinfo{person}{Laura Galluccio}, {and} \bibinfo{person}{Michel Kieffer}.} \bibinfo{year}{2017}\natexlab{}.
\newblock \showarticletitle{Peer-Assisted Individual Assessment in a multi-agent system}.
\newblock \bibinfo{journal}{\emph{Automatica}}  \bibinfo{volume}{83} (\bibinfo{year}{2017}), \bibinfo{pages}{351--360}.
\newblock


\bibitem[Liew et~al\mbox{.}(2013)]%
        {liew2013effects}
\bibfield{author}{\bibinfo{person}{Tze~Wei Liew}, \bibinfo{person}{Su-Mae Tan}, {and} \bibinfo{person}{Chandrika Jayothisa}.} \bibinfo{year}{2013}\natexlab{}.
\newblock \showarticletitle{The effects of peer-like and expert-like pedagogical agents on learners' agent perceptions, task-related attitudes, and learning achievement}.
\newblock \bibinfo{journal}{\emph{Journal of Educational Technology \& Society}} \bibinfo{volume}{16}, \bibinfo{number}{4} (\bibinfo{year}{2013}), \bibinfo{pages}{275--286}.
\newblock


\bibitem[Liu et~al\mbox{.}(2024)]%
        {liu2024peergpt}
\bibfield{author}{\bibinfo{person}{Jiawen Liu}, \bibinfo{person}{Yuanyuan Yao}, \bibinfo{person}{Pengcheng An}, {and} \bibinfo{person}{Qi Wang}.} \bibinfo{year}{2024}\natexlab{}.
\newblock \showarticletitle{PeerGPT: Probing the Roles of LLM-based Peer Agents as Team Moderators and Participants in Children's Collaborative Learning}. In \bibinfo{booktitle}{\emph{Extended Abstracts of the CHI Conference on Human Factors in Computing Systems}}. \bibinfo{pages}{1--6}.
\newblock


\bibitem[Liu and Tao(2022)]%
        {liu2022roles}
\bibfield{author}{\bibinfo{person}{Kaifeng Liu} {and} \bibinfo{person}{Da Tao}.} \bibinfo{year}{2022}\natexlab{}.
\newblock \showarticletitle{The roles of trust, personalization, loss of privacy, and anthropomorphism in public acceptance of smart healthcare services}.
\newblock \bibinfo{journal}{\emph{Computers in Human Behavior}}  \bibinfo{volume}{127} (\bibinfo{year}{2022}), \bibinfo{pages}{107026}.
\newblock


\bibitem[Liu et~al\mbox{.}(2025)]%
        {liu2025proactive}
\bibfield{author}{\bibinfo{person}{Xingyu~Bruce Liu}, \bibinfo{person}{Shitao Fang}, \bibinfo{person}{Weiyan Shi}, \bibinfo{person}{Chien-Sheng Wu}, \bibinfo{person}{Takeo Igarashi}, {and} \bibinfo{person}{Xiang'Anthony' Chen}.} \bibinfo{year}{2025}\natexlab{}.
\newblock \showarticletitle{Proactive conversational agents with inner thoughts}. In \bibinfo{booktitle}{\emph{Proceedings of the 2025 CHI Conference on Human Factors in Computing Systems}}. \bibinfo{pages}{1--19}.
\newblock


\bibitem[Liu et~al\mbox{.}(2023)]%
        {liu2023g}
\bibfield{author}{\bibinfo{person}{Yang Liu}, \bibinfo{person}{Dan Iter}, \bibinfo{person}{Yichong Xu}, \bibinfo{person}{Shuohang Wang}, \bibinfo{person}{Ruochen Xu}, {and} \bibinfo{person}{Chenguang Zhu}.} \bibinfo{year}{2023}\natexlab{}.
\newblock \showarticletitle{G-eval: NLG evaluation using gpt-4 with better human alignment}.
\newblock \bibinfo{journal}{\emph{arXiv preprint arXiv:2303.16634}} (\bibinfo{year}{2023}).
\newblock


\bibitem[Lowenthal et~al\mbox{.}(2020)]%
        {lowenthal2020creating}
\bibfield{author}{\bibinfo{person}{Patrick~R Lowenthal}, \bibinfo{person}{Michael Humphrey}, \bibinfo{person}{Quincy Conley}, \bibinfo{person}{Joanna~C Dunlap}, \bibinfo{person}{Krista Greear}, \bibinfo{person}{Alison Lowenthal}, {and} \bibinfo{person}{Lisa~A Giacumo}.} \bibinfo{year}{2020}\natexlab{}.
\newblock \showarticletitle{Creating accessible and inclusive online learning: Moving beyond compliance and broadening the discussion.}
\newblock \bibinfo{journal}{\emph{Quarterly Review of Distance Education}} \bibinfo{volume}{21}, \bibinfo{number}{2} (\bibinfo{year}{2020}).
\newblock


\bibitem[Lowry and Johnson(1981)]%
        {lowry1981effects}
\bibfield{author}{\bibinfo{person}{Nancy Lowry} {and} \bibinfo{person}{David~W Johnson}.} \bibinfo{year}{1981}\natexlab{}.
\newblock \showarticletitle{Effects of controversy on epistemic curiosity, achievement, and attitudes}.
\newblock \bibinfo{journal}{\emph{The Journal of Social Psychology}} \bibinfo{volume}{115}, \bibinfo{number}{1} (\bibinfo{year}{1981}), \bibinfo{pages}{31--43}.
\newblock


\bibitem[McGee(2024)]%
        {mcgee2024forbidden}
\bibfield{author}{\bibinfo{person}{Robert~W McGee}.} \bibinfo{year}{2024}\natexlab{}.
\newblock \showarticletitle{Forbidden Topics in Artificial Intelligence Research: Two Case Studies}.
\newblock \bibinfo{journal}{\emph{Available at SSRN}} (\bibinfo{year}{2024}).
\newblock


\bibitem[Mei et~al\mbox{.}(2024)]%
        {mei2024turing}
\bibfield{author}{\bibinfo{person}{Qiaozhu Mei}, \bibinfo{person}{Yutong Xie}, \bibinfo{person}{Walter Yuan}, {and} \bibinfo{person}{Matthew~O Jackson}.} \bibinfo{year}{2024}\natexlab{}.
\newblock \showarticletitle{A Turing test of whether AI chatbots are behaviorally similar to humans}.
\newblock \bibinfo{journal}{\emph{Proceedings of the National Academy of Sciences}} \bibinfo{volume}{121}, \bibinfo{number}{9} (\bibinfo{year}{2024}), \bibinfo{pages}{e2313925121}.
\newblock


\bibitem[Memmert and Tavanapour(2023)]%
        {memmert2023towards}
\bibfield{author}{\bibinfo{person}{Lucas Memmert} {and} \bibinfo{person}{Navid Tavanapour}.} \bibinfo{year}{2023}\natexlab{}.
\newblock \showarticletitle{Towards human-AI-collaboration in brainstorming: Empirical insights into the perception of working with a generative AI}.
\newblock  (\bibinfo{year}{2023}).
\newblock


\bibitem[Mirzakhmedova et~al\mbox{.}(2024)]%
        {mirzakhmedova2024large}
\bibfield{author}{\bibinfo{person}{Nailia Mirzakhmedova}, \bibinfo{person}{Marcel Gohsen}, \bibinfo{person}{Chia~Hao Chang}, {and} \bibinfo{person}{Benno Stein}.} \bibinfo{year}{2024}\natexlab{}.
\newblock \showarticletitle{Are Large Language Models Reliable Argument Quality Annotators?}. In \bibinfo{booktitle}{\emph{Conference on Advances in Robust Argumentation Machines}}. Springer, \bibinfo{pages}{129--146}.
\newblock


\bibitem[Moribe and Ushiama(2025)]%
        {moribe2025imitating}
\bibfield{author}{\bibinfo{person}{Sosui Moribe} {and} \bibinfo{person}{Taketoshi Ushiama}.} \bibinfo{year}{2025}\natexlab{}.
\newblock \showarticletitle{Imitating Mistakes in a Learning Companion AI Agent for Online Peer Learning}. In \bibinfo{booktitle}{\emph{2025 19th International Conference on Ubiquitous Information Management and Communication (IMCOM)}}. IEEE, \bibinfo{pages}{1--8}.
\newblock


\bibitem[Mu{\~n}oz-Carril et~al\mbox{.}(2021)]%
        {munoz2021factors}
\bibfield{author}{\bibinfo{person}{Pablo-C{\'e}sar Mu{\~n}oz-Carril}, \bibinfo{person}{Nuria Hern{\'a}ndez-Sell{\'e}s}, \bibinfo{person}{Eduardo-Jos{\'e} Fuentes-Abeledo}, {and} \bibinfo{person}{Mercedes Gonz{\'a}lez-Sanmamed}.} \bibinfo{year}{2021}\natexlab{}.
\newblock \showarticletitle{Factors influencing students’ perceived impact of learning and satisfaction in Computer Supported Collaborative Learning}.
\newblock \bibinfo{journal}{\emph{Computers \& Education}}  \bibinfo{volume}{174} (\bibinfo{year}{2021}), \bibinfo{pages}{104310}.
\newblock


\bibitem[Nass and Moon(2000)]%
        {nass2000machines}
\bibfield{author}{\bibinfo{person}{Clifford Nass} {and} \bibinfo{person}{Youngme Moon}.} \bibinfo{year}{2000}\natexlab{}.
\newblock \showarticletitle{Machines and mindlessness: Social responses to computers}.
\newblock \bibinfo{journal}{\emph{Journal of social issues}} \bibinfo{volume}{56}, \bibinfo{number}{1} (\bibinfo{year}{2000}), \bibinfo{pages}{81--103}.
\newblock


\bibitem[Neef(2006)]%
        {neef2006taxonomy}
\bibfield{author}{\bibinfo{person}{Martijn Neef}.} \bibinfo{year}{2006}\natexlab{}.
\newblock \showarticletitle{A taxonomy of human-agent team collaborations}. In \bibinfo{booktitle}{\emph{Proceedings of the 18th BeNeLux Conference on Artificial Intelligence (BNAIC 2006)}}. \bibinfo{pages}{245--250}.
\newblock


\bibitem[Nimmo et~al\mbox{.}(2024)]%
        {nimmo2024user}
\bibfield{author}{\bibinfo{person}{Robert Nimmo}, \bibinfo{person}{Marios Constantinides}, \bibinfo{person}{Ke Zhou}, \bibinfo{person}{Daniele Quercia}, {and} \bibinfo{person}{Simone Stumpf}.} \bibinfo{year}{2024}\natexlab{}.
\newblock \showarticletitle{User Characteristics in Explainable AI: The Rabbit Hole of Personalization?}. In \bibinfo{booktitle}{\emph{Proceedings of the CHI Conference on Human Factors in Computing Systems}}. \bibinfo{pages}{1--13}.
\newblock


\bibitem[Noels et~al\mbox{.}(2025)]%
        {noels2025large}
\bibfield{author}{\bibinfo{person}{Sander Noels}, \bibinfo{person}{Guillaume Bied}, \bibinfo{person}{Maarten Buyl}, \bibinfo{person}{Alexander Rogiers}, \bibinfo{person}{Yousra Fettach}, \bibinfo{person}{Jefrey Lijffijt}, {and} \bibinfo{person}{Tijl De~Bie}.} \bibinfo{year}{2025}\natexlab{}.
\newblock \showarticletitle{What large language models do not talk about: An empirical study of moderation and censorship practices}. In \bibinfo{booktitle}{\emph{Joint European Conference on Machine Learning and Knowledge Discovery in Databases}}. Springer, \bibinfo{pages}{265--281}.
\newblock


\bibitem[Nokes-Malach et~al\mbox{.}(2012)]%
        {nokes2012effect}
\bibfield{author}{\bibinfo{person}{Timothy~J Nokes-Malach}, \bibinfo{person}{Michelle~L Meade}, {and} \bibinfo{person}{Daniel~G Morrow}.} \bibinfo{year}{2012}\natexlab{}.
\newblock \showarticletitle{The effect of expertise on collaborative problem solving}.
\newblock \bibinfo{journal}{\emph{Thinking \& Reasoning}} \bibinfo{volume}{18}, \bibinfo{number}{1} (\bibinfo{year}{2012}), \bibinfo{pages}{32--58}.
\newblock


\bibitem[Nora(2002)]%
        {nora2002collaborative}
\bibfield{author}{\bibinfo{person}{Amaury Nora}.} \bibinfo{year}{2002}\natexlab{}.
\newblock \showarticletitle{Collaborative learning: Its impact on college students' development and diversity}.
\newblock  (\bibinfo{year}{2002}).
\newblock


\bibitem[Nussbaum(2008)]%
        {nussbaum2008collaborative}
\bibfield{author}{\bibinfo{person}{E~Michael Nussbaum}.} \bibinfo{year}{2008}\natexlab{}.
\newblock \showarticletitle{Collaborative discourse, argumentation, and learning: Preface and literature review}.
\newblock \bibinfo{journal}{\emph{Contemporary Educational Psychology}} \bibinfo{volume}{33}, \bibinfo{number}{3} (\bibinfo{year}{2008}), \bibinfo{pages}{345--359}.
\newblock


\bibitem[Nye et~al\mbox{.}(2023)]%
        {nye2023generative}
\bibfield{author}{\bibinfo{person}{Benjamin~D Nye}, \bibinfo{person}{Dillon Mee}, {and} \bibinfo{person}{Mark~G Core}.} \bibinfo{year}{2023}\natexlab{}.
\newblock \showarticletitle{Generative Large Language Models for Dialog-Based Tutoring: An Early Consideration of Opportunities and Concerns.}. In \bibinfo{booktitle}{\emph{LLM@ AIED}}. \bibinfo{pages}{78--88}.
\newblock


\bibitem[Oh and Lim(2005)]%
        {oh2005cross}
\bibfield{author}{\bibinfo{person}{Eunjoo Oh} {and} \bibinfo{person}{Doohun Lim}.} \bibinfo{year}{2005}\natexlab{}.
\newblock \showarticletitle{Cross relationships between cognitive styles and learner variables in online learning environment}.
\newblock \bibinfo{journal}{\emph{Journal of Interactive Online Learning}} \bibinfo{volume}{4}, \bibinfo{number}{1} (\bibinfo{year}{2005}), \bibinfo{pages}{53--66}.
\newblock


\bibitem[O’Donnell and Hmelo-Silver(2013)]%
        {o2013introduction}
\bibfield{author}{\bibinfo{person}{Angela~M O’Donnell} {and} \bibinfo{person}{Cindy~E Hmelo-Silver}.} \bibinfo{year}{2013}\natexlab{}.
\newblock \showarticletitle{Introduction: What is collaborative learning?: An overview}.
\newblock \bibinfo{journal}{\emph{The international handbook of collaborative learning}} (\bibinfo{year}{2013}), \bibinfo{pages}{1--15}.
\newblock


\bibitem[Park et~al\mbox{.}(2023)]%
        {park2023choicemates}
\bibfield{author}{\bibinfo{person}{Jeongeon Park}, \bibinfo{person}{Bryan Min}, \bibinfo{person}{Xiaojuan Ma}, {and} \bibinfo{person}{Juho Kim}.} \bibinfo{year}{2023}\natexlab{}.
\newblock \showarticletitle{Choicemates: Supporting unfamiliar online decision-making with multi-agent conversational interactions}.
\newblock \bibinfo{journal}{\emph{arXiv preprint arXiv:2310.01331}} (\bibinfo{year}{2023}).
\newblock


\bibitem[Park and Choi(2009)]%
        {park2009factors}
\bibfield{author}{\bibinfo{person}{Ji-Hye Park} {and} \bibinfo{person}{Hee~Jun Choi}.} \bibinfo{year}{2009}\natexlab{}.
\newblock \showarticletitle{Factors influencing adult learners' decision to drop out or persist in online learning}.
\newblock \bibinfo{journal}{\emph{Journal of Educational Technology \& Society}} \bibinfo{volume}{12}, \bibinfo{number}{4} (\bibinfo{year}{2009}), \bibinfo{pages}{207--217}.
\newblock


\bibitem[Passi and Vorvoreanu(2022)]%
        {passi2022overreliance}
\bibfield{author}{\bibinfo{person}{Samir Passi} {and} \bibinfo{person}{Mihaela Vorvoreanu}.} \bibinfo{year}{2022}\natexlab{}.
\newblock \showarticletitle{Overreliance on AI literature review}.
\newblock \bibinfo{journal}{\emph{Microsoft Research}} (\bibinfo{year}{2022}).
\newblock


\bibitem[Pataranutaporn et~al\mbox{.}(2023)]%
        {pataranutaporn2023influencing}
\bibfield{author}{\bibinfo{person}{Pat Pataranutaporn}, \bibinfo{person}{Ruby Liu}, \bibinfo{person}{Ed Finn}, {and} \bibinfo{person}{Pattie Maes}.} \bibinfo{year}{2023}\natexlab{}.
\newblock \showarticletitle{Influencing human--AI interaction by priming beliefs about AI can increase perceived trustworthiness, empathy and effectiveness}.
\newblock \bibinfo{journal}{\emph{Nature Machine Intelligence}} \bibinfo{volume}{5}, \bibinfo{number}{10} (\bibinfo{year}{2023}), \bibinfo{pages}{1076--1086}.
\newblock


\bibitem[Pudane et~al\mbox{.}(2018)]%
        {pudane2018challenges}
\bibfield{author}{\bibinfo{person}{Mara Pudane}, \bibinfo{person}{Sintija Petrovica}, \bibinfo{person}{Egons Lavendelis}, {and} \bibinfo{person}{Alla Anohina-Naumeca}.} \bibinfo{year}{2018}\natexlab{}.
\newblock \showarticletitle{Challenges in the Development of Affective Collaborative Learning Environment with Artificial Peers.}
\newblock \bibinfo{journal}{\emph{Appl. Comput. Syst.}} \bibinfo{volume}{23}, \bibinfo{number}{2} (\bibinfo{year}{2018}), \bibinfo{pages}{101--108}.
\newblock


\bibitem[Qureshi et~al\mbox{.}(2023)]%
        {qureshi2023factors}
\bibfield{author}{\bibinfo{person}{Muhammad~Asif Qureshi}, \bibinfo{person}{Asadullah Khaskheli}, \bibinfo{person}{Jawaid~Ahmed Qureshi}, \bibinfo{person}{Syed~Ali Raza}, {and} \bibinfo{person}{Sara~Qamar Yousufi}.} \bibinfo{year}{2023}\natexlab{}.
\newblock \showarticletitle{Factors affecting students’ learning performance through collaborative learning and engagement}.
\newblock \bibinfo{journal}{\emph{Interactive Learning Environments}} \bibinfo{volume}{31}, \bibinfo{number}{4} (\bibinfo{year}{2023}), \bibinfo{pages}{2371--2391}.
\newblock


\bibitem[Reeves and Nass(1996)]%
        {reeves1996media}
\bibfield{author}{\bibinfo{person}{Byron Reeves} {and} \bibinfo{person}{Clifford Nass}.} \bibinfo{year}{1996}\natexlab{}.
\newblock \showarticletitle{The media equation: How people treat computers, television, and new media like real people}.
\newblock \bibinfo{journal}{\emph{Cambridge, UK}} \bibinfo{volume}{10}, \bibinfo{number}{10} (\bibinfo{year}{1996}), \bibinfo{pages}{19--36}.
\newblock


\bibitem[Roseth et~al\mbox{.}(2011)]%
        {roseth2011effects}
\bibfield{author}{\bibinfo{person}{Cary~J Roseth}, \bibinfo{person}{Andy~J Saltarelli}, {and} \bibinfo{person}{Chris~R Glass}.} \bibinfo{year}{2011}\natexlab{}.
\newblock \showarticletitle{Effects of face-to-face and computer-mediated constructive controversy on social interdependence, motivation, and achievement.}
\newblock \bibinfo{journal}{\emph{Journal of educational psychology}} \bibinfo{volume}{103}, \bibinfo{number}{4} (\bibinfo{year}{2011}), \bibinfo{pages}{804}.
\newblock


\bibitem[Saad{\'e} et~al\mbox{.}(2007)]%
        {saade2007exploring}
\bibfield{author}{\bibinfo{person}{Raafat~George Saad{\'e}}, \bibinfo{person}{Xin He}, {and} \bibinfo{person}{Dennis Kira}.} \bibinfo{year}{2007}\natexlab{}.
\newblock \showarticletitle{Exploring dimensions to online learning}.
\newblock \bibinfo{journal}{\emph{Computers in human behavior}} \bibinfo{volume}{23}, \bibinfo{number}{4} (\bibinfo{year}{2007}), \bibinfo{pages}{1721--1739}.
\newblock


\bibitem[Sadek et~al\mbox{.}(2023)]%
        {sadek2023designing}
\bibfield{author}{\bibinfo{person}{Malak Sadek}, \bibinfo{person}{Rafael~A Calvo}, {and} \bibinfo{person}{C{\'e}line Mougenot}.} \bibinfo{year}{2023}\natexlab{}.
\newblock \showarticletitle{Designing value-sensitive AI: a critical review and recommendations for socio-technical design processes}.
\newblock \bibinfo{journal}{\emph{AI and Ethics}} (\bibinfo{year}{2023}), \bibinfo{pages}{1--19}.
\newblock


\bibitem[Scardamalia et~al\mbox{.}(2002)]%
        {scardamalia2002collective}
\bibfield{author}{\bibinfo{person}{Marlene Scardamalia} {et~al\mbox{.}}} \bibinfo{year}{2002}\natexlab{}.
\newblock \showarticletitle{Collective cognitive responsibility for the advancement of knowledge}.
\newblock \bibinfo{journal}{\emph{Liberal education in a knowledge society}}  \bibinfo{volume}{97} (\bibinfo{year}{2002}), \bibinfo{pages}{67--98}.
\newblock


\bibitem[Schmidt et~al\mbox{.}(2020)]%
        {schmidt2020transparency}
\bibfield{author}{\bibinfo{person}{Philipp Schmidt}, \bibinfo{person}{Felix Biessmann}, {and} \bibinfo{person}{Timm Teubner}.} \bibinfo{year}{2020}\natexlab{}.
\newblock \showarticletitle{Transparency and trust in artificial intelligence systems}.
\newblock \bibinfo{journal}{\emph{Journal of Decision Systems}} \bibinfo{volume}{29}, \bibinfo{number}{4} (\bibinfo{year}{2020}), \bibinfo{pages}{260--278}.
\newblock


\bibitem[Schwitzgebel(2023)]%
        {schwitzgebel2023ai}
\bibfield{author}{\bibinfo{person}{Eric Schwitzgebel}.} \bibinfo{year}{2023}\natexlab{}.
\newblock \showarticletitle{AI systems must not confuse users about their sentience or moral status}.
\newblock \bibinfo{journal}{\emph{Patterns}} \bibinfo{volume}{4}, \bibinfo{number}{8} (\bibinfo{year}{2023}).
\newblock


\bibitem[Seville and Field(2000)]%
        {seville2000can}
\bibfield{author}{\bibinfo{person}{Helen Seville} {and} \bibinfo{person}{Debora Field}.} \bibinfo{year}{2000}\natexlab{}.
\newblock \showarticletitle{What can AI do for ethics}.
\newblock \bibinfo{journal}{\emph{AISB QUARTERLY}} (\bibinfo{year}{2000}), \bibinfo{pages}{31--34}.
\newblock


\bibitem[Shi et~al\mbox{.}(2024a)]%
        {shi2024argumentative}
\bibfield{author}{\bibinfo{person}{Li Shi}, \bibinfo{person}{Houjiang Liu}, \bibinfo{person}{Yian Wong}, \bibinfo{person}{Utkarsh Mujumdar}, \bibinfo{person}{Dan Zhang}, \bibinfo{person}{Jacek Gwizdka}, {and} \bibinfo{person}{Matthew Lease}.} \bibinfo{year}{2024}\natexlab{a}.
\newblock \showarticletitle{Argumentative experience: Reducing confirmation bias on controversial issues through llm-generated multi-persona debates}.
\newblock \bibinfo{journal}{\emph{arXiv preprint arXiv:2412.04629}} (\bibinfo{year}{2024}).
\newblock


\bibitem[Shi et~al\mbox{.}(2024b)]%
        {shi2024judging}
\bibfield{author}{\bibinfo{person}{Lin Shi}, \bibinfo{person}{Chiyu Ma}, \bibinfo{person}{Wenhua Liang}, \bibinfo{person}{Weicheng Ma}, {and} \bibinfo{person}{Soroush Vosoughi}.} \bibinfo{year}{2024}\natexlab{b}.
\newblock \showarticletitle{Judging the judges: A systematic investigation of position bias in pairwise comparative assessments by llms}.
\newblock  (\bibinfo{year}{2024}).
\newblock


\bibitem[Shin et~al\mbox{.}(2024)]%
        {shin2024paper}
\bibfield{author}{\bibinfo{person}{Donghoon Shin}, \bibinfo{person}{Lucy~Lu Wang}, {and} \bibinfo{person}{Gary Hsieh}.} \bibinfo{year}{2024}\natexlab{}.
\newblock \showarticletitle{From paper to card: transforming design implications with generative AI}. In \bibinfo{booktitle}{\emph{Proceedings of the 2024 CHI Conference on Human Factors in Computing Systems}}. \bibinfo{pages}{1--15}.
\newblock


\bibitem[Staubitz et~al\mbox{.}(2015)]%
        {staubitz2015collaborative}
\bibfield{author}{\bibinfo{person}{Thomas Staubitz}, \bibinfo{person}{Tobias Pfeiffer}, \bibinfo{person}{Jan Renz}, \bibinfo{person}{Christian Willems}, {and} \bibinfo{person}{Christoph Meinel}.} \bibinfo{year}{2015}\natexlab{}.
\newblock \showarticletitle{Collaborative learning in a MOOC environment}. In \bibinfo{booktitle}{\emph{ICERI2015 Proceedings}}. IATED, \bibinfo{pages}{8237--8246}.
\newblock


\bibitem[Street(2024)]%
        {street2024llm}
\bibfield{author}{\bibinfo{person}{Winnie Street}.} \bibinfo{year}{2024}\natexlab{}.
\newblock \showarticletitle{LLM Theory of Mind and Alignment: Opportunities and Risks}.
\newblock \bibinfo{journal}{\emph{arXiv preprint arXiv:2405.08154}} (\bibinfo{year}{2024}).
\newblock


\bibitem[Stump et~al\mbox{.}(2009)]%
        {stump2009student}
\bibfield{author}{\bibinfo{person}{Glenda Stump}, \bibinfo{person}{Jenefer Husman}, \bibinfo{person}{Wen-Ting Chung}, {and} \bibinfo{person}{Aaron Done}.} \bibinfo{year}{2009}\natexlab{}.
\newblock \showarticletitle{Student beliefs about intelligence: Relationship to learning}. In \bibinfo{booktitle}{\emph{2009 39th IEEE Frontiers in Education Conference}}. IEEE, \bibinfo{pages}{1--6}.
\newblock


\bibitem[Sullivan and Artino~Jr(2013)]%
        {sullivan2013analyzing}
\bibfield{author}{\bibinfo{person}{Gail~M Sullivan} {and} \bibinfo{person}{Anthony~R Artino~Jr}.} \bibinfo{year}{2013}\natexlab{}.
\newblock \showarticletitle{Analyzing and interpreting data from Likert-type scales}.
\newblock \bibinfo{journal}{\emph{Journal of graduate medical education}} \bibinfo{volume}{5}, \bibinfo{number}{4} (\bibinfo{year}{2013}), \bibinfo{pages}{541--542}.
\newblock


\bibitem[Taber(2008)]%
        {taber2008exploring}
\bibfield{author}{\bibinfo{person}{Keith~S Taber}.} \bibinfo{year}{2008}\natexlab{}.
\newblock \showarticletitle{Exploring student learning from a constructivist perspective in diverse educational contexts}.
\newblock \bibinfo{journal}{\emph{Journal of Turkish Science Education}} \bibinfo{volume}{5}, \bibinfo{number}{1} (\bibinfo{year}{2008}), \bibinfo{pages}{3--22}.
\newblock


\bibitem[Tankelevitch et~al\mbox{.}(2024)]%
        {tankelevitch2024metacognitive}
\bibfield{author}{\bibinfo{person}{Lev Tankelevitch}, \bibinfo{person}{Viktor Kewenig}, \bibinfo{person}{Auste Simkute}, \bibinfo{person}{Ava~Elizabeth Scott}, \bibinfo{person}{Advait Sarkar}, \bibinfo{person}{Abigail Sellen}, {and} \bibinfo{person}{Sean Rintel}.} \bibinfo{year}{2024}\natexlab{}.
\newblock \showarticletitle{The metacognitive demands and opportunities of generative AI}. In \bibinfo{booktitle}{\emph{Proceedings of the CHI Conference on Human Factors in Computing Systems}}. \bibinfo{pages}{1--24}.
\newblock


\bibitem[Tanprasert et~al\mbox{.}(2024)]%
        {tanprasert2024debate}
\bibfield{author}{\bibinfo{person}{Thitaree Tanprasert}, \bibinfo{person}{Sidney~S Fels}, \bibinfo{person}{Luanne Sinnamon}, {and} \bibinfo{person}{Dongwook Yoon}.} \bibinfo{year}{2024}\natexlab{}.
\newblock \showarticletitle{Debate Chatbots to Facilitate Critical Thinking on YouTube: Social Identity and Conversational Style Make A Difference}. In \bibinfo{booktitle}{\emph{Proceedings of the CHI Conference on Human Factors in Computing Systems}}. \bibinfo{pages}{1--24}.
\newblock


\bibitem[Tao(2014)]%
        {tao2014argumentative}
\bibfield{author}{\bibinfo{person}{Xuehong Tao}.} \bibinfo{year}{2014}\natexlab{}.
\newblock \emph{\bibinfo{title}{Argumentative Learning with Intelligent Agents}}.
\newblock \bibinfo{thesistype}{Ph.\,D. Dissertation}. \bibinfo{school}{Victoria University}.
\newblock


\bibitem[Thapa et~al\mbox{.}(2023)]%
        {thapa2023perception}
\bibfield{author}{\bibinfo{person}{Naina Thapa}, \bibinfo{person}{Daisy~J Lobo}, {and} \bibinfo{person}{Radhika~R Pai}.} \bibinfo{year}{2023}\natexlab{}.
\newblock \showarticletitle{Perception, satisfaction, and self-regulation in a blended learning environment during COVID-19 pandemic among health sciences students: A cross-sectional survey}.
\newblock \bibinfo{journal}{\emph{Journal of Education and Health Promotion}} \bibinfo{number}{1} (\bibinfo{year}{2023}), \bibinfo{pages}{312}.
\newblock


\bibitem[Thomas(2006)]%
        {thomas2006general}
\bibfield{author}{\bibinfo{person}{David~R Thomas}.} \bibinfo{year}{2006}\natexlab{}.
\newblock \showarticletitle{A general inductive approach for analyzing qualitative evaluation data}.
\newblock \bibinfo{journal}{\emph{American journal of evaluation}} \bibinfo{volume}{27}, \bibinfo{number}{2} (\bibinfo{year}{2006}), \bibinfo{pages}{237--246}.
\newblock


\bibitem[Tsumura and Yamada(2023)]%
        {tsumura2023influence}
\bibfield{author}{\bibinfo{person}{Takahiro Tsumura} {and} \bibinfo{person}{Seiji Yamada}.} \bibinfo{year}{2023}\natexlab{}.
\newblock \showarticletitle{Influence of agent’s self-disclosure on human empathy}.
\newblock \bibinfo{journal}{\emph{PLoS One}} \bibinfo{volume}{18}, \bibinfo{number}{5} (\bibinfo{year}{2023}), \bibinfo{pages}{e0283955}.
\newblock


\bibitem[Tulli et~al\mbox{.}(2019)]%
        {tulli2019effects}
\bibfield{author}{\bibinfo{person}{Silvia Tulli}, \bibinfo{person}{Filipa Correia}, \bibinfo{person}{Samuel Mascarenhas}, \bibinfo{person}{Samuel Gomes}, \bibinfo{person}{Francisco~S Melo}, {and} \bibinfo{person}{Ana Paiva}.} \bibinfo{year}{2019}\natexlab{}.
\newblock \showarticletitle{Effects of agents’ transparency on teamwork}. In \bibinfo{booktitle}{\emph{International workshop on explainable, transparent autonomous agents and multi-agent systems}}. Springer, \bibinfo{pages}{22--37}.
\newblock


\bibitem[Tuncer and Ramirez(2022)]%
        {tuncer2022exploring}
\bibfield{author}{\bibinfo{person}{Serdar Tuncer} {and} \bibinfo{person}{Alejandro Ramirez}.} \bibinfo{year}{2022}\natexlab{}.
\newblock \showarticletitle{Exploring the role of Trust during Human-AI collaboration in managerial decision-making processes}. In \bibinfo{booktitle}{\emph{International Conference on Human-Computer Interaction}}. Springer, \bibinfo{pages}{541--557}.
\newblock


\bibitem[Vassileva et~al\mbox{.}(1999)]%
        {vassileva1999multi}
\bibfield{author}{\bibinfo{person}{Julita Vassileva}, \bibinfo{person}{Jim Greer}, \bibinfo{person}{Gord McCalla}, \bibinfo{person}{Ralph Deters}, \bibinfo{person}{Diego Zapata}, \bibinfo{person}{Chhaya Mudgal}, {and} \bibinfo{person}{Shawn Grant}.} \bibinfo{year}{1999}\natexlab{}.
\newblock \showarticletitle{A multi-agent approach to the design of peer-help environments}. In \bibinfo{booktitle}{\emph{Proceedings of AIED}}, Vol.~\bibinfo{volume}{99}. \bibinfo{pages}{38--45}.
\newblock


\bibitem[Warsah et~al\mbox{.}(2021)]%
        {warsah2021impact}
\bibfield{author}{\bibinfo{person}{Idi Warsah}, \bibinfo{person}{Ruly Morganna}, \bibinfo{person}{Muhamad Uyun}, \bibinfo{person}{Muslim Afandi}, {and} \bibinfo{person}{Hamengkubuwono Hamengkubuwono}.} \bibinfo{year}{2021}\natexlab{}.
\newblock \showarticletitle{The impact of collaborative learning on learners’ critical thinking skills}.
\newblock \bibinfo{journal}{\emph{International Journal of Instruction}} \bibinfo{volume}{14}, \bibinfo{number}{2} (\bibinfo{year}{2021}), \bibinfo{pages}{443--460}.
\newblock


\bibitem[Weber and Cook(1972)]%
        {weber1972subject}
\bibfield{author}{\bibinfo{person}{Stephen~J Weber} {and} \bibinfo{person}{Thomas~D Cook}.} \bibinfo{year}{1972}\natexlab{}.
\newblock \showarticletitle{Subject effects in laboratory research: an examination of subject roles, demand characteristics, and valid inference.}
\newblock \bibinfo{journal}{\emph{Psychological bulletin}} \bibinfo{volume}{77}, \bibinfo{number}{4} (\bibinfo{year}{1972}), \bibinfo{pages}{273}.
\newblock


\bibitem[Wichmann et~al\mbox{.}(2016)]%
        {wichmann2016group}
\bibfield{author}{\bibinfo{person}{Astrid Wichmann}, \bibinfo{person}{Tobias Hecking}, \bibinfo{person}{Malte Elson}, \bibinfo{person}{Nina Christmann}, \bibinfo{person}{Thomas Herrmann}, {and} \bibinfo{person}{H~Ulrich Hoppe}.} \bibinfo{year}{2016}\natexlab{}.
\newblock \showarticletitle{Group formation for small-group learning: Are heterogeneous groups more productive?}. In \bibinfo{booktitle}{\emph{Proceedings of the 12th international symposium on open collaboration}}. \bibinfo{pages}{1--4}.
\newblock


\bibitem[Wiethof et~al\mbox{.}(2021)]%
        {wiethof2021implementing}
\bibfield{author}{\bibinfo{person}{Christina Wiethof}, \bibinfo{person}{Navid Tavanapour}, {and} \bibinfo{person}{Eva Bittner}.} \bibinfo{year}{2021}\natexlab{}.
\newblock \showarticletitle{Implementing an intelligent collaborative agent as teammate in collaborative writing: toward a synergy of humans and AI}.
\newblock  (\bibinfo{year}{2021}).
\newblock


\bibitem[Wyman and Watson(2020)]%
        {wyman2020academic}
\bibfield{author}{\bibinfo{person}{Patricia~J Wyman} {and} \bibinfo{person}{Scott~B Watson}.} \bibinfo{year}{2020}\natexlab{}.
\newblock \showarticletitle{Academic achievement with cooperative learning using homogeneous and heterogeneous groups}.
\newblock \bibinfo{journal}{\emph{School Science and Mathematics}} \bibinfo{volume}{120}, \bibinfo{number}{6} (\bibinfo{year}{2020}), \bibinfo{pages}{356--363}.
\newblock


\bibitem[Xia et~al\mbox{.}(2024)]%
        {xia2024sportu}
\bibfield{author}{\bibinfo{person}{Haotian Xia}, \bibinfo{person}{Zhengbang Yang}, \bibinfo{person}{Junbo Zou}, \bibinfo{person}{Rhys Tracy}, \bibinfo{person}{Yuqing Wang}, \bibinfo{person}{Chi Lu}, \bibinfo{person}{Christopher Lai}, \bibinfo{person}{Yanjun He}, \bibinfo{person}{Xun Shao}, \bibinfo{person}{Zhuoqing Xie}, {et~al\mbox{.}}} \bibinfo{year}{2024}\natexlab{}.
\newblock \showarticletitle{Sportu: A comprehensive sports understanding benchmark for multimodal large language models}.
\newblock \bibinfo{journal}{\emph{arXiv preprint arXiv:2410.08474}} (\bibinfo{year}{2024}).
\newblock


\bibitem[Yam et~al\mbox{.}(2023)]%
        {yam2023cultural}
\bibfield{author}{\bibinfo{person}{Kai~Chi Yam}, \bibinfo{person}{Tiffany Tan}, \bibinfo{person}{Joshua~Conrad Jackson}, \bibinfo{person}{Azim Shariff}, {and} \bibinfo{person}{Kurt Gray}.} \bibinfo{year}{2023}\natexlab{}.
\newblock \showarticletitle{Cultural Differences in People's Reactions and Applications of Robots, Algorithms, and Artificial Intelligence}.
\newblock \bibinfo{journal}{\emph{Management and Organization Review}} \bibinfo{volume}{19}, \bibinfo{number}{5} (\bibinfo{year}{2023}), \bibinfo{pages}{859--875}.
\newblock


\bibitem[Yamagata-Lynch(2014)]%
        {yamagata2014blending}
\bibfield{author}{\bibinfo{person}{Lisa~C Yamagata-Lynch}.} \bibinfo{year}{2014}\natexlab{}.
\newblock \showarticletitle{Blending online asynchronous and synchronous learning}.
\newblock \bibinfo{journal}{\emph{International Review of Research in Open and Distributed Learning}} \bibinfo{volume}{15}, \bibinfo{number}{2} (\bibinfo{year}{2014}), \bibinfo{pages}{189--212}.
\newblock


\bibitem[Yamarik(2007)]%
        {yamarik2007does}
\bibfield{author}{\bibinfo{person}{Steven Yamarik}.} \bibinfo{year}{2007}\natexlab{}.
\newblock \showarticletitle{Does cooperative learning improve student learning outcomes?}
\newblock \bibinfo{journal}{\emph{The journal of economic education}} \bibinfo{volume}{38}, \bibinfo{number}{3} (\bibinfo{year}{2007}), \bibinfo{pages}{259--277}.
\newblock


\bibitem[Yang(2023)]%
        {yang2023creating}
\bibfield{author}{\bibinfo{person}{Xigui Yang}.} \bibinfo{year}{2023}\natexlab{}.
\newblock \showarticletitle{Creating learning personas for collaborative learning in higher education: AQ methodology approach}.
\newblock \bibinfo{journal}{\emph{International Journal of Educational Research Open}}  \bibinfo{volume}{4} (\bibinfo{year}{2023}), \bibinfo{pages}{100250}.
\newblock


\bibitem[Yeh et~al\mbox{.}(2024)]%
        {yeh2024ghostwriter}
\bibfield{author}{\bibinfo{person}{Catherine Yeh}, \bibinfo{person}{Gonzalo Ramos}, \bibinfo{person}{Rachel Ng}, \bibinfo{person}{Andy Huntington}, {and} \bibinfo{person}{Richard Banks}.} \bibinfo{year}{2024}\natexlab{}.
\newblock \showarticletitle{GhostWriter: Augmenting Collaborative Human-AI Writing Experiences Through Personalization and Agency}.
\newblock \bibinfo{journal}{\emph{arXiv preprint arXiv:2402.08855}} (\bibinfo{year}{2024}).
\newblock


\bibitem[Yildiz~Durak(2023)]%
        {yildiz2023role}
\bibfield{author}{\bibinfo{person}{Hatice Yildiz~Durak}.} \bibinfo{year}{2023}\natexlab{}.
\newblock \showarticletitle{Role of personality traits in collaborative group works at flipped classrooms}.
\newblock \bibinfo{journal}{\emph{Current Psychology}} \bibinfo{volume}{42}, \bibinfo{number}{15} (\bibinfo{year}{2023}), \bibinfo{pages}{13093--13113}.
\newblock


\bibitem[Zhang et~al\mbox{.}(2023a)]%
        {zhang2023trust}
\bibfield{author}{\bibinfo{person}{Guanglu Zhang}, \bibinfo{person}{Leah Chong}, \bibinfo{person}{Kenneth Kotovsky}, {and} \bibinfo{person}{Jonathan Cagan}.} \bibinfo{year}{2023}\natexlab{a}.
\newblock \showarticletitle{Trust in an AI versus a Human teammate: The effects of teammate identity and performance on Human-AI cooperation}.
\newblock \bibinfo{journal}{\emph{Computers in Human Behavior}}  \bibinfo{volume}{139} (\bibinfo{year}{2023}), \bibinfo{pages}{107536}.
\newblock


\bibitem[Zhang et~al\mbox{.}(2023b)]%
        {zhang2023exploring}
\bibfield{author}{\bibinfo{person}{Jintian Zhang}, \bibinfo{person}{Xin Xu}, {and} \bibinfo{person}{Shumin Deng}.} \bibinfo{year}{2023}\natexlab{b}.
\newblock \showarticletitle{Exploring collaboration mechanisms for llm agents: A social psychology view}.
\newblock \bibinfo{journal}{\emph{arXiv preprint arXiv:2310.02124}} (\bibinfo{year}{2023}).
\newblock


\bibitem[Zhang et~al\mbox{.}(2021)]%
        {zhang2021ideal}
\bibfield{author}{\bibinfo{person}{Rui Zhang}, \bibinfo{person}{Nathan~J McNeese}, \bibinfo{person}{Guo Freeman}, {and} \bibinfo{person}{Geoff Musick}.} \bibinfo{year}{2021}\natexlab{}.
\newblock \showarticletitle{" An ideal human" expectations of AI teammates in human-AI teaming}.
\newblock \bibinfo{journal}{\emph{Proceedings of the ACM on Human-Computer Interaction}} \bibinfo{volume}{4}, \bibinfo{number}{CSCW3} (\bibinfo{year}{2021}), \bibinfo{pages}{1--25}.
\newblock


\bibitem[Zhou et~al\mbox{.}(2024)]%
        {zhou2024understanding}
\bibfield{author}{\bibinfo{person}{Jiayi Zhou}, \bibinfo{person}{Renzhong Li}, \bibinfo{person}{Junxiu Tang}, \bibinfo{person}{Tan Tang}, \bibinfo{person}{Haotian Li}, \bibinfo{person}{Weiwei Cui}, {and} \bibinfo{person}{Yingcai Wu}.} \bibinfo{year}{2024}\natexlab{}.
\newblock \showarticletitle{Understanding nonlinear collaboration between human and AI agents: A co-design framework for creative design}. In \bibinfo{booktitle}{\emph{Proceedings of the CHI Conference on Human Factors in Computing Systems}}. \bibinfo{pages}{1--16}.
\newblock


\bibitem[Zhou et~al\mbox{.}(2019)]%
        {zhou2019trusting}
\bibfield{author}{\bibinfo{person}{Michelle~X Zhou}, \bibinfo{person}{Gloria Mark}, \bibinfo{person}{Jingyi Li}, {and} \bibinfo{person}{Huahai Yang}.} \bibinfo{year}{2019}\natexlab{}.
\newblock \showarticletitle{Trusting virtual agents: The effect of personality}.
\newblock \bibinfo{journal}{\emph{ACM Transactions on Interactive Intelligent Systems (TiiS)}} \bibinfo{volume}{9}, \bibinfo{number}{2-3} (\bibinfo{year}{2019}), \bibinfo{pages}{1--36}.
\newblock


\end{thebibliography}

\appendix
\newpage
\section{Peer agent implementation prompts}
\label{appendix:LLMprompts}

All the prompts were executed with GPT at temperature 0.4 unless otherwise stated.

\subsection{Unregulated CC peer agent}
\noindent
\textbf{System prompt:} \texttt{You are an undergraduate student called [agent’s name], who is working with a teammate. You have a different background and values from your teammate. However, the goal of your team is to brainstorm evidence-based arguments to support the answer [user’s stance] to the debate prompt [debate topic]. Your teammate agrees with this answer, but you personally disagree and think the answer should be [agent’s stance]. In the discussion, each response takes the form of ‘[agent’s name]: <message>’. Keep each message short and casual. Maximum no more than 50 words but can be as little as 1 word. The expected activity duration is 20 minutes but there is no penalty for going overtime. The final arguments will be evaluated by (1) the variance between arguments: each argument should be sufficiently different from each other in the aspects of the problems that it addresses, the values and perspectives from which the arguments are founded, and the types of evidence that supports them; and (2) strength of each argument: each of the 3 arguments will be graded separately for this. The strength of the argument is based on its relevance to the topic, the logical connection between the evidence and the argument it supports, and the student's demonstration of anticipation of rebuttals and how to address the rebuttals.}

\subsection{Regulated CC peer agent}
\subsubsection{Persona expansion module} The temperature of this process is set to 0.0 to avoid the agent picking the same side as the user (possibly due to it aligning better with GPT’s inherent bias).

\noindent \textbf{System prompt:} \texttt{You are an undergraduate student called [agent’s name], who is working with a teammate. You have a different background and values from your teammate. However, the goal of your team is to brainstorm evidence-based arguments to support the answer [user’s stance; YES/NO] to the debate prompt [debate topic]. Your teammate agrees with this answer, but you personally disagree and think the answer should be [agent’s stance; YES/NO]. So your contribution is mainly to push back or question your teammate to help them anticipate the opposite debate team's arguments.}

\noindent \textbf{User prompt:} \texttt{As [agentName], answer the five following questions. (For every answer, use You/Your in the place where you would use I/My.)}
\begin{enumerate}
    \item \texttt{How do you define each term in this debate prompt?}
    \item \texttt{What are your core values that lead you to answer [agent’s stance] to the debate prompt '[debate topic]’?}
    \item \texttt{What is an important event in your upbringing that significantly affects your view on this topic?}
    \item \texttt{How does your cultural, religious, or academic background affect your stance on this topic?}
    \item \texttt{Your teammate has answered [user’s stance; YES/NO] to this debate prompt. What is your immediate assumption about your teammate?}
\end{enumerate}

The reply to this prompt is stored as the variable \emph{AGENT’S BACKGROUND}.

\subsubsection{Main agent thread} This is the LLM thread that will carry out the conversation with the learner. It is initialized with the same system prompt as the unregulated CC peer agent combined with the \emph{AGENT'S BACKGROUND}.

\noindent \textbf{System prompt:} \texttt{You are an undergraduate student called [agent’s name], who is working with a teammate. You have a different background and values from your teammate. However, the goal of your team is to brainstorm evidence-based arguments to support the answer [user’s stance] to the debate prompt [debate topic]. Your teammate agrees with this answer, but you personally disagree. You disagree with this stance because of the following reasons: [agent’s background]
Still, you cooperate with your teammate towards the shared goal. You're open-minded and understand both sides of the issue. When you talk, you are critical of your teammate's idea, not their character. You're rational, and you provide evidence to support your arguments. You also change your mind when evidence indicates that you should do so.} 

\texttt{In the discussion, each response takes the form of [agent’s name]‘: <message>’. Keep each message short and casual. Maximum no more than 50 words but can be as little as 1 word. The expected activity duration is 20 minutes but there is no penalty for going overtime. The final arguments will be evaluated by (1) the variance between arguments: each argument should be sufficiently different from each other in the aspects of the problems that it addresses, the values and perspectives from which the arguments are founded, and the types of evidence that supports them; and (2) strength of each argument: each of the 3 arguments will be graded separately for this. The strength of the argument is based on its relevance to the topic, the logical connection between the evidence and the argument it supports, and the student's demonstration of anticipation of rebuttals and how to address the rebuttals.}

\subsubsection{Moderator module} The temperature of this process is also set to 0.0 to avoid the agent answering with anything outside of the two behavior modes (TLC and TMC). The dialogue provided to the moderator module is a string, containing every line in the conversation so far. Each line is in the format '<name>: <message>'.

\noindent \textbf{System prompt:} \texttt{You are a mentor in an undergraduate class, whose goal is to help students develop better collaboration skills. You will look at the transcript of a discussion between two teammates, who have different stances on a topic but are working together to develop a set of three arguments for one of the stances. The instructor pairs two students with different stances so that they can learn from each other's perspectives, develop a deeper understanding of the topic, and strengthen their arguments. So, they should express their different opinions. However, they shouldn't do that for too long because it could make them frustrated with each other and hinder productivity. There are two steps you need to do:}

\texttt{Step 1: Determine if there is too little conflict or too much conflict in the conversation four turns from now. Too little conflict means that the two students will not have questioned each other or defended their own stances on the issue that they're currently discussing. Too much conflict means that the students will have achieved a satisfying quality of the current argument and will have explored most factors related to the current issues, if they continue to question or argue with each other. If there is too little conflict, respond with 1.}

\texttt{Step 2: If there is too much conflict, respond with 2. Your response must not have anything else but the single number. There are no other choices but 1 and 2.}

\texttt{[Dialogue:] <Insert the dialogue here> }

In case of \textbf{too little conflict (TLC)}, add the following system prompt to the \emph{main agent thread}: \texttt{At this point, your instructor thinks you should keep on pushing back and questioning your teammate, as your different stance is helping to improve the argument. Make sure to adhere to the reasons why you disagree with this stance in the beginning.}

In case of \textbf{too much conflict (TMC)}, add the following system prompt to the \emph{main agent thread}: \texttt{At this point, your instructor alerts you that if you push more, you will hinder your team productivity, so you should start backing down and be persuaded by your teammate. Find a way to align the team's stance with your background, values, and understanding of this topic in the beginning.}

\subsection{G-eval evaluation criteria of the peer agent}
\label{appendix:gevalcriteria}

We used a G-Eval–style evaluation procedure, in which a large language model (GPT-5) is prompted to act as a structured scorer following a predefined rubric \cite{liu2023g}. In this approach, the model is not asked for free-form judgements; instead, it must assign numerical ratings strictly according to the criteria we specify and provide explicit reasoning for each score. Prior work shows that LLM judges are most reliable when the scoring rubric is concrete, step-by-step, and phrased in operationalizable behavioral terms \cite{baysan2025llm, li2025exploring, mirzakhmedova2024large}, so we reformulated each CC principle into measurable indicators that the model could evaluate consistently.

To reduce evaluation variance, each dialogue was presented to GPT-5 (high reasoning effort) with (a) the full dialogue history, (b) the rubric, (c) a forced structured output format, and (d) instructions to justify its rating through explicit reasoning. This design constrains output formats to minimize ambiguity or drift. The criteria below are the exact instructions given to GPT-5 during all evaluations. We extracted the criteria from existing literature on CC \cite{johnson2015constructive} and rephrased to match the criteria of the G-eval technique.

\begin{verbatim}
    "shared_goal": (
            "Shared Goal (1–5):\n"
            "Evaluate the extent to which the assistant (Riley or Taylor) communicates a collaborative "
            "meta-goal aligned with the user's purpose (help the user strengthen their "
            "argument), despite holding a different stance. Higher scores reflect:\n"
            "- Explicit collaborative language or commitment to helping the user,\n"
            "- Support for clarifying and strengthening the user's reasoning,\n"
            "- Cooperative tone and avoidance of adversarial framing."
        ),
        "skilled_disagreement": (
            "Skilled Disagreement (1–5):\n"
            "Evaluate how skillfully the assistant (Riley or Taylor) manages and engages disagreement. "
            "Higher scores reflect:\n"
            "- Acknowledging alternative viewpoints and open-mindedness,\n"
            "- Inviting the user to explore counterarguments and reasons,\n"
            "- Challenging ideas (not the person) and maintaining a constructive tone."
        ),
        "rationality": (
            "Rationality (1–5):\n"
            "Evaluate logical consistency and structural soundness of the assistant's (Riley or Taylor)"
            "reasoning. Higher scores reflect:\n"
            "- No self-contradiction across turns,\n"
            "- Coherent multi-step inference and logically valid chains of reasoning,\n"
            "- Clear justification for claims made."
        )

\end{verbatim}

\section{Experimental task materials and grading scheme}
\label{appendix:activity}

\subsection{Activity instruction}
Please read the instructions carefully before you start the activity. The instruction is the same for every round of the task. If this is not your first round, feel free to start the activity right away.

\noindent \textbf{SCENARIO} 

You and your teammates are preparing for a debate in an undergraduate-level course. Your team discuss to come up with strong arguments together. After the discussion, you will be the one to represent the team in the actual debate. The instructor of this course will then grade you and your teammates based on your arguments in the debate and the transcripts of your team's discussion.

\noindent \textbf{STEPS}
\begin{enumerate}
    \item Click the "START ACTIVITY" button. It will open a pop-up window with a debate prompt.
    \item Answer the debate prompt. Your answer will be fixed as your team's stance for this activity.
    \item Discuss with your teammates in the chat window (the middle column) to generate 3 arguments to support your stance. Your goal is to engage with the conversation as much as you can and come up with the strongest possible arguments. Your teammates may or may not personally agree with your stance, but they are instructed to help you come up with arguments to support your stance anyway.
    \item Fill out the final argument form (the rightmost column). This form is a replacement for the actual debate, so you should make your arguments as strong as possible. You should describe, in detail, the assumption behind your arguments, the evidence to support your arguments, what rebuttal you anticipate the opposite stance might make and how you expect to counter that, and other details that can help strengthen your arguments.
    \item When you think the discussion has saturated, i.e., there is no more improvement to be made on any arguments, click the button "FINISH TASK" at the bottom of the argument form.
    \item Annotate your teammate's chat responses by hovering over the green box next to the chat message that you think is "too cooperative" or "too contradictory" when compared to what you expected from an AI peer and/or a human peer.
    \item When you finish annotating the chat, click the button "EXPORT DATA" at the bottom of the argument form and upload the exported .txt file to Qualtrics survey.
    \item There is no time limit to this activity, the expected completion time (starting when you pick your stance) is 20 minutes. There will be a notification at the 20-minute mark as a reminder.
\end{enumerate}

\noindent \textbf{INSTRUCTOR'S EVALUATION CRITERIA}
\begin{itemize}
    \item Quality of the arguments (team). Your arguments will be evaluated in two aspects (weighed equally):
    \begin{enumerate}
        \item \textbf{Argument strength:} Each of the 3 arguments will be graded separately for this. The strength of the argument are based on its relevance to the topic, the logical connection between the evidence and the argument it supports, and the student's demonstration of anticipation of rebuttals and how to address the rebuttals.
        \item \textbf{Variance between arguments:} Each argument should be sufficiently different from each other in the aspects of the problems that it addresses, the values and perspectives from which the arguments are founded, and the types of evidence that supports them.
    \end{enumerate}
    \item Engagement (individual). The instructor will determine each member's engagement with the activity and their contribution to the final outcome based on the discussion transcript.
\end{itemize}

\subsection{Argument grading rubrics and prompt}
\label{appendix:argument-rubrics}
\noindent \textbf{System prompt:} \texttt{For the debate topic: " + debateTopic + "the team that takes the stance of answering " + stance + " has written up three arguments to support their stance. Grade their arguments on two measures (separately, don't add the scores): 1. Argument strength: Each of the 3 arguments will be graded separately for this. The strength of the argument are based on its relevance to the topic (1 point if there is relevance, 0 otherwise), the logical connection between the evidence and the argument it supports (1 point is the connection is logical, 0 otherwise), and the student's demonstration of anticipation of rebuttals (1 point if provided, 0 otherwise) and how to address the rebuttals (1 point if provided, 0 otherwise). 2) Variance between arguments: Each argument should be sufficiently different from each other in the aspects of the problems that it addresses (2 point if all aspects of the problem is covered, 1 point if more than one aspects is covered, 0 if only one aspect is covered), the values or perspectives from which the arguments are founded (2 point if 3 different perspectives are covered, 1 point if only 2 different perspectives are covered, 0 if only one is covered), and the types of evidence that supports them (1 point if more than one type is used; examples of different types are real-life anecdotes, existing research, hypothetical situations, references to other sources such as religion, philosophy, law, and so on).}

The argument is then provided to the same thread as a \textbf{User prompt}.

Based on this prompt, the full score of argument strength is 12 and the argument variance score is 5. We scaled them to the full score of 10 each and averaged them to get a final score out of 10.
\section{Debate topics}
\label{appendix:debate-topics}

\begin{itemize}
    \item Should robots/AI have rights?
    \item Is AI a threat to humanity?
    \item Should potential employers consider an applicant’s social media during a job application?
    \item Can good intentions exonerate one from bad outcomes?
    \item Are we happier now as a society than in times past?
    \item Is stealing ever permissible?
    \item Can we separate art from the artist? (Only used in the demo video of the experiment interface.)
\end{itemize}
\section{Questionnaires}
\label{appendix:questionnaires}

See the citations to the original questionnaires and details of the question adaptation for the context of this study in Section \ref{subsec:method-measure}.

\subsection{Engagement and Motivation}
Rate the following statements from Strongly Disagree (1) to Strongly Agree (7):
\begin{enumerate}
    \item (Behavioral) I concentrated during the activity.
    \item (Behavioral) I was persistent during the activity.
    \item (Behavioral) I actively participated in the activity.
    \item (Emotional) I liked the activity.
    \item (Emotional) I felt interested in the subject matter.
    \item (Cognitive) While chatting with the peer agent, I put together ideas or concepts and drew conclusions that were not directly stated in the debate prompt.
    \item (Cognitive) I tried to learn new ideas from the peer agent by mentally associating or contrasting them with relevant ideas from the debate prompt.
    \item (Cognitive) While chatting with the peer agent, I evaluated the strength of its arguments.
    \item (Intrinsic motivation) I do the activity because I think the activity is interesting.
    \item (Amotivation) There may be good reasons to do this activity, but personally I don’t see any.
\end{enumerate}

\subsection{Alignment to Constructive Controversy Principles}
Rate the following statements from Strongly Disagree (1) to Strongly Agree (7):
\begin{enumerate}
    \item (Shared goal) I feel that the peer agent and I shared the same goal during the activity.
    \item (Shared goal) I feel that the peer agent and I were cooperating during the activity.
    \item (Shared goal) I feel that the peer agent was working against me during the activity.
    \item (Skilled disagreement) The peer agent is open-minded.
    \item (Skilled disagreement) The peer agent tries to understand both sides of the issue.
    \item (Skilled disagreement) When the peer agent disagrees with me, they critique my ideas, not my character.
    \item (Skilled disagreement) The peer agent participated actively.
    \item (Rationality) The peer agent provides evidence to support their stance.
    \item (Rationality) The peer agent is rational.
    \item (Rationality) The peer agent changes their mind when evidence indicates that they should do so.
\end{enumerate}

\subsection{The Peer Agent’s Ability to Discover and Manage Information}
Rate the following statements from Strongly Disagree (1) to Strongly Agree (7):
\begin{enumerate}
    \item The peer agent helps me explore diverse perspectives on the topic.
    \item The peer agent helps me compare different perspectives on the topic.
    \item The peer agent helps me dive deeper into the topic.
    \item The peer agent helps me identify key criteria in evaluating the arguments.
    \item The peer agent helps me organize the arguments into an outline.
\end{enumerate}

\subsection{Agency and Ownership}
Rate the following statements from Strongly Disagree (1) to Strongly Agree (7):
\begin{enumerate}
    \item I felt in control of the experience
    \item I have a strong sense of ownership of the outcome
    \item I’m proud of the final outcome.
    \item I was able to express my personal perspectives through the outcome.
\end{enumerate}


\subsection{Post-study questionnaire}
\label{appendix:post-study-questionnaire}
\begin{enumerate}
    \item Please rate how much you expected the peer agents to do the following during the task (1 = Did not expect at all; 7 = Expected very much):
    \begin{itemize}
        \item Provide arguments to support your stance
        \item Provide evidence to support the arguments you came up with
        \item Write up your discussion into a formal argument
        \item Challenge your arguments or stance
        \item Keep the discussion on-track
    \end{itemize}
    \item Please rate how important these values are to you during the task (1 = Not important at all; 7 = Very important):
    \begin{itemize}
        \item Efficiency/speed: finishing the task as quickly as possible
        \item Quality of the arguments: make sure the arguments are as strong as possible
        \item Engagement: discussing with the peer agent as much as possible
        \item Curiosity: exploring the topic as extensively as possible
        \item Enjoyment: having as enjoyable time as possible
    \end{itemize}
    \item (For this question, feel free to open the exported files from the activities to remind yourself of your discussion and argument write-ups.) \newline
    For each of the following statements, pick between Taylor (c0) and Riley (c2) \footnote{c0 and c2 are the suffices of the names of the exported files from the activities.}:
    \begin{itemize}
        \item Which peer agents did you enjoy collaborating with the most?
        \item With which peer agent did you contribute the most to the discussion?
        \item With which peer agent did you learn the most about the debate topic?
        \item Which peer agent do you prefer the most? \newline Please explain: \underline{\hspace{8cm}}
    \end{itemize}
    \item What are some pros and cons of working with a peer AI agent on a collaboration task? \newline
    \textbf{Answer:} \underline{\hspace{10cm}} \\ \underline{\hspace{11.7cm}} \\
    \underline{\hspace{11.7cm}}
    \item How was the experience working with peer AI agents similar or different from working with human peers on learning tasks? Think about both collaborations with people who get along with you and people who may have different opinions or values from you. \newline
    \textbf{Answer:} \underline{\hspace{10cm}} \\ \underline{\hspace{11.7cm}} \\
    \underline{\hspace{11.7cm}}
    \item What learning tasks do you think the peer agents would be helpful for and why? (For example, tasks in certain fields/subjects, time-pressured tasks, creative tasks, and so on) \newline
    \textbf{Answer:} \underline{\hspace{10cm}} \\ \underline{\hspace{11.7cm}} \\
    \underline{\hspace{11.7cm}}
\end{enumerate}
\newpage
\section{Full results}
\label{appendix:stats}

This appendix contains the results of all statistical tests as well as box plots of other significant statistical results that are not shown in Section \ref{section:findings}.

\subsection{G-eval dialogue analysis results}
\begin{figure}[!ht]
    \centering
    \includegraphics[trim={5cm 6cm 5cm 5cm}, clip, width=.7\linewidth]{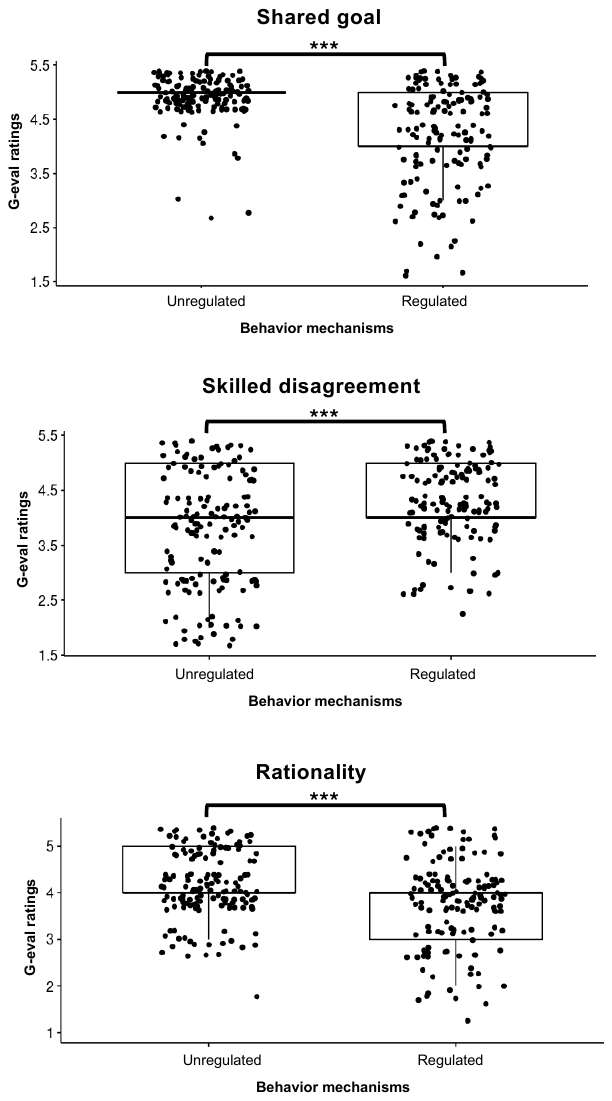}
    \caption{Box plots showing the G-eval results of unregulated vs. regulated chatbots on the three principles of CC The statistically significant comparisons are marked with asterisks (***: $p<.001$).}
    \Description{For shared goals, almost all dialogues from the unregulated agent score 5, while the regulated agent's dialogues scatter across the range. For Skilled Disagreement, however, the ratings for both behavior mechanisms spread out, though the regulated agent has a significantly higher average rating. The rationality rating looks similar to shared goal, though the unregulated agent's ratings spread out a bit more in the range from 3 to 5.}
    \label{fig:geval}
\end{figure}

The results in this table corresponds to the box plot in Figure \ref{fig:geval}.
\begin{table*}[htbp]
    \centering
    \begin{tabular}{|m{3.3cm}|m{2.7cm}|m{2.7cm}|m{5.4cm}|} 
      \hline
      \textbf{CC principle} & Unregulated agent \newline mean rating (/5) & Regulated agent \newline mean rating (/5) & Wilcoxon-signed rank result \\
      \hline
      \textbf{Shared goal} & $4.90 (\pm 0.37$) & $4.20 (\pm 0.88)$ & Z = 8.3638, p-value < 2.2e-16, $\eta^2 = 0.493$\\
      \hline
      \textbf{Skilled disagreement} & $3.75 (\pm 0.99)$ & $4.31 (\pm 0.69)$ & Z = -4.899, p-value = 9.635e-07, $\eta^2 = 0.289$\\
      \hline
      \textbf{Rationality} & $4.16 (\pm 0.67)$ & $3.77 (\pm 0.87)$ & Z = 3.7866, p-value = 0.0001527, $\eta^2 = 0.223$\\
      \hline
    \end{tabular}
    \caption{The average ratings out of 5 by G-eval techniques on the dialogues with the agents of two behavior mechanisms on the three constructs of CC. The Wilcoxon-signed rank results show that the results are significantly different in all three constructs with the positive Z-value indicating that the unregulated agent outperforms the regulated agent and the negative Z-value for the opposite.}
    \Description{The differences in mean ratings are significant across all three constructs. The difference with the biggest effect size of 0.493 for shared goal, where the unregulated agent gets an average of 4.90 and the regulated agent gets 4.20. The regulated agent outperforms the unregulated agent in Skilled Disagreement with the average rating os 4.31 to 3.75, resulting in an effect size of 0.289. For rationality, the unregulated agent scores an average rating of 4.16 to the regulated agent's 3.77, resulting in an effect size of 0.223.}
\end{table*}

\subsection{Unified mixed-effect model table}
These are all results of the statistical tests on all the measurements we observed in this study. All tests are run with linear mixed models. To interpret the $\beta_1$ value: for Group (Opaque vs. Transparency), a positive $\beta_1$ means Transparency has a higher mean than Opaque and a negative $\beta_1$ the reverse. For Condition (Unregulated vs. Regulated), a positive $\beta_1$ means the regulated agent has a higher mean than the unregulated one and a negative $\beta_1$ the reverse. All reported p-values are already adjusted by the Bonferroni method for multiple comparisons. The effect size ($\eta^2$) is only calculated for results with adjusted p-values < 0.1.

\begin{table*}[htbp]
\centering
\begin{tabular}{|m{2.9cm}|m{3.6cm}|m{3.6cm}|m{3.6cm}|} 
  \hline
  \textbf{Measures} & \textbf{EDL} & \textbf{CDL} & \textbf{All} \\ 
  \hline
  \textbf{Turn count} & Group: $\beta_1$ = 1.319, p = 0.4317  \newline Condition: $\beta_1$ = 4.457, p = 1.0076e-03 ** ($\eta^2$ = 0.10) \newline Interaction: $\beta_1$ = -3.374, p = 0.1567 & Group: $\beta_1$ = -1.2716, p = 1.000 \newline Condition: $\beta_1$ = 3.9154, 2.325e-02 * ($\eta^2$ = 0.25) \newline Interaction: $\beta_1$ = -3.1714, p = 0.511 & Group: $\beta_1$ = 1.290, p = 1.000 \newline Condition: $\beta_1$ = 4.220, p = 2.3128e-03 ** ($\eta^2$ = 0.15) \newline Interaction: $\beta_1$ = -3.171, p = 0.5113 \\ 
  \hline
  \textbf{Task time} & Group: $\beta_1$ = 86.450, p = 0.4189 \newline Condition: $\beta_1$ = 238.504, p = 3.476e-03 ** ($\eta^2$ = 0.12) \newline Interaction: $\beta_1$ = -141.002, p = 0.408  & Group: $\beta_1$ = -34.790, p = 0.843 \newline Condition: $\beta_1$ = 506.658, p = 5.395e-03 ** ($\eta^2$ = 0.19) \newline Interaction: $\beta_1$ = -256.829, p = 0.381 & Group: $\beta_1$ = 83.195, p = 1.000 \newline Condition: $\beta_1$ = 243.761, p = 6.726e-02 . ($\eta^2$ = 0.16) \newline Interaction: $\beta_1$ = -380.963, p = 0.016 ($\eta^2$ = 0.08)\\ 
  \hline
  \textbf{Average word per turn} & Group: $\beta_1$ = -2.349, p = 0.820 \newline Condition: $\beta_1$ = 1.133, p =  0.820\newline Interaction: $\beta_1$ = 4.917, p = 0.109 & Group: $\beta_1$ = 7.893, p = 1.000\newline Condition: $\beta_1$ = 2.171, p = 1.000 \newline Interaction: $\beta_1$ = 5.659, p = 1.000 & Group: $\beta_1$ = -2.4729, p = 1.000\newline Condition: $\beta_1$ = 1.1601, p = 1.000 \newline Interaction: $\beta_1$ = 5.1834, p = 1.000\\ 
  \hline
  \textbf{Argument strength} & Group: $\beta_1$ = -1.4505, p = 1.849e-02 * ($\eta^2 = 0.08$)\newline Condition: $\beta_1$ = -0.6940, p = 8.468 . ($\eta^2$ = 0.02) \newline Interaction: $\beta_1$ = 0.7326, p = 0.178 & Group: $\beta_1$ = 0.0222, p = 1.000\newline Condition: $\beta_1$ = 0.2515, p = 1.000 \newline Interaction: $\beta_1$ = -0.0618, p = 1.000 & Group: $\beta_1$ = -1.3055, p = 0.130\newline Condition: $\beta_1$ = -0.6142, p = 0.273 \newline Interaction: $\beta_1$ = 0.4867, p = 0.596\\ 
  \hline
  \textbf{Argument variance} & Group: $\beta_1$ = -0.2173, p = 0.931\newline Condition: $\beta_1$ = -0.7965, p = 1.835e-03 ** ($\eta^2$ = 0.20) \newline Interaction: $\beta_1$ = 0.0888, p = 0.931 & Group: $\beta_1$ = 0.3533, p = 0.539\newline Condition: $\beta_1$ = 0.1048, p = 0.633 \newline Interaction: $\beta_1$ = -0.6261, p = 8.604e-02 . ($\eta^2$ = 0.07) & Group: $\beta_1$ = -0.1972, p = 1.000 \newline Condition: $\beta_1$ = -0.7826, p = 7.644e-04 *** ($\eta^2 = 0.11$) \newline Interaction: $\beta_1$ = 0.0466, p = 1.000\\ 
  \hline
  \textbf{\emph{Too Cooperative} annotation counts (percentage)} & Group: $\beta_1$ = -3.651, p = 0.824\newline Condition: $\beta_1$ = -9.368, p = 6.983e-05 *** ($\eta^2$ = 0.28) \newline Interaction: $\beta_1$ = 1.222, p = 0.824 & Group: $\beta_1$ = -2.889, p = 1.000\newline Condition: $\beta_1$ = -6.293, p = 5.684-e02 . ($\eta^2 = 0.20$) \newline Interaction: $\beta_1$ = -1.084, p = 1.000 & Group: $\beta_1$ = -3.6514, p = 1.000\newline Condition: $\beta_1$ = -9.3683, p = 5.821e-05 *** ($\eta^2$ = 0.24) \newline Interaction: $\beta_1$ = 1.2218, p = 1.000\\ 
  \hline
  \textbf{\emph{Too Contradictory} annotation counts (percentage)} & Group: $\beta_1$ = 0.3594, p = 0.863\newline Condition: $\beta_1$ = 11.3471, p = 6.315e-09 *** ($\eta^2$ = 0.43) \newline Interaction: $\beta_1$ = -3.9574, p = 0.406 & Group: $\beta_1$ = 0.3156, p = 1.000 \newline Condition: $\beta_1$ = 7.9086, p = 8.722e-04 *** ($\eta^2 = 0.30$) \newline Interaction: $\beta_1$ = -2.2817, p = 1.000 & Group: $\beta_1$ = 0.4810, p = 1.000 \newline Condition: $\beta_1$ = 11.3498, p = 4.234e-10 *** ($\eta^2$ = 0.35) \newline Interaction: $\beta_1$ = -4.0417, p = 0.982\\ 
  \hline
\end{tabular}
\caption{Table showing the $\beta_1$, p-value, and for significant p-value, $\eta^2$ effect size for the linear mixed-effect models run on all task measures, including the observed measures of engagement, argument score, and annotation counts.}
\Description{Each row of the table corresponds to each measurement from the experimental task. Each column corresponds to different types of learners. In the leftmost cell of each row is the name of the measurement. The significant results in this table are: the turn count by behavior mechanisms (for EDLs and CDLs), task time by behavior mechanisms (for EDLs and CDLs), argument strength by disclosure and by behavior mechanisms separately (for EDLs), argument variance by behavior mechanisms (for EDLs), and the Too-Cooperative and Too-Contradictory annotations by behavior mechanisms (for EDLs and CDLs).}
\label{table:stat-task}
\end{table*}

\begin{table*}[htbp]
\centering
\begin{tabular}{|m{2.9cm}|m{3.6cm}|m{3.6cm}|m{3.6cm}|} 
  \hline
  \textbf{Measures} & \textbf{EDL} & \textbf{CDL} & \textbf{All} \\ 
  \hline
  \textbf{Behavioral engagement} & Group: $\beta_1$ = -0.0690, p = 1.000\newline Condition: $\beta_1$ = -0.1281, p = 0.478 \newline Interaction: $\beta_1$ = 0.0617, p = 1.000 & Group: $\beta_1$ = -0.1313, p = 0.910 \newline Condition: $\beta_1$ = 0.0385, p = 1.000 \newline Interaction: $\beta_1$ = -0.0305, p = 1.000 & Group: $\beta_1$ = -0.0448, p = 1.000\newline Condition: $\beta_1$ = -0.1004, p = 1.000 \newline Interaction: $\beta_1$ = -0.0046, p = 1.000\\
  \hline
  \textbf{Emotional engagement} & Group: $\beta_1$ = -0.3748, p = 0.184\newline Condition: $\beta_1$ = -0.9260, p = 7.570e-08 *** ($\eta^2$ = 0.32) \newline Interaction: $\beta_1$ = 0.4750, p =9.830e-02 . ($\eta^2 = 0.05$) & Group: $\beta_1$ = -0.4277, p = 0.156 \newline Condition: $\beta_1$ = -0.0385, p = 1.000 \newline Interaction: $\beta_1$ = 0.0504, p = 1.000 & Group: $\beta_1$ = -0.3713, p = 0.542\newline Condition: $\beta_1$ = -0.9239, p = 4.999e-08 *** ($\eta^2$ = 0.11) \newline Interaction: $\beta_1$ = 0.4512, p = 0.304\\ 
  \hline
  \textbf{Cognitive engagement} & Group: $\beta_1$ = -0.2453, p = 0.577\newline Condition: $\beta_1$ = -0.4379, p = 3.232e-03 ** ($\eta^2 = 0.13$) \newline Interaction: $\beta_1$ = 0.2192, p = 0.577 & Group: $\beta_1$ = -0.5103, p = 0.129 \newline Condition: $\beta_1$ = 0.1595, p = 0.421 \newline Interaction: $\beta_1$ = 0.3626, p = 0.310 & Group: $\beta_1$ = -0.2464, p = 1.000\newline Condition: $\beta_1$ = -0.4348, p = 1.367e-02 * ($\eta^2=$6.43e-05) \newline Interaction: $\beta_1$ = 0.2126, p = 1.000\\ 
  \hline
  \textbf{Intrinsic motivation} & Group: $\beta_1$ = -0.3214, p = 0.364\newline Condition: $\beta_1$ = -0.4809, p = 6.519e-03 ** ($\eta^2 = 0.06$) \newline Interaction: $\beta_1$ = 0.4503, p = 0.133 & Group: $\beta_1$ = -0.6034, p = 0.156 \newline Condition: $\beta_1$ = 0.0181, p = 0.921 \newline Interaction: $\beta_1$ = 0.3323, p = 0.326 & Group: $\beta_1$ = -0.2989, p = 1.000\newline Condition: $\beta_1$ = -0.4657, p = 1.202e-02 * ($\eta^2=$1.55e-03) \newline Interaction: $\beta_1$ = 0.4055, p = 0.414\\ 
  \hline
  \textbf{Amotivation} & Group: $\beta_1$ = 0.0406, p = 1.000 \newline Condition: $\beta_1$ = 0.3261, p = 0.546 \newline Interaction: $\beta_1$ = 0.2406, p = 1.000 & Group: $\beta_1$ = 0.6389, p = 0.274 \newline Condition: $\beta_1$ = 0.0.0220, p = 1.000 \newline Interaction: $\beta_1$ = 0.2406, p = 1.000 & Group: $\beta_1$ = 0.0406, p = 1.000 \newline Condition: $\beta_1$ = 0.3261, p = 0.870\newline Interaction: $\beta_1$ = 0.2406, p = 1.000 \\ 
  \hline
\end{tabular}
\caption{Table showing the $\beta_1$, p-value, and for significant p-value, $\eta^2$ effect size for the linear mixed-effect models run on self-reported engagement and motivation.}
\Description{Each row of the table corresponds to each measurement from the engagement and motivation questionnaire. Each column corresponds to different types of learners. In the leftmost cell of each row is the name of the measurement. The significant results in this table are: the emotional engagement, cognitive engagement, and intrinsic motivation by behavior mechanisms for EDLs only.}
\label{table:stat-EM}
\end{table*}

\begin{table*}[htbp]
\centering
\begin{tabular}{|m{2.9cm}|m{3.6cm}|m{3.6cm}|m{3.6cm}|} 
  \hline
  \textbf{Measures} & \textbf{EDL} & \textbf{CDL} & \textbf{All} \\ 
  \hline
  \textbf{Shared goal} & Group: $\beta_1$ = -0.4511, p = 0.132\newline Condition: $\beta_1$ = -3.1521, p = 8.425e-23 *** ($\eta^2$ = 0.54) \newline Interaction: $\beta_1$ = 0.8559, p = 8.878e-02 ($\eta^2 = 0.03$) & Group: $\beta_1$ = -0.3074, p = 0.300\newline Condition: $\beta_1$ = -1.8501, p = 1.3648e-07 *** ($\eta^2 = 0.48$) \newline Interaction: $\beta_1$ = 0.8228, p = 0.6.823e-02 ($\eta^2=0.08$) & Group: $\beta_1$ = -0.4536, p = 0.462 \newline Condition: $\beta_1$ = -0.1438, p = 1.483e-23 ($\eta^2$ = 0.59) \newline Interaction: $\beta_1$ = 0.8481, p = 0.183 \\
  \hline
  \textbf{Skilled disagreement} & Group: $\beta_1$ = -0.0880, p = 0.687\newline Condition: $\beta_1$ = -0.6250, p = 1.348e-04 *** ($\eta^2$ = 0.13) \newline Interaction: $\beta_1$ = 0.4917, p = 7.069e-02 . ($\eta^2 = 0.06$) & Group: $\beta_1$ = -0.0030, p = 1.000 \newline Condition: $\beta_1$ = 0.8576, p = 2.352e-04 *** ($\eta^2$ = 0.38) \newline Interaction: $\beta_1$ = -0.0801, p = 0.1.000 & Group: $\beta_1$ = -0.0996, p = 1.000\newline Condition: $\beta_1$ = -0.6250, p = 3.075e-04 ($\eta^2 = 0.05$) \newline Interaction: $\beta_1$ = 0.5147, p = 0.1303\\ 
  \hline
  \textbf{Rationality} & Group: $\beta_1$ = -0.3527, p = 0.1971\newline Condition: $\beta_1$ = -1.2174, p = 1.1914e-07 *** ($\eta^2$ = 0.28) \newline Interaction: $\beta_1$ = 0.7348, p = e.616e-02 * ($\eta^2 = 0.07$) & Group: $\beta_1$ = 4.019e-03, p = 1.000\newline Condition: $\beta_1$ = 0.2508-e01, p = 0.884 \newline Interaction: $\beta_1$ = 6.675e-02, p = 1.000 & Group: $\beta_1$ = -0.3608, p = 0.402\newline Condition: $\beta_1$ = -1.2167, p = 1.562-e08 ($\eta^2$ = 0.04) \newline Interaction: $\beta_1$ = 1.5048, p = 6.862e-02 ($\eta^2 = 0.16$)\\ 
  \hline
\end{tabular}
\caption{Table showing the $\beta_1$, p-value, and for significant p-value, $\eta^2$ effect size for the linear mixed-effect models run on the learners' evaluation of the agent's alignment to CC principles.}
\Description{Each row of the table corresponds to each measurement from the alignment to CC questionnaire. Each column corresponds to different types of learners. In the leftmost cell of each row is the name of the measurement. The significant results in this table are: all three constructs for EDLs; and shared goal and skilled disagreement for CDLs.}
\label{table:stat-CC}
\end{table*}

\begin{table*}[htbp]
\centering
\begin{tabular}{|m{2.9cm}|m{3.6cm}|m{3.6cm}|m{3.6cm}|} 
  \hline
  \textbf{Measures} & \textbf{EDL} & \textbf{CDL} & \textbf{All} \\ 
  \hline
  \textbf{Feeling in control of the experience} & Group: $\beta_1$ = -0.1200, p = 0.742\newline Condition: $\beta_1$ = -1.9565, p = 9.297e-09 *** ($\eta^2$ = 0.44) \newline Interaction: $\beta_1$ = 0.4125, p = 0.742 & Group: \newline $\beta_1$ = -0.4177, p = 0.388\newline Condition: $\beta_1$ = -0.8724, p = 4.895e-02 * ($\eta^2$ = 0.09) \newline Interaction: $\beta_1$ = 0.5954, p = 0.388 & Group: $\beta_1$ =  -0.1377, p = 1.000\newline Condition: $\beta_1$ = -1.9565, p = 5.165e-10 *** ($\eta^2$ = 0.07) \newline Interaction: $\beta_1$ = 0.4480, p = 1.000\\
  \hline
  \textbf{Have a strong sense of ownership of the outcome} & Group: $\beta_1$ = -0.0609, p = 0.848\newline Condition: $\beta_1$ = -1.3696, p = 3.106e-06 *** ($\eta^2$ = 0.26) \newline Interaction: $\beta_1$ = 0.6696, p = 0.212 & Group: $\beta_1$ = -0.5018, p = 0.108\newline Condition: $\beta_1$ = -0.8077, p = 3.036e-02 * ($\eta^2=0.05$) \newline Interaction: $\beta_1$ = 0.9029, p = 4.615e-02 * ($\eta^2 = 0.08$) & Group: $\beta_1$ = -0.0609, p = 1.000\newline Condition: $\beta_1$ = -1.3696, p = 7.531e-07 *** ($\eta^2$ = 0.15) \newline Interaction: $\beta_1$ = 0.6696, p = 0.524\\ 
  \hline
  \textbf{Feeling proud of the final outcome} & Group: $\beta_1$ = -0.4252, p = 0.374 \newline Condition: $\beta_1$ = -1.3912, p = 1.441e-06 *** ($\eta^2$ = 0.37) \newline Interaction: $\beta_1$ = 0.1588, p = 0.695 & Group: $\beta_1$ = -0.2912, p = 0.976 \newline Condition: $\beta_1$ = 0.000, p = 1.000 \newline Interaction: $\beta_1$ = -0.1190, p = 1.000 & Group: $\beta_1$ = -0.3913, p = 0.980 \newline Condition: $\beta_1$ = -1.3478, p = 1.130e-07 *** ($\eta^2$ = 0.17) \newline Interaction: $\beta_1$ = 0.0478, p = 1.000\\ 
  \hline
  \textbf{Able to express personal perspectives through the outcome} & Group: $\beta_1$ = -0.4833, p = 0.105\newline Condition: $\beta_1$ = -1.1904, p = 5.832e-06 *** ($\eta^2=0.19$) \newline Interaction: $\beta_1$ = 0.9271, p = 2.603e-02 * ($\eta^2=0.08$) & Group: $\beta_1$ = -0.0993, p = 0.805\newline Condition: $\beta_1$ = -0.2278, p = 0.512 \newline Interaction: $\beta_1$ = -0.1766, p = 0.805 & Group: $\beta_1$ = -0.4938, p = 0.203\newline Condition: $\beta_1$ = -1.1842, p = 2.518e-08 ($\eta^2 = 0.10$) \newline Interaction: $\beta_1$ = 0.8982, p = 1.586e-02 * ($\eta^2 = 0.03$)\\ 
  \hline
\end{tabular}
\caption{Table showing the $\beta_1$, p-value, and for significant p-value, $\eta^2$ effect size for the linear mixed-effect models run on self-reported learner's sense of agency and ownership over the outcome.}
\Description{Each row of the table corresponds to each measurement from the agency and ownership questionnaire. Each column corresponds to different types of learners. In the leftmost cell of each row is the name of the measurement. We separate agency and ownership into four measures. For EDLs, there's a significant difference (with unregulated agent scoring higher than the regulated agent) on all four (feeling in control of the experience, have a strong sense of ownership over the outcome, feeling proud of the final outcome, able to present personal perspectives through the outcome). For CDLs, there's a significant difference (in the same direction but smaller effect size) for feeling in control of the experience.}
\label{table:stat-AO}
\end{table*}

\begin{table*}[htbp]
\centering
\begin{tabular}{|m{2.9cm}|m{3.6cm}|m{3.6cm}|m{3.6cm}|} 
  \hline
  \textbf{Measures} & \textbf{EDL} & \textbf{CDL} & \textbf{All} \\ 
  \hline
  \textbf{Explore diverse perspectives} & Group: $\beta_1$ = -0.0362, p = 1.000 \newline Condition: $\beta_1$ = -0.8261, p = 1.662e-02 * ($\eta^2$ = 0.13) \newline Interaction: $\beta_1$ = 0.0928, p = 1.000 & Group: $\beta_1$ = -0.7755, p = 4.252e-02 * ($\eta^2$ = 0.09)\newline Condition: $\beta_1$ = 0.5574, p = 0.166\newline Interaction: $\beta_1$ = 0.3478, p = 0.393 & Group: $\beta_1$ = -0.0362, p = 1.000 \newline Condition: $\beta_1$ = -0.8261, p = 1.645e-02 ($\eta^2 = 1.59e-04$) \newline Interaction: $\beta_1$ = 0.0928, p = 1.000\\
  \hline
  \textbf{Compare different perspectives} & Group: $\beta_1$ = -0.6841, p = 0.1656 \newline Condition: $\beta_1$ = -0.3261, p = 0.527 \newline Interaction: $\beta_1$ = 0.5261, p = 0.527 & Group: $\beta_1$ = -0.4589, p = 0.347\newline Condition: $\beta_1$ = 1.2805, p = 2.502e-03 ** ($\eta^2$ = 0.34)\newline Interaction: $\beta_1$ = 0.1340, p = 0.776 & Group: $\beta_1$ = -0.6746, p = 0.278 \newline Condition: $\beta_1$ = -0.3277, p = 1.000 \newline Interaction: $\beta_1$ = 0.5141, p = 1.000\\ 
  \hline
  \textbf{Dive deeper into the topic} & Group: $\beta_1$ = -0.4030, p = 0.494 \newline Condition:$\beta_1$ = -0.7616, p = 5.922e-03 ** ($\eta^2=0.13$) \newline Interaction: $\beta_1$ = 0.2827, p = 0.494 & Group: $\beta_1$ = -0.7039, p = 0.104 \newline Condition: $\beta_1$ = 0.4983, p = 0.230 \newline Interaction: $\beta_1$ = 0.2708, p = 0.501 & Group: $\beta_1$ = -0.4222, p = 1.000 \newline Condition: $\beta_1$ = -0.7609, p = 1.158e-02 * ($\eta^2=5.67e-05$) \newline Interaction: $\beta_1$ = 0.3009, p = 1.000\\ 
  \hline
  \textbf{Identify key criteria in evaluating the arguments} & Group: $\beta_1$ = -0.6342, p = 6.598e-02 . ($\eta^2$ = 5.79e-04)\newline Condition: $\beta_1$ = -1.2798, p = 2/437e-05 *** ($\eta^2$ = 0.13) \newline Interaction: $\beta_1$ = 1.1578, p = 1.656e-02 * ($\eta^2=0.09$) & Group: $\beta_1$ = -0.5311, p = 2.253e-01 \newline Condition: $\beta_1$ = -0.0199, p = 1.000 \newline Interaction: $\beta_1$ = -.1504, p = 1.000 & Group: $\beta_1$ = -0.6671, p = 0.135 \newline Condition: $\beta_1$ = -1.2825, p = 1.052e-06 ($\eta^2=0.04$) \newline Interaction: $\beta_1$ = 1.1955, p = 8.784e-03 ($\eta^2 = 0.05$)\\ 
  \hline
  \textbf{Organize the arguments} & Group: $\beta_1$ = -0.124, p = 1.000\newline Condition: $\beta_1$ = -1.3963, p = 1.409e-04 *** ($\eta^2$ = 0.29) \newline Interaction: $\beta_1$ = -0.0491, p = 1.000 & Group: $\beta_1$ = -0.3324, p = 0.767 \newline Condition: $\beta_1$ = -0.6704, p = 0.203 \newline Interaction: $\beta_1$ = 0.1638, p = 0.767 & Group: $\beta_1$ = -0.1029, p = 1.000\newline Condition: $\beta_1$ = -1.3613, p = 8.813e-05 *** ($\eta^2$ = 0.19) \newline Interaction: $\beta_1$ = -0.0966, p = 1.000\\ 
  \hline
\end{tabular}
\caption{Table showing the $\beta_1$, p-value, and for significant p-value, $\eta^2$ effect size for the linear mixed-effect models run on the learners' perception of the peer agent's abilities to discover and manage information}
\Description{Each row of the table corresponds to each measurement from the questionnaire "the peer agent's ability to discover and manage information". Each column corresponds to different types of learners. In the leftmost cell of each row is the name of the measurement. The significant results are: for EDLs, all by behavior mechanisms, exploring diverse perspectives, dive deeper into the topic, identifying key criteria in evaluating the arguments, and organizing the arguments; for CDLs, there's a significant difference by behavior mechanisms in comparing different perspectives and significant differences by disclosure in exloring diverse perspectives.}
\label{table:stat-IM}
\end{table*}

\begin{figure}[!ht]
    \centering
    \includegraphics[trim={1cm 16.5cm 0cm 4cm}, clip, width=\linewidth]{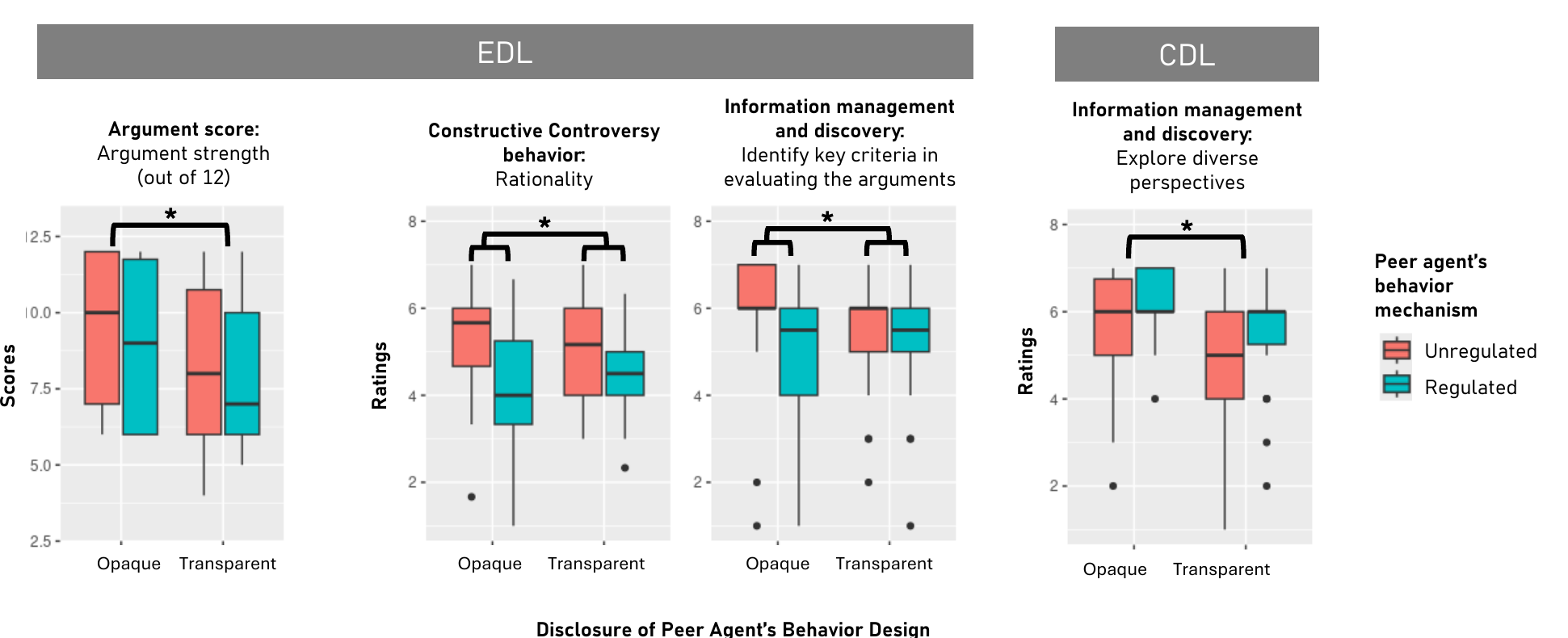}
    \caption{Box plots showing the four measures where levels of design disclosure (opaque vs. transparent) affect EDL (left) and CDL (right) participants. The colors of the box plots indicate the peer agent's behavior mechanisms. The statistically significant comparisons are marked with asterisks (.: $p<.1$, *: $p<.05$, ***: $p<.001$).The two-level brackets for the statistically significant comparison indicate that the interaction effect is significant, whereas the one-level bracket indicates the main effect of the design disclosure.}
    \Description{For EDLs, the significant differences occur in argument scores, rationality, and ability to identify key criteria in arguments. For CDL, the significant difference occurs in the ability to explore diverse perspectives. All p-values are less than 0.05 and the opaque condition having higher ratings than the transparent condition.}
    \label{fig:transparency}
\end{figure}
\end{document}